\documentclass[twocolumn]{aastex631}

\newcommand{\Xlan}{X$_{\rm lan}$}

\usepackage{amsfonts,amsmath,graphicx,natbib,subfigure,color,gensymb,longtable,verbatim,graphicx,float,soul,multirow}

\begin{document}

\title{Impact of Systematic Modeling Uncertainties on Kilonova Property Estimation}

\author[0000-0001-6415-0903]{D. Brethauer}
\affiliation{Department of Astronomy, University of California, Berkeley, CA 94720-3411, USA}

\author[0000-0002-5981-1022]{D. Kasen}
\affiliation{Department of Physics, University of California, 366 Physics North MC 7300, Berkeley, CA 94720, USA}

\author[0000-0003-4768-7586]{R. Margutti}
\affiliation{Department of Astronomy, University of California, Berkeley, CA 94720-3411, USA}
\affiliation{Department of Physics, University of California, 366 Physics North MC 7300, Berkeley, CA 94720, USA}

\author[0000-0002-7706-5668]{R. Chornock}
\affiliation{Department of Astronomy, University of California, Berkeley, CA 94720-3411, USA}



\begin{abstract}

The precise atomic structure and therefore the wavelength-dependent opacities of lanthanides are highly uncertain. This uncertainty introduces systematic errors in modeling transients like kilonovae and estimating key properties such as mass, characteristic velocity, and heavy metal content. Here, we quantify how atomic data from across the literature as well as choices of thermalization efficiency of r-process radioactive decay heating impact the light curve and spectra of kilonovae. Specifically, we analyze the spectra of a grid of models produced by the radiative transfer code \texttt{Sedona} that span the expected range of kilonova properties to identify regions with the highest systematic uncertainty. Our findings indicate that differences in atomic data have a substantial impact on estimates of lanthanide mass fraction, spanning approximately one order of magnitude for lanthanide-rich ejecta, and demonstrate the difficulty in precisely measuring the lanthanide fraction in lanthanide-poor ejecta. Mass estimates vary typically by 25-40$\%$ for differing atomic data. Similarly, the choice of thermalization efficiency can affect mass estimates by 20$\%$ to 50$\%$. Observational properties such as color and decay rate are \textit{highly} model-dependent. Velocity estimation, when fitting solely based on the light curve, can have a typical error of $\sim 100\%$. Atomic data of light r-process elements can strongly affect blue emission. Even for well-observed events like GW170817, the total lanthanide production estimated using different atomic datasets can vary by a factor of $\sim6$. 

\end{abstract}

\keywords{Kilonovae ---}


\section{Introduction} \label{sec:intro}

Compact-object mergers involving a neutron star (NS) have long been at the forefront of high-energy phenomena, from short gamma ray bursts (e.g., \citealt{Eichler89,Narayan92}) to sites of \textit{rapid neutron capture} (r-process) nucleosynthesis \citep{Lattimer&Schramm74,Lattimer&Schramm76,Symbalisty&Schramm82,Eichler89,Freiburghaus99,Rosswog99}. The discovery of gravitational wave source GW170817 and its electromagnetic counterparts AT2017gfo, GRB170817A, and the resulting afterglow \citep{Abbott17,Goldstein17,Savchenko17,Coulter17,Arcavi17,Lipunov17,Soares-Santos17,Tanvir17,Valenti17,Alexander17,Evans17,Hallinan17,Margutti17,Sugita18,Troja17,Chornock17,Andreoni17,Cowperthwaite17,Diaz17,Drout17,Hu17,Kasliwal17,Pian17,Shappee17,Smartt17,Utsumi17,Pozanenko18} confirmed the connection between compact objects and some short GRBs, ushering in a new era transient multi-messenger astronomy (see \citealt{Margutti&Chornock21} and \citealt{Nakar20} for detailed reviews). The multi-wavelength electromagnetic radiation from AT2017gfo was the kilonova (KN), which is mainly powered by the radioactive decay of r-process material, and lasted between $\sim$ days and $\sim$ weeks depending on the wavelength observed at (\citealt{Villar17} and references therein). AT2017gfo provided an unprecedented amount of data, and represents one realization of KN emission in the broader phase space of potential KN emission. Theory predicts and observations have shown a wide diversity of KN emission as a direct result of the three main properties of the ejecta launched during and after the merger event: the mass, characteristic velocity, and heavy metal content (typically denoted as the mass fraction of lanthanides, \Xlan, a product of r-process nucleosynthesis). 

Encoded within the KN spectra is information about the compact objects that created them and the fate of the remnant (e.g., \citealt{Kasen17,Radice20,Radice23}) as well as the evolution of the heavy metal content of the Universe (e.g., \citealt{Qian&Wasserburg07,Hotokezaka15,Wallner15,Ji16,Rosswog18}). In order to make meaningful progress on extracting this information, both high accuracy and high precision measurements are required. However, to construct simulations and models of KNe, astronomers are required to make numerous model-dependent assumptions concerning the heating rate from the decay of r-process material \citep{Lippuner15,Zhu21,Sarin24}, r-process decay products \citep{Zhu21}, the velocity profile of the ejecta (e.g., \citealt{Fryer24}), how efficiently the radioactive decay products of r-process material deposit their energy into the ejecta \citep{Barnes16,Rosswog17,Bulla24}, the multi-dimensional structure of the ejecta (e.g., \citealt{Kawaguchi18,Wollaeger21,Shingles23,Bulla24}), or the distribution of lanthanide-rich material within the ejecta as a function of radius \citep{Kasen17}.

Another such assumption regards how the products of r-process nucleosynthesis interact with the radiation of the KN. The presence of a high lanthanide fraction manifests as an infrared-bright transient, due to the extremely high opacity ($\gtrsim 10^2$ cm$^2$/g)  in the optical and UV from the valence f-shell electron of lanthanide elements (i.e., \citealt{Kasen13,Tanaka20,Fontes20}). Despite the importance of the lanthanides in creating the characteristic r-process infrared excess, the precise atomic structure of these elements is highly uncertain, resulting in highly uncertain opacities \citep{Kasen13}. This uncertainty introduces an additional systematic error source into modeling KNe that we quantify here by using three sets of atomic structure data produced from various codes (further explained in Section \ref{Subsec:AtomicData}).

Additionally, as radioactive isotopes generated from r-process decay, they release energy in the form of neutrinos, alpha particles, beta particles, gamma rays, and fission fragments. Each species deposits a varying fraction of its energy into the ejecta. To analytically describe this variable energy deposition, we use two different prescriptions: a global mass-averaged efficiency where all zones have the same thermalization efficiency (i.e., \citealt{Barnes16}), and a local density-based efficiency (i.e., \citealt{Barnes16,Bulla24}) where the thermalization efficiency is based on the local conditions in each zone (further explained in Section \ref{Subsec:therm}). Each approach leads to a different energy budget for the KN to emit, which alters the resulting spectra and similarly creates a systematic error source that we quantify here.

In Section \ref{sec:Sedona} we discuss the setup of our radiative transfer code simulations and the chosen range of ejecta parameters. In Section \ref{sec:models} we discuss the different atomic data we vary as well as the thermalization-prescriptions.  In Section \ref{Sec:LightCurve} we present the variance in the resulting light curves and spectra from the different atomic datasets and thermalization-prescriptions. Section \ref{Sec:Error} we quantify the error in parameter estimation based on atomic dataset and thermalization-prescription. Finally, we summarize our findings in Section \ref{Sec:Conc}.

\section{Sedona and Initial Conditions Setup} \label{sec:Sedona}

We use the Monte Carlo Radiative transfer code \texttt{Sedona} \citep{Kasen06,Roth15} to generate synthetic spectra, from which we derive light curves and bolometric luminosities. We explore a grid of models using M $\in$ [0.001, 0.01, 0.1] M$_\odot$, $v_k \in$ [0.1,0.3] c, and $\log_{10} (X_{\rm lan}) \in$ [-9, -4, -2] (where M is the total ejecta mass, $v_k$ is the characteristic velocity of the ejecta defined as $v_k = \sqrt{\frac{2E_k}{M}}$ and $E_k$ is the total kinetic energy, and c is the speed of light). The range of this grid is motivated by observational KN parameter estimates of AT2017gfo (e.g., \citealt{Villar17,Coughlin18,Ristic23}), GRMHD simulations of compact object merger events (e.g., \citealt{Radice20}), and nucleosynthetic yields of r-process ejecta (e.g., \citealt{Lippuner15}). We generate synthetic spectra for each combination of atomic data and thermalization efficiency prescription to explore the impact on a wide extent of expected theoretical KN emission. 

We run each simulation as a spherically symmetric model with 80 zones that are expanding homologously defined by a temperature, density, velocity, composition, and size. We initialize the model at time $t_0$= 0.25\,d after the merger, early enough that radiative diffusion will not yet have caused substantial energy loss, and \texttt{Sedona} homologously expands the initial conditions while dynamically evolving the properties of each zone (or contracts in the case of simulating earlier times). Homologous expansion sets in on a dynamical timescale which is $\approx$ 1 s, which is well before we initialize each simulation. The density of each zone is defined by a broken power law

\begin{equation}
        \rho (v) =
        \Biggl\{ \begin{array}{ll}
            \eta_\rho \dfrac{M}{v_t^3 t^3} \left(\dfrac{v}{v_t}\right)^{-\delta} & v \leq v_t \\
            \\
             \eta_\rho \dfrac{M}{v_t^3 t^3} \left(\dfrac{v}{v_t}\right)^{-n} & v > v_t, 
        \end{array} 
\end{equation}
where $\delta$ and $n$ are the power law index of the inner and outer ejecta, respectively, $v_t$ is the transition velocity between the two power law indices, $M$ is the total mass, $t$ is time, and $\eta_\rho$ is the normalization constant. Following \cite{Kasen17} we adopt $\delta = 1$ and $n = 10$ as the broken power-law density profile arises from disc models (e.g., the velocity distributions studied in \citealt{Fryer24}). The transitional velocity, $v_t$, is defined as

\begin{equation}
    v_t = \eta_v v_k = \eta_v \sqrt{\frac{2E_k}{M}}, 
\end{equation}

where $\eta_v$ is the normalization constant to ensure the ejecta has total kinetic energy $E_k$. The maximum velocity of each model is 3$v_t$ for 0.1c models and 2$v_t$ for 0.3c models, as after the transition velocity the steep $n = 10$ power law decline in density rapidly makes the outermost ejecta mass negligible.

The composition of each model is based on solar abundance patterns (or meteoric in cases where solar abundances are not available) presented in \cite{Asplund09}, and r-process residuals from \cite{Simmerer04} for elements with atomic number Z = 31-70. This differs from the models presented in \cite{Kasen17}, which use an even distribution of non-lanthanide material as opposed to solar abundance patterns. The composition is then normalized by mass fraction such that all elements of Z = 58-70 have a total mass fraction of \Xlan \, and all other elements have a total mass fraction of 1-\Xlan. We do not consider any elements of Z $\geq$ 71. 

The level populations of each element are determined by Local Thermodynamic Equilibrium (LTE). We expect non-LTE (NLTE) effects to begin affecting the spectra on a timescale by which the majority of the mass in the ejecta is optically thin as determined by

\begin{equation}
    t_{\rm NLTE} \approx 7.2  \Big(\frac{M_{ej}}{0.03\, M_\odot}\Big)^{0.5} \Big(\frac{\kappa}{1 \rm \, cm^2 \, g^{-1} }\Big)^{0.5} \Big(\frac{v_k}{0.1c}\Big)^{-1} \rm{d},
    \label{Eq:NLTE}
\end{equation}

where $M_{ej}$ is the total mass ejected, $\kappa$ is the opacity of the material, and $v_k$ is the characteristic velocity. Beyond this timescale, the spectra will become less accurate, though for the purposes of comparing atomic data can still be illuminating.

Each spectrum is calculated every 0.1\,d starting from t = 0.1\,d to 35\,d with 1524 logarithmically spaced frequency points between $10^{13}$ and $2 \times 10^{16}$ Hz. Following \cite{Kasen17}, we limit hydrodynamical time steps to 10\% of the elapsed time, which is sufficient to resolve the expansion evolution of the ejecta. However, we do not include the physics of free neutron decay or shock breakout in order to isolate the effects of atomic structure and thermalization efficiency, and so caution about the use of these models at times t $\lesssim$ 0.5\,d.

At each time step, Monte Carlo packets (effectively bundles of photons of a given wavelength that total up to a specified energy amount) are released in accordance to the r-process heating rate convolved with the thermalization efficiency and interact with a zone through scattering and absorption. Monte Carlo photons that reach the outer edge of the simulation (defined by the maximum velocity) escape the ejecta and are collected and binned in time and frequency to generate the spectral time series of the model, with all relevant Doppler shift and light travel-time effects taken into account for an observer infinitely far away.

\section{Model Experimental Variables} \label{sec:models}

We consider three sets of atomic data and two thermalization efficiency prescriptions for a total of six combinations. While bound-bound transitions from lanthanide species dominate the opacity, we still consider free-free and electron scattering opacities in each model as they become more important at longer wavelengths where lanthanides no longer dominate as strongly and are simple to include from their analytic formula. We do not consider bound-free absorption due to the dominance of bound-bound absorption from lanthanides (i.e., $\kappa_{bf, Nd} = 0$ at $\lesssim 15$ eV while $\kappa_{bb, Nd} \approx 50$ cm$^2$ g$^{-1}$ at line locations).

\subsection{Atomic Data} \label{Subsec:AtomicData}

We consider three atomic data sets that are commonly found across KN modeling literature and each take a unique approach to atomic modeling, which we label as follows:

\begin{itemize}
    \item Dataset \texttt{HULLAC} - Z = 31-70 data presented in \cite{Tanaka20};
    \item Dataset \texttt{LASER} - Lanthanide data presented in \cite{Fontes20};
    \item Dataset \texttt{Autostructure} - Code used in \cite{Kasen17};
\end{itemize}

Both Atomic Datasets \texttt{HULLAC} and \texttt{Autostructure} employ the ``Sobolev expansion opacity" for binning the large number of lines from lanthanides. The ``Sobolev expansion opacity" utilizes the Sobolev approximation \citep{Sobolev60} for bound-bound transitions, which is applicable when the thermal line width of a given line is negligible compared to that of the expansion velocity. This is true for KN ejecta, as the expansion velocity is typically of the order 10$^3$ km/s while the thermal velocities are of the order 1 km/s \citep{Kasen13}. The lines are then binned within the broader frequency bins of the simulation. Atomic Dataset \texttt{LASER} also bins the lines, but instead employs the ``line-smearing" approach which artificially broadens the lines to the resolution of the frequency grid of a simulation while preserving the area under the opacity curve \citep{Fontes20}.
 
\subsubsection{Atomic Dataset \texttt{HULLAC}}

\cite{Tanaka18,Tanaka20} generated theoretical atomic data using the Hebrew University Lawrence Livermore Atomic Code (\texttt{HULLAC}, \citealt{Bar-Shalom01}) for elements Z = 26-88, up to triply ionized species, in a self-consistent and systematic way for a large number of elements. For this work, we use elements Z = 31-70.

\texttt{HULLAC} assumes that the electron orbital functions are represented well by a single electron Dirac equation with a central-field potential that arises from spherically-averaged electron-electron interactions plus that of the nuclear charge. Due to the aspherical nature of $p, d,$ and $f$ orbitals, this may alter the energy level of orbitals.

\subsubsection{Atomic Dataset \texttt{LASER}} \label{Subsubsec:LASER}

All data for elements Z=58-70 are derived from opacity tables provided at NIST-LANL \citep{NIST-LANL}, which provide the electron scattering, bound-bound, bound-free, and free-free opacities over the same grid of $\frac{h\nu}{k_bT}$ where $h$ is the Planck constant, $\nu$ is the frequency of the photon, $k_b$ is the Boltzmann constant, and $T$ is the temperature. Each opacity is binned to the frequency grid of our \texttt{Sedona} simulation, then bilinearly interpolated between (logarithmic) density and (linear) temperature points to calculate the opacity at any given temperature and density. The opacity contribution from each atomic species is then summed.

It is important to note that the opacities derived in the table assume a pure composition of each element, which will have a systematic offset from a mixed composition as the free electron density is set by the ionization structure of multiple elements in the mixed composition. The line-smeared approach for bound-bound transitions presented in the opacity tables agree with the Sobolev approximation in the optically thin limit \citep{Fontes20}, though tend to produce higher opacities in the optically thick regime as the artificial broadening of lines results in photons interacting with lines at frequencies that should not interact, thereby artificially inflating the opacity. NIST-LANL only includes lanthanides and actinides,  so we use the non-lanthanide elements binned as described in the section above. 

When calculating the energy levels of low ionization states, \texttt{LASER} also considers the difference between the calculated ionization potential and the experimental ionization potential provided by NIST. The energy levels of the calculated atoms are then shifted by the difference between the two.

\subsubsection{Atomic Dataset \texttt{Autostructure}}

Dataset \texttt{Autostructure} is the same atomic structure code that is utilized in \cite{Kasen17}, and used to calculate KN spectra when it was the first considered that lanthanides were an important source of opacity in neutron star mergers \citep{Kasen13}. All lanthanide element data were produced with the \texttt{Autostructure} code \citep{Badnell11} up to quadruple ionized species, while all other elements were produced by \texttt{cmfgen} \citep{Hillier01}. The atomic data produced by \texttt{Autostructure} have been optimized to produce the correct ground and first two excited levels for the singly ionized element Nd, while for all other elements only the ground state was optimized. Similarly to \cite{Kasen17}, we approximate elements heavier than Z = 28 as lighter elements (while maintaining their original total mass) since the behavior of the valence electron should remain similar across any individual column on the periodic table (e.g., Os and Fe as shown in \citealt{Kasen17}). 

\subsection{Thermalization Efficiency Prescriptions} \label{Subsec:therm}

Each simulation receives an input of energy from the r-process heating rate that is convolved with a thermalization efficiency prescription. This thermalization-prescription is either applied on a by-zone basis (local) or shared among all zones (global). We consider one r-process heating rate per unit mass for all models defined by:

\begin{equation}
    \dot{Q_r} = At^\alpha + B_1 e^{-t/\beta_1} + B_2 e^{-t/\beta_2}
\end{equation}

where A = 8.49$\times10^{9}$ erg g$^{-1}$ s$^{-1}$, $\alpha$ is -1.36, $B_1$ is 8.34$\times10^{9}$ erg g$^{-1}$ s$^{-1}$, $\beta_1$ is 3.63 d, $B_2$ is 8.86$\times10^{8}$ erg g$^{-1}$ s$^{-1}$, and $\beta_2$ is 10.8\,d as defined by the heating rate per unit mass for material of electron fraction $Y_e$ = 0.13, entropy per baryon of 32$k$, and expansion timescale 0.84 ms from \cite{Lippuner15}.

While \cite{Sarin24} have shown that variable heating rates can also introduce a considerable source of error in modeling KNe, we restrict our models to only one r-process heating rate to isolate the effects of thermalization efficiency on KN parameter estimation.

\subsubsection{Global Thermalization Efficiency Prescription}
For the global thermalization-prescription, we use the analytical formula presented in \cite{Barnes16} (their Equation 34) to describe the thermalization efficiency of all radioactive decay products in the r-process material: 

\begin{equation}
    f_{\rm tot} (t) = 0.36 \bigg[e^{-at}+\frac{\ln(1+2bt^d)}{2bt^d}\bigg],
\end{equation}
where $a,b$, and $d$ are determined by the total mass and characteristic velocity of the ejecta and $f_{\rm tot} (t)$ is the fraction of decay energy deposited in the ejecta divided by the total decay energy emitted. The values derived in \cite{Barnes16} are an analytical fit to the total thermalization fraction determined from numerical simulations of KNe with different masses and velocities. Specifically, we use the values shown in Table 1 for random magnetic fields in our simulations. For any values and velocities between the presented values of the table, we bilinearly interpolate each parameter. 

The formula includes neutrino losses and \cite{Barnes16} find that the radioactive decay products carry approximately 20\%, 45\%, and 35\% of the decay energy for beta particles, gamma rays, and neutrinos, respectively. Though, due to the functional form, the maximum efficiency at t = 0\,d is 72\% which is in rough agreement with other estimates from \cite{Wollaeger18}. 

\subsubsection{Local Thermalization Efficiency Prescription}

For the local thermalization-prescription, we break down the thermalization efficiency of each component of r-process decay products according to the formula presented in \cite{Barnes16,Rosswog17,Bulla24}:

\begin{equation}
    f_j (\textbf{r},t) = \frac{\ln (1+2\eta^2)}{2\eta^2},
\end{equation}

where \textbf{r} is the position in the ejecta, $2\eta^2 = \frac{2A_j}{t \rho (\textbf{r},t)}$ and $A_j$ is [1.2, 1.3, and 0.2] $\times 10^{-11}$ g cm$^{-3}$ s for alpha particles, beta particles, and fission fragments, respectively, and $f_j (\textbf{r},t)$ is the equivalent of $f_{\rm tot}$ but as a function of position and for a decay product $j$. Following \cite{Bulla24}, we assume that the radioactive decay products carry 35\%, 40\%, 20\%, 5\%, and 0\% of the decay energy for neutrinos, gamma rays, beta particles, alpha particles, and fission fragments, respectively as supported by the findings of \cite{Barnes16,Wollaeger18} for the distribution of decay energy.

Neutrinos are assumed to escape immediately and do not contribute to heating the ejecta due to low densities ($\rho$ $\lesssim 10^{-8}$ g cm$^{-3}$ at all times simulated). This implies an instantaneous loss of 35\% of r-process heating energy at all times and causing a maximum efficiency of 65\% at t = 0\,d. Gamma rays are simulated directly by injecting them into the zone of r-process decay as 1 MeV photons where they can be absorbed or scattered.

\section{Kilonova Light Curve and Spectra \label{Sec:LightCurve}}

We first discuss the impact of atomic dataset and thermalization-prescription on the observables of a fiducial KN model of M = $10^{-2}\ $M$_\odot$, v = 0.1c, $\log_{10} (X_{\rm lan})$ = $-2$.

\begin{figure*}[h]
    \centering
  \includegraphics[width=0.87\textwidth,trim={3cm 4.5cm 3cm 5.5cm},clip]{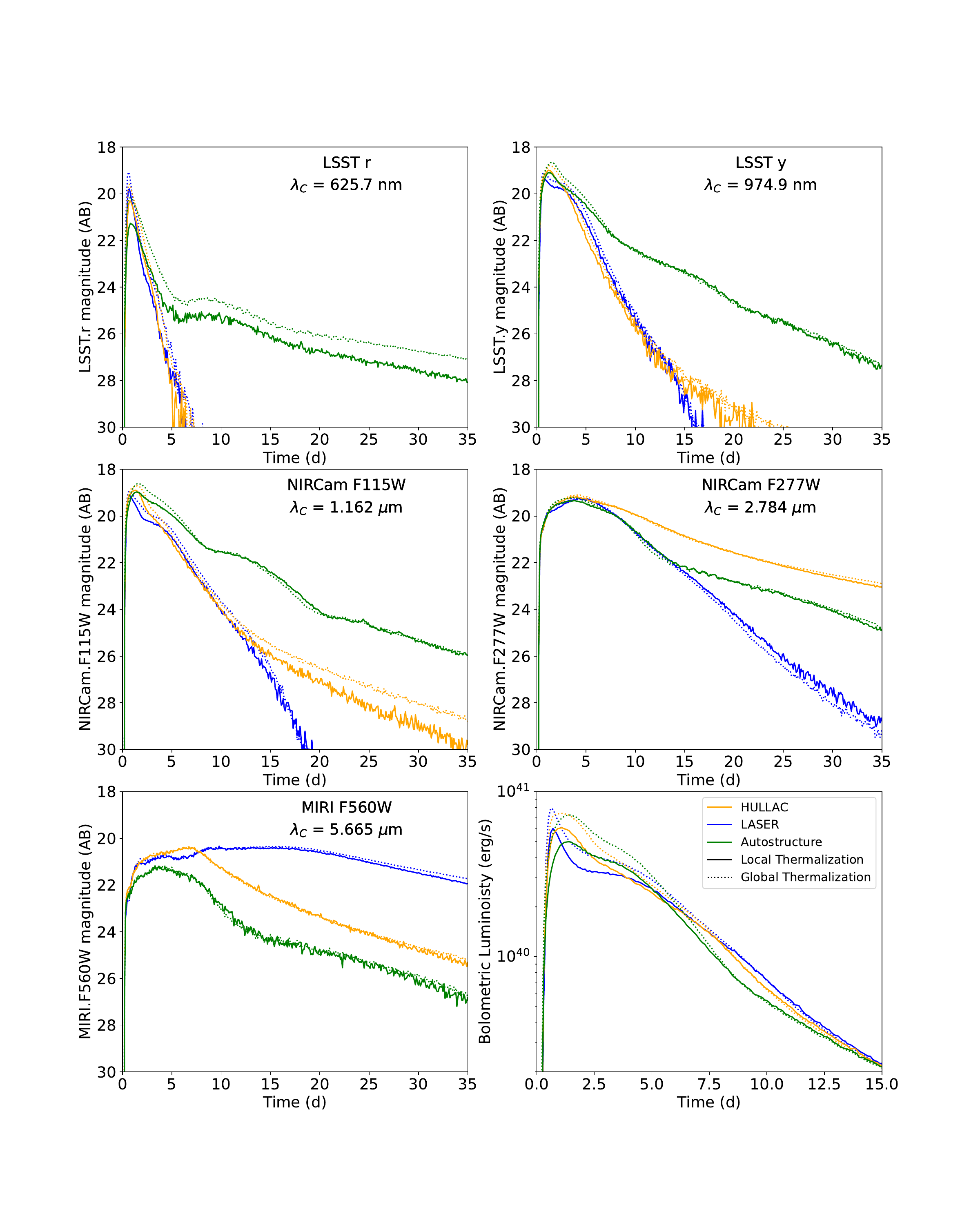}
    \caption{Comparing the fiducial model of M = $10^{-2}\ $M$_\odot$, v = 0.1c, $\log_{10} (X_{\rm lan})$ = -2 across the LSST $r$ and $y$ filters, the JWST F115W, F277W, F560W filters, and bolometric luminosity curve with Atomic Datasets \texttt{HULLAC}, \texttt{LASER}, and \texttt{Autostructure} in orange, blue, and green, respectively at a distance of 40.7 Mpc. Models using the local thermalization-prescription are solid lines, while the global thermalization-prescription is represented by the dotted line. The choice of atomic data for light r-process elements (31 $\leq$ $Z$ $\leq$ 50) has a dramatic impact on optical emission as seen by the differences between Atomic Datasets \texttt{HULLAC} and \texttt{LASER} against \texttt{Autostructure}. Differences in lanthanide atomic data result in a $\gtrsim 3$ mag difference at NIR and MIR wavelengths by $\delta t \sim$  2 weeks that only continues to grow. The global thermalization-prescription tends to increase emission across filters by 0.5-1 mags, especially at bluer filters and at early times while the ejecta is optically thick. Notably, we would expect NLTE effects to become important at $\delta t \sim 13$\,d as per Equation \ref{Eq:NLTE}.}
    \label{Fig:LightCurveComp}
\end{figure*}

Figure \ref{Fig:LightCurveComp} illustrates the differences in light curves caused by the choice of atomic data and thermalization-prescription for the fiducial model; despite having the same ejecta parameters, the LSST $r$ and $y$ filters, the JWST F115W, F277W, F560W filters, and bolometric luminosity curves, while initially within $\sim$ 1 mag of each other, rapidly diverge as the ejecta becomes optically thin. The optical filters, $r$ and $y$, are strongly impacted by the choice of atomic data for light r-process elements (31 $\leq$ $Z$ $\leq$ 50), as seen by comparing the differences between Atomic Datasets \texttt{HULLAC} and \texttt{LASER} (which share light r-process atomic data) against \texttt{Autostructure}. The differences in lanthanide atomic data most strongly impact the light curves in the NIR and MIR JWST filters as each atomic dataset creates an entirely unique light curve with typical offsets of 4 to 6 mags by $\delta t$ $\sim 2$ weeks. 

The dotted lines and solid lines in Figure \ref{Fig:LightCurveComp} represent the global and local thermalization-prescriptions, respectively. The thermalization-prescription has the greatest impact in optical filters and at early times, with the global prescription increasing the emission by 0.5 to 1 magnitudes compared to that of the local prescription. The local prescription accounts for the fact that thermalization will be relatively lower in the low density outer layers of ejecta, from which much of the early emission arises. Eventually, the bolometric luminosity curves of the two prescriptions converge once the ejecta becomes optically thin and the light curves maintain an approximately constant offset as the KN fades.

\subsection{Spectra} \label{SubSec:Spectra}
\begin{figure*}
    \centering
  \includegraphics[width=0.87\textwidth,trim={2.5cm 6.2cm 3.5cm 7.25cm},clip]{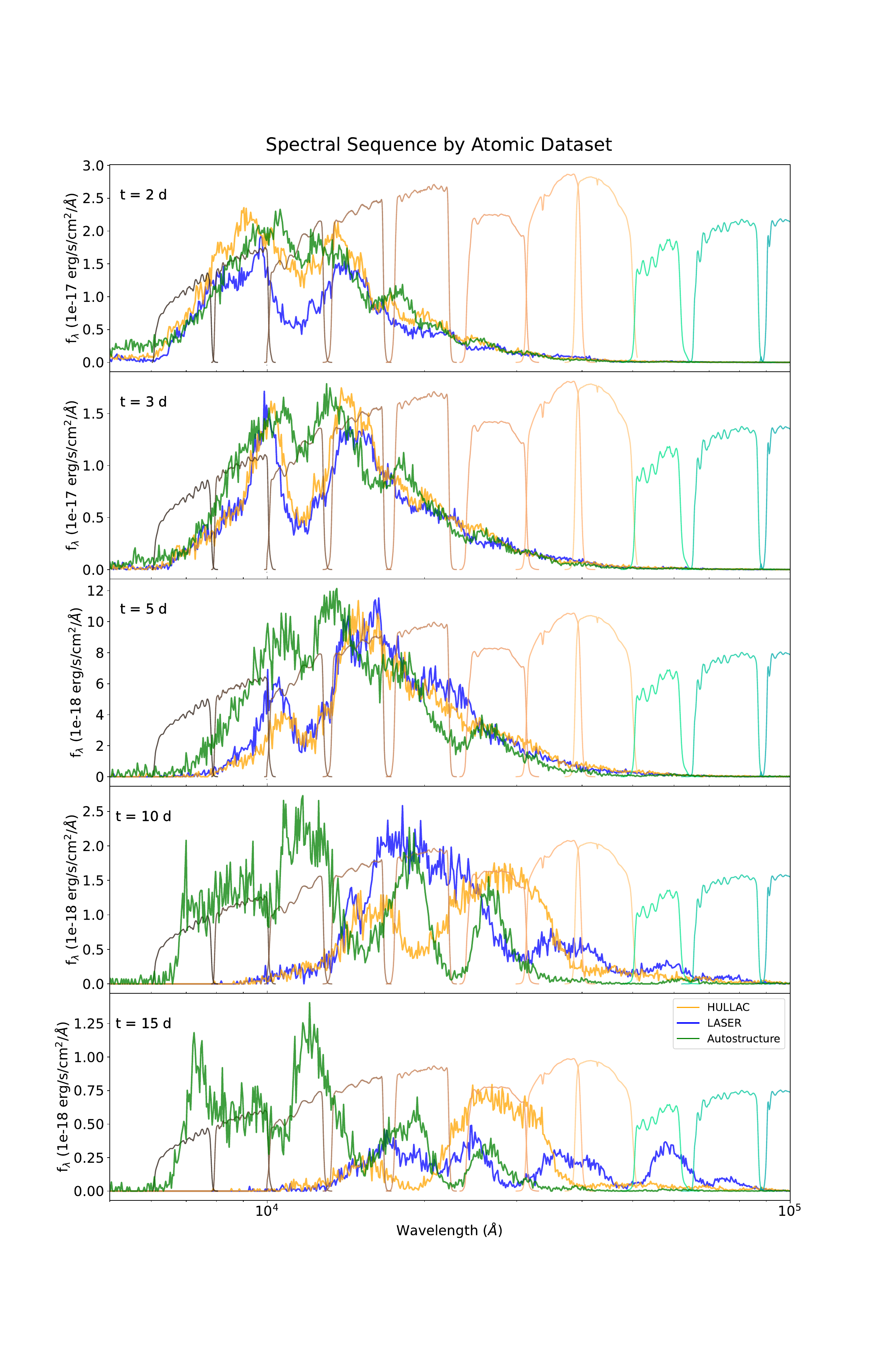}
    \caption{Spectral sequence of a M = $10^{-2} $M$_\odot$, v = 0.1c, $\log_{10} (X_{\rm lan})$ = -2 model using Atomic Dataset \texttt{HULLAC} (orange), \texttt{LASER} (blue), and \texttt{Autostructure} (green) at $t = 2, 3, 5, 10,$ and $15$\,d overlaid on top of the transmission curves of JWST photometric filters. At early times, the predictions from each Atomic Dataset agree relatively well, but rapidly diverge (particularly in the IR) and result in vastly different colors. The $\delta t$ = 15\,d spectral comparison, while likely in a regime where NLTE will affect the precise shape of features, particularly shows the uncertainty in the specific locations of lines between Atomic Datasets from the qualitative differences in shape which should remain unaffected by NLTE effects.}
    \label{Fig:ComparisonSpecSeq}
\end{figure*}

While the different Atomic Datasets generate remarkably different light curves, they all agree on the same general spectral features at early times: a doubly peaked spectrum with the bluer peak at $\sim 1 \mu$m and the second at $\sim 1.5 \mu$m (though this may largely be due to the peak blackbody wavelength for temperatures typical of KNe, $\sim$ few $\times$ 10$^3$ K, and qualitatively similar high opacity feature between the peaks) with a rapidly decaying blue flux and a long-lasting red flux. Figure \ref{Fig:ComparisonSpecSeq} shows a spectral sequence from 2 to 15 days post merger of the fiducial M = $10^{-2}$ M$_\odot$, v = 0.1c, and $\log_{10} (X_{\rm lan})$ = -2 model using the local thermalization efficiency prescription for each Atomic Dataset. 

The differences in line locations and strengths between Atomic Datasets becomes most obvious in the late time spectra as distinct features from lines or line blends emerge. While effects such as NLTE would likely alter the precise shape of the spectra at these late times, the disagreement in the placements of the lines illustrates the level of uncertainty in lanthanide atomic structure. Furthermore, the pattern of lines is qualitatively different; the pattern of features is entirely unique and not mirrors of other models shifted redward or blueward. Except in cases where a line wavelength has been experimentally measured (i.e., Sr II from \cite{Sneppen23a}, possible Y II from \citealt{Sneppen23b}, Te III from \citealt{Hotokezaka23}), extracting information based on the locations of a spectral features should be done with a high level of caution.



\subsection{Model Grid Light Curve Properties}
Figure \ref{Fig:LightCurveComp} clearly demonstrates that each atomic dataset results in substantial changes in light curve properties such as peak magnitude and rise time to peak, color, decay rate post peak, and time above half peak luminosity. In the following sections, we use those metrics to quantify the differences between models in more detail by expanding our analysis to the grid of KN models with M $\in$ [0.001, 0.01, 0.1] M$_\odot$, $v_k \in$ [0.1,0.3] c, and $\log_{10} (X_{\rm lan}) \in$ [-9, -4, -2], to identify the impact of atomic data and thermalization-prescription on observational properties and how that impact changes with ejecta properties.

\subsubsection{Peak Magnitudes and Peak Times}

\begin{figure*}
    \centering
  \includegraphics[width=0.97\textwidth,trim={0 6cm 0 6.7cm},clip]{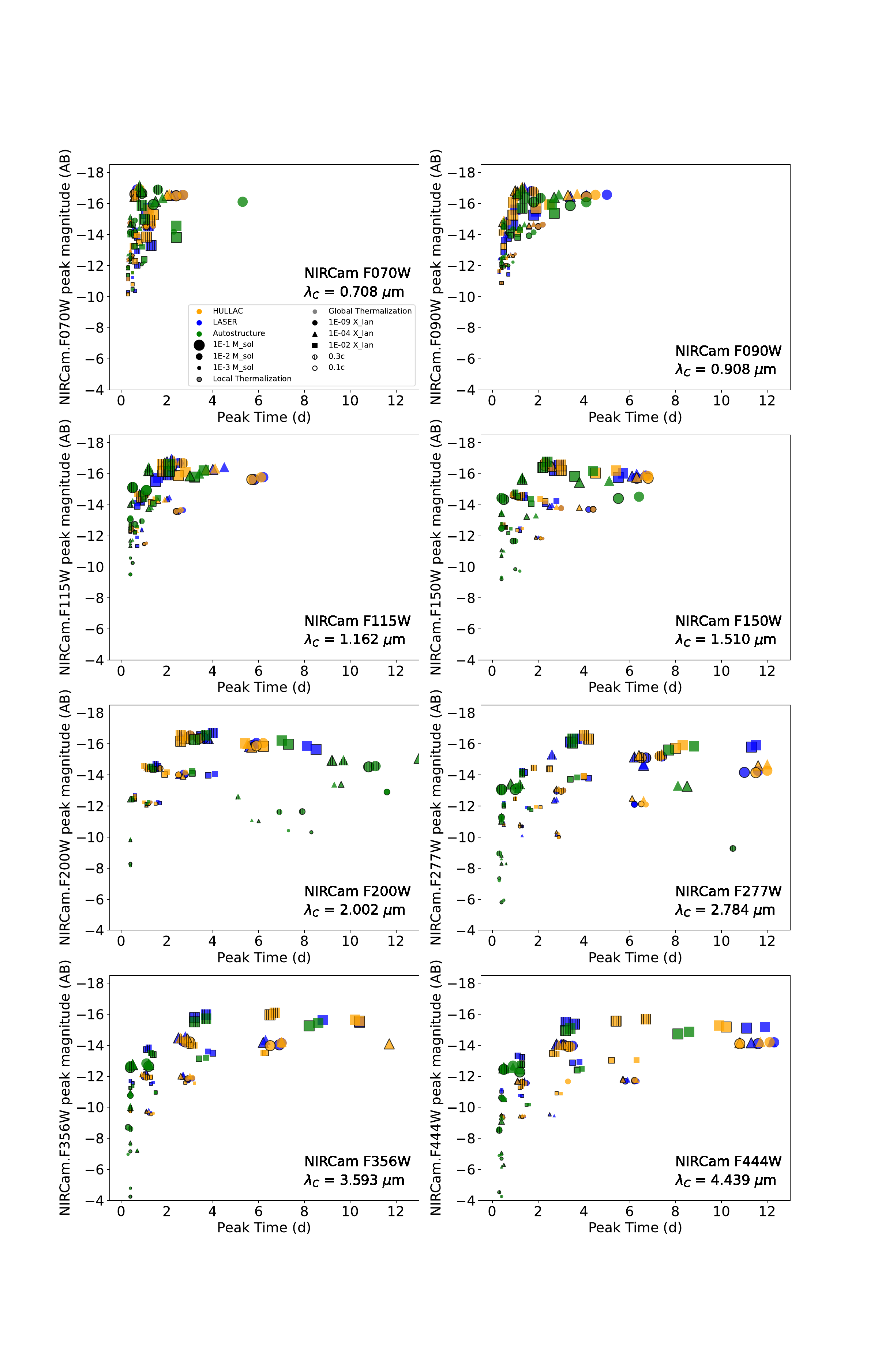}
    \caption{Parameter space for peak absolute magnitude in each JWST NIRCam filter vs. peak time. Models using Atomic Dataset \texttt{HULLAC} are in orange, Dataset \texttt{LASER} in blue, and Dataset \texttt{Autostructure} in green. Larger size corresponds to larger mass models. A black outline indicates that the local thermalization efficiency prescription was used, while no boundary indicates global thermalization-prescription. Circles, triangles, and squares indicate $\log_{10}$ \Xlan\, = -9, -4, and -2, respectively. Hatches inside the markers indicate a model with characteristic velocity of 0.3c, while those without 0.1c. The effects of choice of Atomic Dataset already becomes apparent in the simple observable of time of peak and peak magnitude and especially so at redder filters; for the same ejecta parameters, while the peak magnitude is approximately constant, the time at which the KN achieves its peak shifts by almost four days in the highest \Xlan\, models in the reddest filters.}
    \label{Fig:JWSTPeakMag}
\end{figure*}

\begin{figure*}
    \centering
  \includegraphics[width=0.97\textwidth,trim={0 6cm 0 7cm},clip]{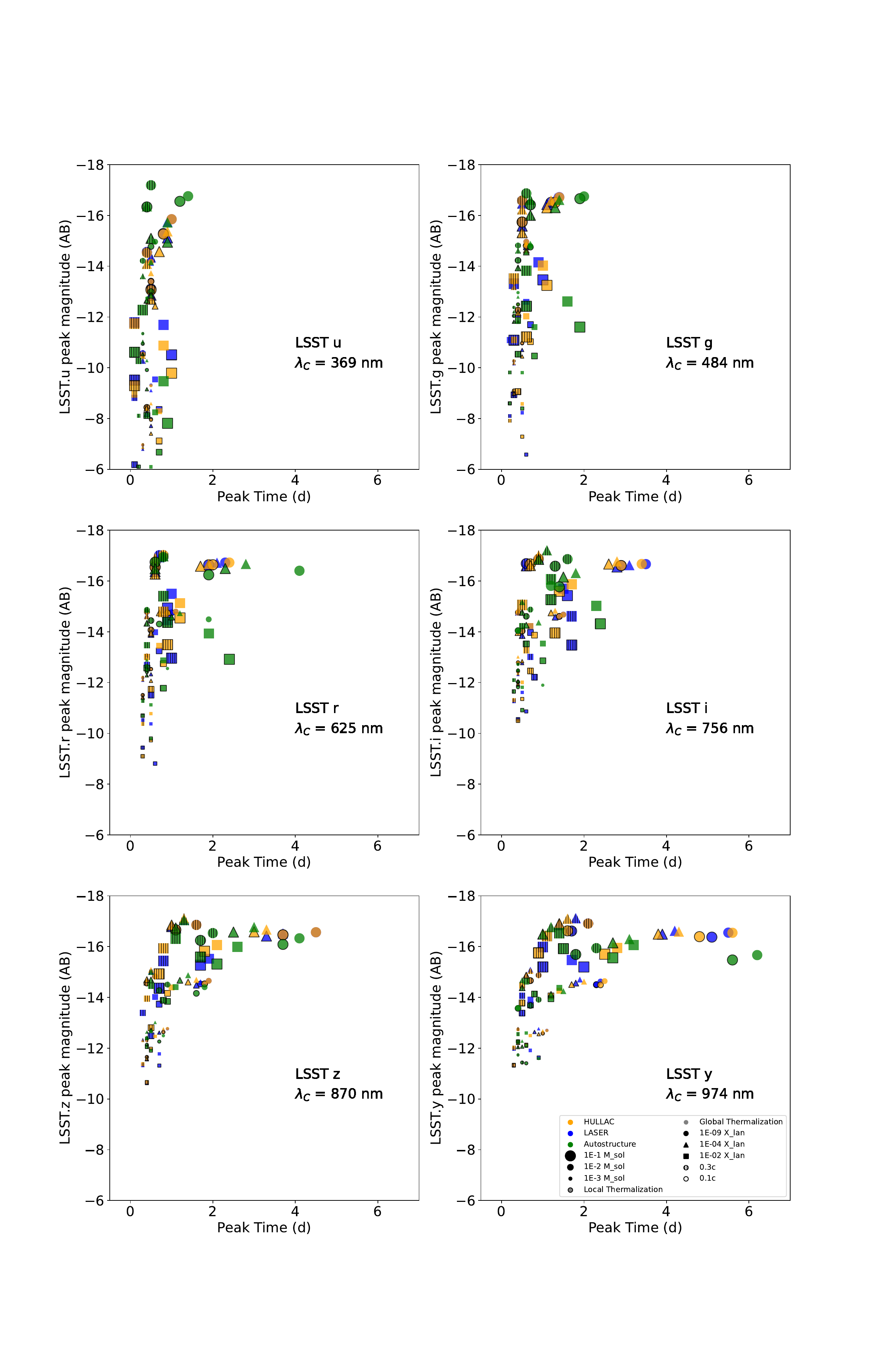}
    \caption{Same as Figure \ref{Fig:JWSTPeakMag}, except for each LSST filter $u, g, r, i, z,$ and $y$. Similarly to the JWST filters, the reddest filters show the largest spread in days until peak among the different Atomic Datasets. As many of the LSST filters are blueward of JWST, this highlights the rapid decline of KNe at blue wavelengths as the majority of models achieve their peak in the $u, g,$ and $r$ filters within two days of merger.}
    \label{Fig:LSSTPeakMag}
\end{figure*}

Figures \ref{Fig:JWSTPeakMag} and \ref{Fig:LSSTPeakMag} show the peak magnitude and time since merger for JWST and LSST filters, respectively, for each model in the grid. We select these filters for their practical and discriminating power; LSST carried out on the Rubin Observatory, if equipped with Target of Opporutnity capabilities, has large discovery potential, thanks to the combination of exquisite sensitivity and large field of view (i.e., \citealt{Margutti18,Andreoni22}) and JWST to put forward quantitative expectations of KN brightness for purposes like determining exposure time as well as the capability to constrain atomic dataset. For both LSST and JWST, the reddest bands are the brightest and most useful for discriminating between different model parameters. Furthermore, the effects of choice of Atomic Dataset already becomes apparent in the simple observable of time of peak and peak magnitude and especially so at redder filters; for the same ejecta parameters, while the peak magnitude is approximately constant, the time at which the KN achieves its peak shifts by almost four days in the highest \Xlan\, models in the reddest JWST filters. Similarly, the choice of thermalization efficiency prescription changes the peak magnitude in most LSST filters by at least a magnitude.

\subsubsection{Color}

\begin{figure*}
    \centering
  \includegraphics[width=0.97\textwidth,trim={0 6cm 0 7cm},clip]{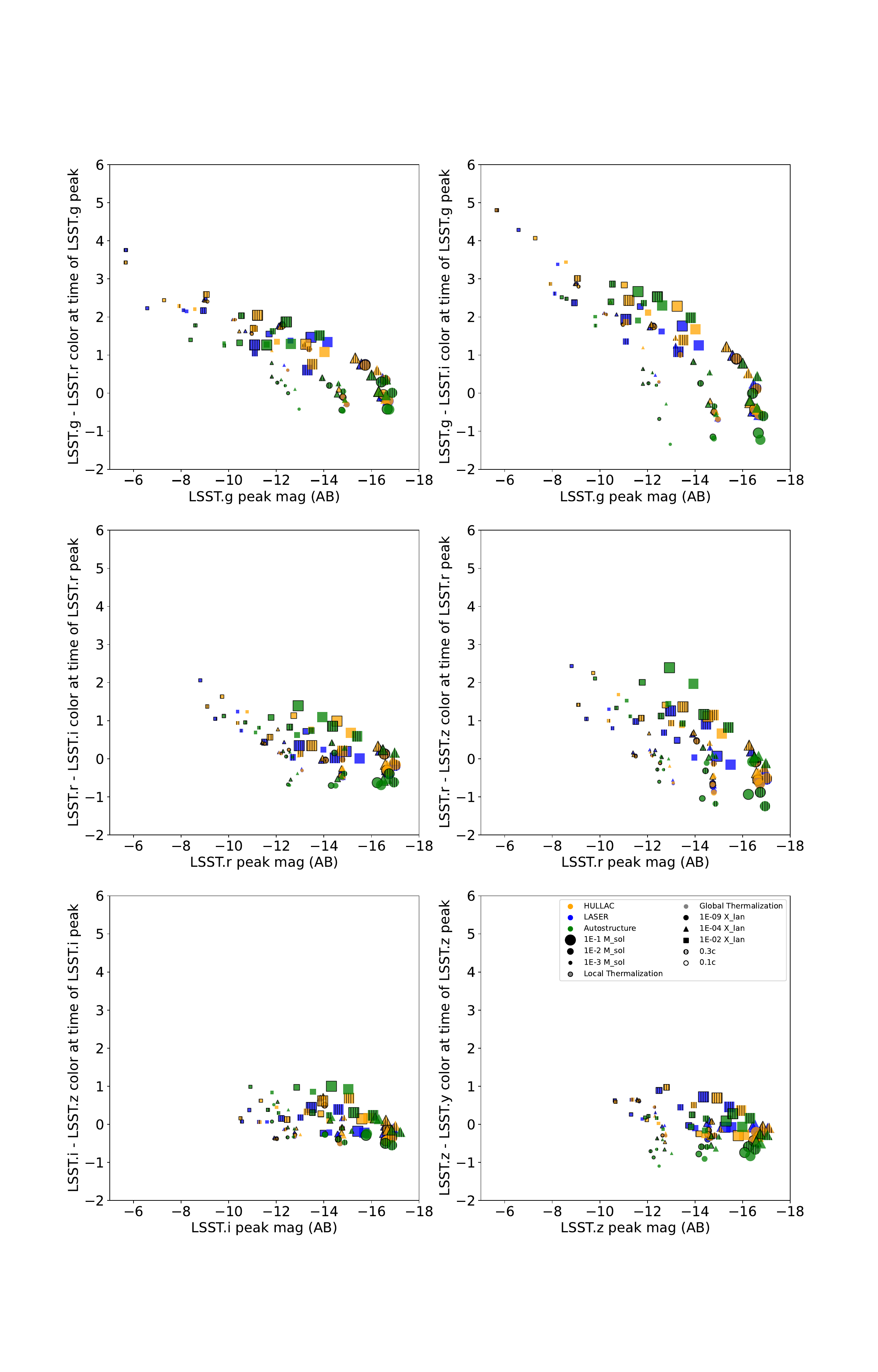}
    \caption{Same legend as Figure \ref{Fig:LSSTPeakMag}. Parameter space of LSST $g-r, g-i, r-i, r-z, i-z,$ and $z-y$ color at time of peak of the first filter. Due to the spacing of the grid, it is possible to see that each unique mass forms an arc of models in this parameter space, with more massive models pushing to higher peak magnitudes and higher \Xlan\, pushing models to redder colors and fainter peak magnitudes, creating an observable arc.}
    \label{Fig:LSSTColorPeak}
\end{figure*}

Figure \ref{Fig:LSSTColorPeak} shows the color at the time of peak of the shorter-wavelength filter. There is clear correlation of color with ejecta mass, velocity, lanthanide fraction, atomic dataset, and thermalization-prescription. Higher \Xlan\, models produce redder emission, more massive models reach a more luminous peak magnitude, and models with higher characteristic velocities achieve similar or bluer colors than their slower counterparts. Effectively, each mass of the model grid forms arcs in the color-peak magnitude space where the mass sets the starting point along the peak magnitude axis and the \Xlan\, determines where along the arc the model sits. Choice of Atomic Dataset sets at what color value the arc ends, with Atomic Dataset \texttt{LASER} models generally ending at the reddest colors for the highest \Xlan models. As a second-order effect, the choice in thermalization efficiency prescription shifts the KN along the arc with the global prescription resulting in a bluer KN and the local prescription a redder KN. The farther apart the filters are in wavelength space, the easier it becomes to distinguish KN ejecta parameters based on where they fall in this parameter space. However, observational limitations such as limiting magnitudes will place constraints on the feasibility of sampling maximally wavelength separated like $g-y$, and so colors with smaller wavelength separation like $g-z$ may prove more fruitful. 

\subsubsection{Decay Rate and Time Above Half Peak Luminosity}

\begin{figure*}
    \centering
  \includegraphics[width=0.97\textwidth,trim={0 6cm 0 6.7cm},clip]{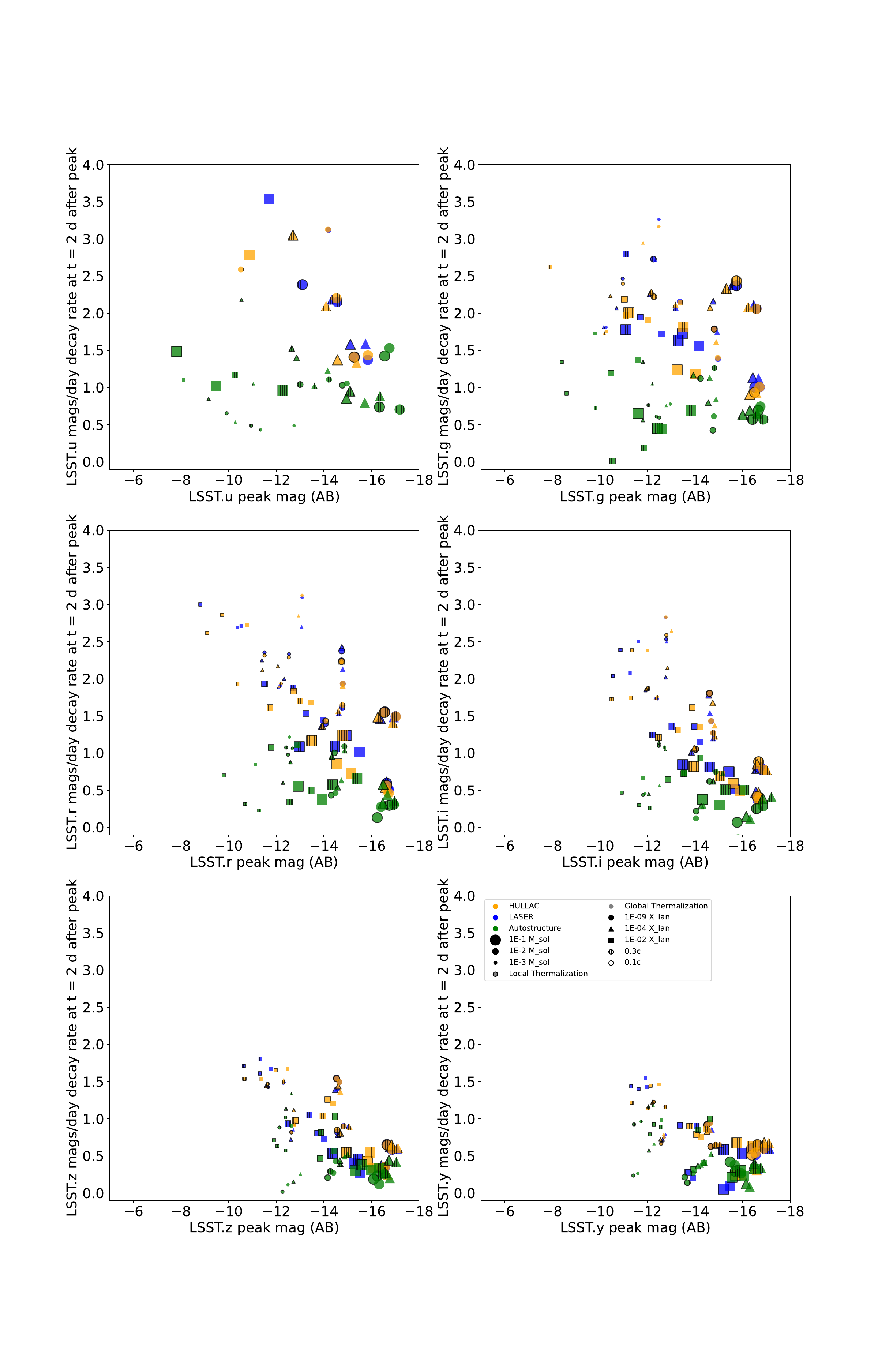}
    \caption{Parameter space of the decay rate in each LSST filter vs. the magnitude at $\delta t$ = 2 days after peak of a given filter. Atomic Dataset \texttt{Autostructure} stands out as the slowest declining across all LSST filters, especially the bluest filters like $u$, $g$, and $r$.}
    \label{Fig:LSSTPeakRiseDelta}
\end{figure*}

Figure \ref{Fig:LSSTPeakRiseDelta} shows the rise rate in magnitudes per day in each LSST filter as a function of magnitude at t = $t_{peak} + 2$\,d. Bluer filters like $u,\,g,$ and $r$ are much more rapidly decaying than redder filters like $i, z,$ and $y$ with some models having minimal decay. The light curves resulting from the Atomic Datasets begin to diverge in characteristics with Atomic Dataset \texttt{Autostructure} models as the slowest declining in the bluest filters while light curves from Atomic Dataset \texttt{LASER} models tend to be more rapidly declining and therefore redder than the light curves of other Atomic Datasets, though the divergence is not sufficient to distinguish between Atomic Datasets \texttt{HULLAC} and \texttt{LASER} by decay rate alone.

\begin{figure*}
    \centering
  \includegraphics[width=0.97\textwidth,trim={0 6cm 0 7cm},clip]{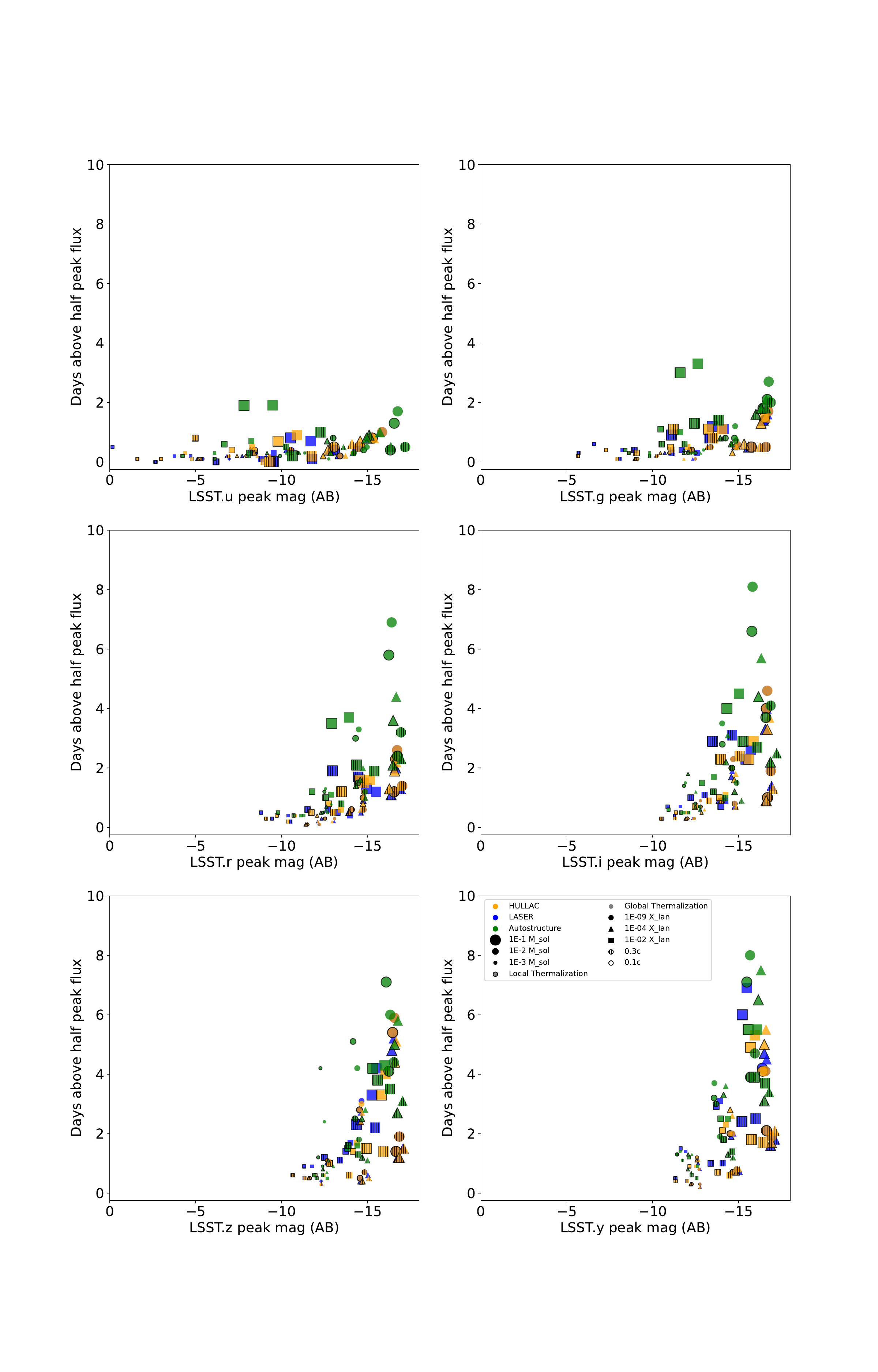}
    \caption{Parameter space of time above half peak vs the peak magnitude of that filter. Due to the slower decay rate in LSST filters of Atomic Dataset \texttt{Autostructure}, those models last above half maximum luminosity by as much as a factor of 2 longer.}
    \label{Fig:LSSTTimeHalf}
\end{figure*}

Figure \ref{Fig:LSSTTimeHalf} shows the time spent in each filter above half the peak luminosity. The emission in bluer filters rarely last more than a few days above half maximum while the emission in red filters can spend more than a week above half maximum. The emission from Atomic Dataset \texttt{Autostructure} models stands out due to the longer times above half peak across all filters by up to a factor of 2 as a result of the slower decay rate.

\subsubsection{Bolometric Luminosity}

Figure \ref{Fig:BolometricThermRatio} shows the ratio of the bolometric luminosity curves of the global to the local thermalization efficiency prescriptions. The ratio has been smoothed then rescaled by a factor of 0.65/0.72 to correct for the disagreement in maximum efficiency to ensure that any differences are due to the intrinsic analytical formulae instead of a disagreement about the assumed fraction of decay energy that escapes in neutrinos. There is a clear difference in luminosity of factors between 0.8 and $\sim 1.5$ that evolve with time. This is indicative that the two thermalization-prescriptions deposit a substantially different amount of r-process decay energy in the ejecta, leading to discrepancies in the bolometric luminosity.

Choice of Atomic Dataset largely affects the shape of the bolometric luminosity curve, as shown in the bottom right panel of Figure \ref{Fig:LightCurveComp}. The fiducial Atomic Dataset \texttt{LASER} model has a higher opacity and therefore traps more energy at early times, resulting in a lower bolometric luminosity. The energy then emerges at late times as the ejecta becomes optically thin, which creates the brighter late-time bolometric luminosity curve than the models using Atomic Datasets \texttt{HULLAC} and \texttt{Autostructure}.

\subsection{Atomic Data Systematic Uncertainty} \label{Subsec:Atomic}
The above figures show that the different Atomic Datasets produce a range of KN colors for the same ejecta properties, with Atomic Dataset \texttt{Autostructure} producing the bluest, Atomic Dataset \texttt{LASER} producing the reddest, and Atomic Dataset \texttt{HULLAC} being in between the two. At smaller \Xlan, the differences between Atomic Dataset \texttt{HULLAC} and \texttt{LASER} are minimal, as expected since they share non-lanthanide atomic data, but are substantial for the highest \Xlan\, models where lanthanide opacities dominate. For example, the highest \Xlan\, and most massive models exhibit a scatter of almost 3 magnitudes in $u$ band while peaking at approximately the same time, while in the $y$ band the scatter shrinks to less than a magnitude in brightness and approximately a day in peak time (Figure \ref{Fig:LSSTPeakMag}). 

The scatter of observational properties within an Atomic Dataset is typically quite similar, though offset from one another. The offset for each Atomic Dataset is most prominent in the colors of each model (Figure \ref{Fig:LSSTColorPeak}), namely the $g - i$, $r - i$, and $r-z$ colors which span between 1.5 and 2 magnitudes difference in colors for the highest \Xlan\, models. This is also true of the decay rate in each LSST filter, especially the bluest filters (Figure \ref{Fig:LSSTPeakRiseDelta}). Many of the models with Atomic Datasets \texttt{HULLAC} and \texttt{LASER} so rapidly decay in $u$ band that most models have dropped by more than 5 mags in a day, while Atomic Dataset \texttt{Autostructure} models are declining much more slowly at $\sim 1.5$ mags/day. The substantial difference in atomic data means that the bluest filter that achieves a scatter across Atomic Datasets of decay rate of less than $\pm$0.1 mags/day is $z$ band, though this is likely due to the fact that the overall extent of the color and decay rate phase space covered by the models shrinks when looking at redder filters. The spread between Atomic Datasets only grows for lower mass models, with a spread of $z$ band decay rate of approximately 1 mag/day.

Therefore there is a tension in determining the optimal filter for discerning KN ejecta properties; bluer optical filters like $u, g,$ and $r$ have a higher intrinsic spread that can be utilized to identify ejecta properties but are the most model-dependent. Redder optical filters like $i, z,$ and $y$ are less model-dependent but have a lower intrinsic spread and are more difficult to discern ejecta properties.

At late times, the difference between Atomic Datasets is most apparent in the IR colors. In Figure \ref{Fig:ComparisonSpecSeq}, the remaining features at 15 days dominate the predicted colors for each model, and their different placements manifest as vastly different colors. Atomic Dataset \texttt{HULLAC} is dominated by a feature at $\sim 2.5 \mu$m, while Atomic Dataset \texttt{Autostructure} has a number of features spanning most JWST NIR filters. Atomic Dataset \texttt{LASER} is the only dataset to have a long-lasting MIRI feature, resulting in an extremely long-lived MIRI lightcurve and red color. However, we caution using the LTE models presented here as line lists are incomplete for MIR wavelengths and the nebular MIR spectra may be able to resolve individual atomic lines. 

\subsection{Thermalization Efficiency Systematic Uncertainty} \label{Subsec:Therm}

The impact of thermalization-prescription on observational properties manifests as a systematic offset in a direction that mimics more massive models across all Atomic Datasets and ejecta parameters. For example, in Figures \ref{Fig:JWSTPeakMag} and \ref{Fig:LSSTPeakMag}, the global thermalization-prescription drives points to brighter and later peaks. Similarly, the global thermalization-prescription pushes a given KN to brighter and bluer colors in Figure \ref{Fig:LSSTColorPeak}. 

\begin{figure*}
    \centering
  \includegraphics[width=0.87\textwidth,trim={1cm 1.5cm 1cm 1.5cm},clip]{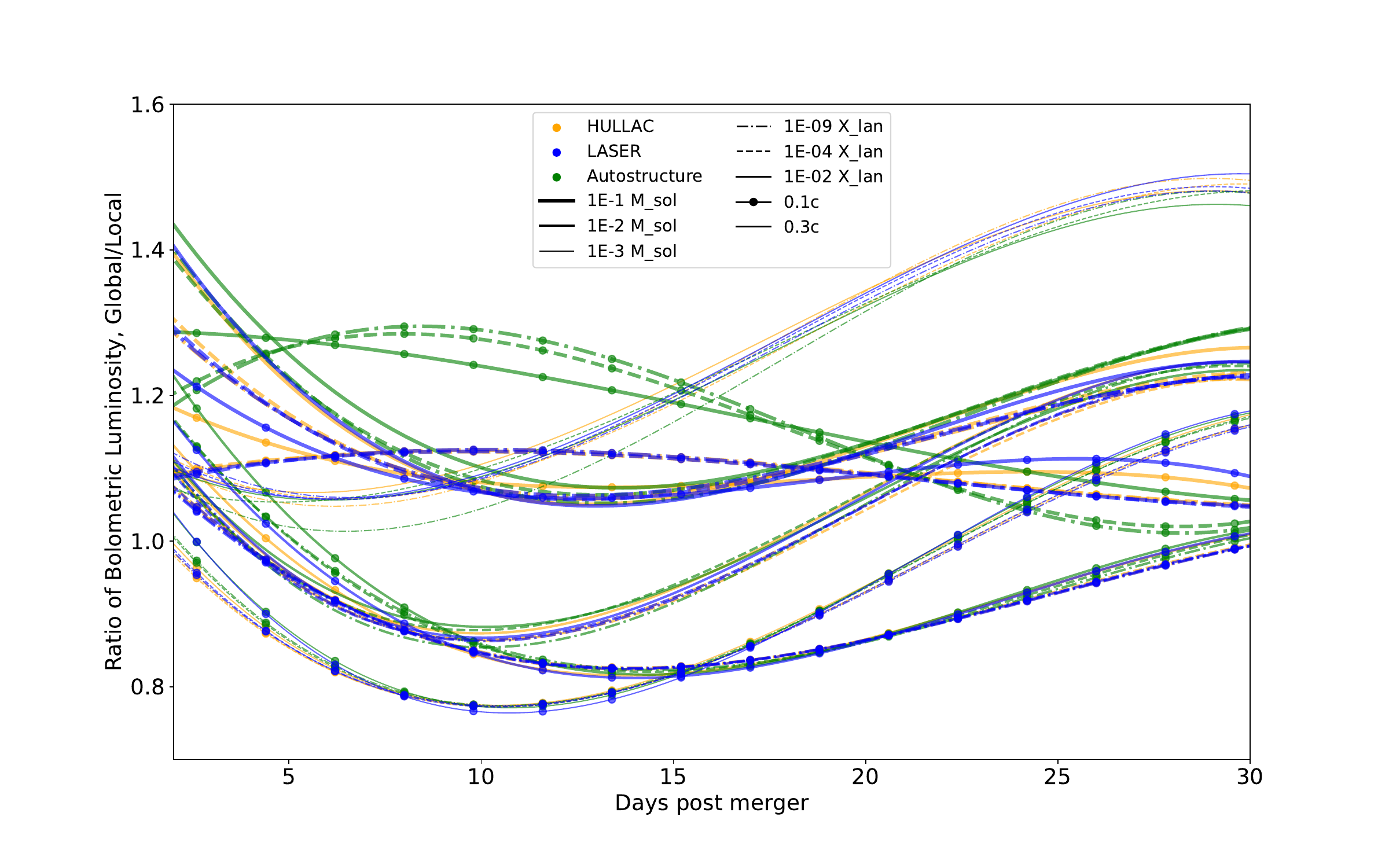}
    \caption{Ratio of the global thermalization-prescription bolometric luminosity to the local thermalization-prescription bolometric luminosity curve as a function of time for each model (smoothed to remove numerical noise). The predictions from global prescription models tend to differ from the local prescription predictions by $\sim$20-50\%. Excluding the least massive models limits the offset to up to $\sim 30\%$, regardless of Atomic Dataset.}
    \label{Fig:BolometricThermRatio}
\end{figure*}

Figure \ref{Fig:BolometricThermRatio} shows the systematic brightness offset of the global thermalization-prescription models compared to the local thermalization-prescription models. Luminosity is most tightly correlated with the ejecta mass through the r-process heating rate, as an increase in mass results in an increase in r-process decay energy. The effect of thermalization efficiency prescription on inferring the correct velocity measurement is negligible as it largely affects the overall normalization of the spectra at each epoch.

Roughly, the 20-50\% offset in luminosity can correspond to a difference in ejecta mass estimation of those factors. More massive models appear to be less affected, with uncertainties up to $\sim30\%$. This may be because the more massive models remain optically thick for longer and the fraction of energy lost by the outermost ejecta is more negligible compared to that of the less massive models. Figure \ref{Fig:BolometricThermRatio} additionally makes clear that this systematic difference happens regardless of atomic data as models with the same ejecta parameters eventually converge. This is to be expected, as at early times when the ejecta is optically thick and blackbody-like the difference in spectra is a result of differing strong absorption features that allow a varying luminosity to escape. As the ejecta becomes optically thin, the luminosity of the KN becomes approximately the r-process heating rate as all radiation can “instantly” escape the ejecta and so becomes the same across all models regardless of Atomic Dataset. Thus, the difference in luminosity between the models reflects the underlying difference in thermalization factors.

From Figures \ref{Fig:JWSTPeakMag} - \ref{Fig:BolometricThermRatio}, it becomes immediately clear that treating the thermalization efficiency prescription globally as opposed to locally causes a systematic offset in model observable properties approximately in the same direction across all atomic datasets, masses, velocities, and lanthanide fractions for each observable property. Generally, the global thermalization efficiency prescription retains more energy in the ejecta than the local thermalization efficiency prescription, resulting in brighter KNe across all wavelengths, though especially at optical wavelengths. Additionally, using a local thermalization efficiency prescription allows for more accurate modeling of ejecta structures beyond those used in the numerical calculations of \cite{Barnes16}.

\section{Effects on Parameter Estimation}
\label{Sec:Error}

The previous sections have shown that choices of atomic data and thermalization-prescription can lead to large systematic offsets in observable properties and therefore in the estimation of KN parameters. Here, we quantify the systematic uncertainty due to thermalization efficiency prescription and atomic dataset.

In the era of LSST, there may be serendipitous discoveries of KNe that are only discovered well after the transient has faded. In these cases, the limited LSST dataset collected can still constrain valuable ejecta properties based on the characteristics of the lightcurve (e.g., \citealt{Cowperthwaite19,Andreoni22a}). In the following sections, we discuss the impact the various atomic datasets and thermalization-prescription may have on parameter estimation for cases in which there may only be a handful of photometric points across a few bands.

\subsection{Error Estimation with GW170817} \label{Subsec:170817}

To illustrate the impact of the systematic uncertainty caused by atomic data, we refit GW170817/AT2017gfo to derive properties of the ejecta using each atomic dataset. We fit the lightcurves by adding together two 1D models, a low \Xlan\, component and a high \Xlan\, component to mimic the polar, blue ejecta and the red, dynamical ejecta interpretation of ejecta components. While summing the emission does not take into account effects of interaction between the two components, it is sufficient to quantify the difference in parameter estimation between the various atomic datasets and commonly used in other analyses (i.e., \citealt{Coughlin18}). We present these ``fits" as relative estimates to capture the systematic error between atomic datasets, not as absolute fits with statements about the true parameters of GW170817.

We use a grid with masses M $\in$ [0.01, 0.02, 0.03, 0.04, 0.05] M$_\odot$, $v_k \in$ [0.05, 0.1, 0.2, 0.3] c, and $\log_{10} (X_{\rm lan}) \in$ [-9, -5, -4, -3, -2] as components for each model. Additionally, we include the models described in Section \ref{sec:Sedona} for model matching purposes. We fit the lightcurve of $g, r, i, z, y, J, H,$ and $Ks$ filters presented in \cite{Villar17} and references therein at times $\delta t \geq 1$ day to the equivalent LSST and 2MASS filters by minimizing the weighted $\chi^2$ function

\begin{equation}
    \chi^2 = \sum^{N}_i \frac{w_{\rm obs, i}^2(y_{\rm obs, i} - y_{\rm m})^2}{\sigma_{\rm m}^2},
    \label{Eq:ChiSquare}
\end{equation}

where $w_{\rm obs}$, $y_{\rm obs}$, $y_{\rm m}$, and $\sigma_{\rm obs}$ are the weights, observed magnitude, model magnitude, and observed uncertainty, respectively and N is the number of observations. Once the best-fitting model within the grid was identified, we then iteratively tested a finer resolution grid around the lanthanide-rich component. At each step, we generated 16 more models that were half the logarithmic distance in \Xlan\, half the linear distance in mass, and half the linear distance in velocity to the surrounding grid points.

However, since we use the equivalent LSST or 2MASS filter to compare our models to GW170817, there are two additional error sources that are not included in the photometric data provided by \cite{Villar17}: the systematic offset of the LSST or 2MASS filter against the similar filter of the observation, and the scatter of points caused by the subtle differences in filters across the telescopes used to observe GW170817. As such, we artificially inflate the uncertainties of each observation by a factor equal to the scatter of of observed magnitudes across each filter to account for the extra variance of the data set. We then weight each point by dividing by the number of data points with that filter to normalize such that each filter holds an equal weight on the fit and prevent filters with a large number of data points from dominating the fit.

\begin{figure*}[h]
    \centering
  \includegraphics[width=0.87\textwidth,trim={3cm 1cm 3cm 3cm},clip]{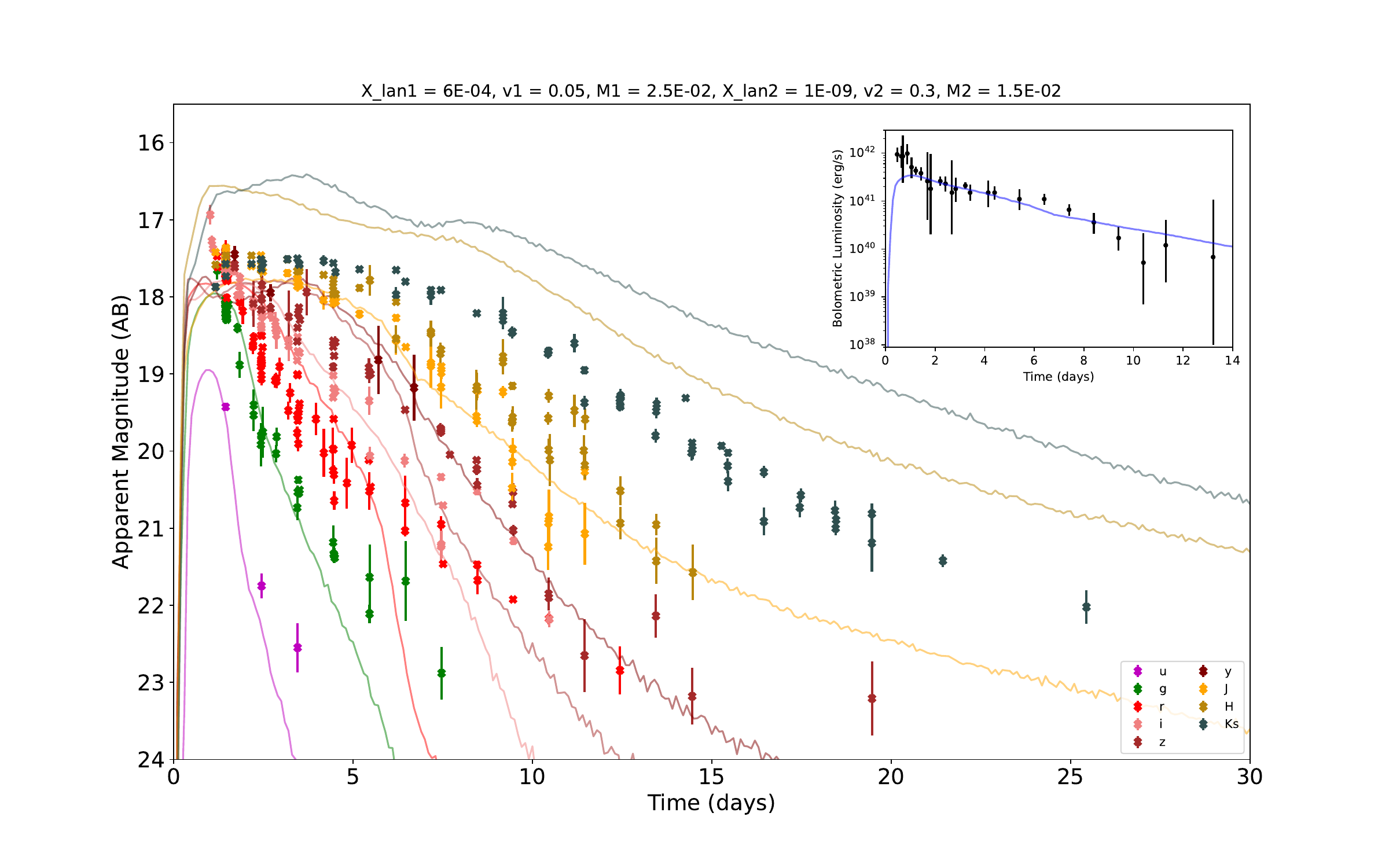}
    \caption{Best-fitting light curve model for Atomic Dataset \texttt{HULLAC} with bolometric luminosity points from \cite{Coughlin18} and model bolometric luminosity in the inset. \cite{Coughlin18} extrapolates the SED to redward of 2.5 $\mu$m by fitting a blackbody to photometric points, which tends to overestimate the true bolometric luminosities of the models presented in this work. The model tends to under-produce optical flux until $z$ band, agree for $J$ band, then over-produce $H$ and $Ks$ bands while having a similar bolometric luminosity. This is likely indicative of too much reprocessing of optical light into IR and demonstrates the difficulty of simplified 1D models fitting AT2017gfo.}
    \label{Fig:GWTanakaFit}
\end{figure*}

\begin{figure*}[h]
    \centering
  \includegraphics[width=0.87\textwidth,trim={3cm 1cm 3cm 3cm},clip]{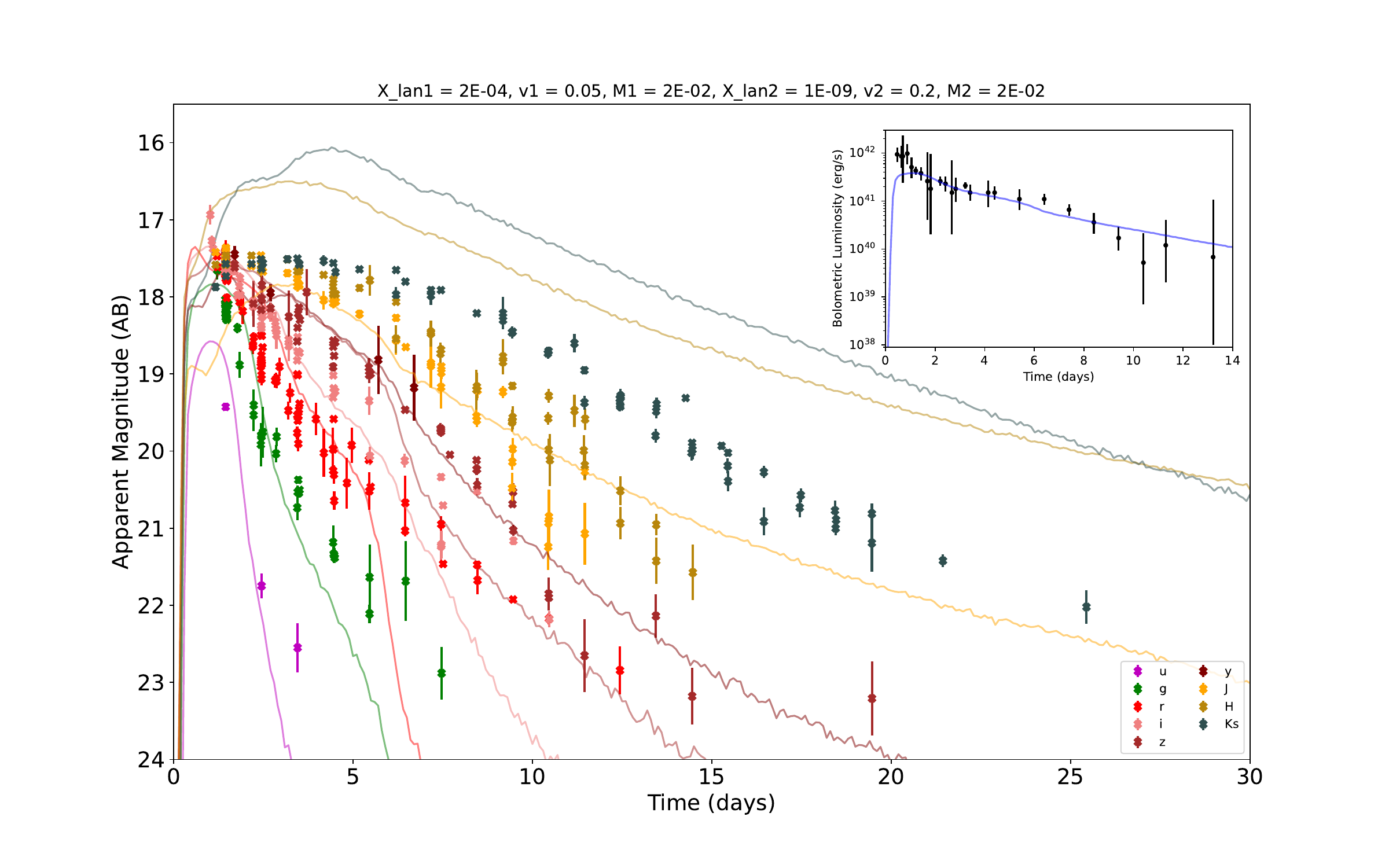}
    \caption{Best-fitting light curve model for Atomic Dataset \texttt{LASER}. Similarly to Figure \ref{Fig:GWTanakaFit}, \texttt{LASER} over-produces $H$ and $Ks$ bands while under-producing optical emission, especially at $\delta t \gtrsim 5$ d, though the bolometric luminosity curve is consistent.}
    \label{Fig:GWAlamosFit}
\end{figure*}

\begin{figure*}[h]
    \centering
  \includegraphics[width=0.87\textwidth,trim={3cm 1cm 3cm 3cm},clip]{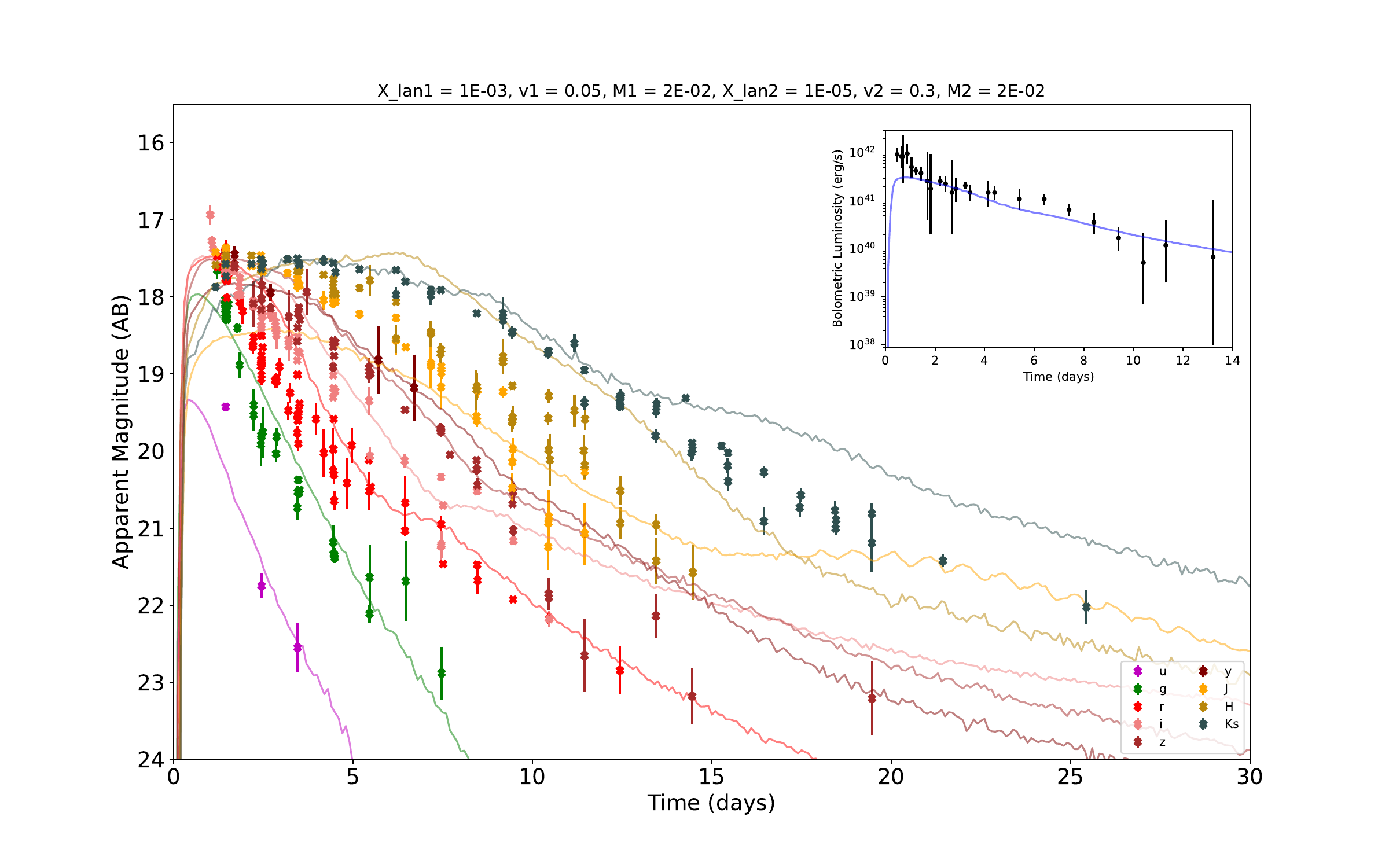}
    \caption{Best-fitting light curve model for Atomic Dataset \texttt{Autostructure}. Atomic Dataset \texttt{Autostructure} is the only dataset presented here that was capable of producing sufficiently long-lasting optical bands and not over-producing IR filters by more than a magnitude at early times, likely due to its treatment of light r-process elements.}
    \label{Fig:GWAutoFit}
\end{figure*}

The best-fitting model matching parameters for each atomic dataset are listed in Table \ref{Tab:BestFit}, and shown in Figures \ref{Fig:GWTanakaFit}, \ref{Fig:GWAlamosFit}, and \ref{Fig:GWAutoFit}. Importantly, the errors provided in Table \ref{Tab:BestFit} are representative of the spacing of the grid and do not reflect the true statistical uncertainties of fitting the AT2017gfo dataset; the errors are provided to represent our coarse grid and of the meaningfulness of the difference in parameter estimates between the Atomic Datasets.

The best-fitting models using Atomic Datasets \texttt{HULLAC} and \texttt{LASER} are overluminous in the $H$ and $Ks$ bands while tending to be slightly underluminous in optical bands (Figures \ref{Fig:GWTanakaFit}, \ref{Fig:GWAlamosFit}, and \ref{Fig:GWAutoFit}). However, the bolometric luminosity (shown in the upper right corner of each Figure as calculated by \citealt{Coughlin18} with the bolometric luminosity of the model in blue) is approximately correct. Taken together, one explanation is that this indicates these models produce approximately the correct luminosity but reprocess too much of the optical emission into the IR. Additionally, Atomic Datasets \texttt{HULLAC} and \texttt{LASER} have a long-lasting feature that aligns with the $H$ and $Ks$ filters but has a steep decline in flux just blueward the $J$ filter, causing the large color difference that does not exist in Atomic Dataset \texttt{Autostructure} models. 

However, the large IR output is needed to achieve the late time IR color observed in GW170817. \cite{Kasliwal22} present \textit{Spitzer} IR photometric points and limits at $\delta t$ = 43 and 74 days for 3.6 and 4.5 $\mu$m. While we recommend caution when extrapolating the LTE models presented in this work to such late times, the data presented in \cite{Kasliwal22} can help to constrain models. At $\delta t$ = 43 d, GW170817 has a 3.6 $\mu$m - 4.5 $\mu$m color of $> 1.3$ mags. By $\delta t$ = 74 d, the same color is $> -0.8$ mags. While high \Xlan\, and high mass models using Atomic Datasets \texttt{HULLAC} and \texttt{LASER} are capable of achieving the brightness of the 4.5 $\mu$m photometry point, the $\delta t$ = 43\,d minimal color is more constraining and only achieved by models using Atomic Dataset \texttt{LASER} under LTE assumptions in the equivalent JWST filters. While effects like NLTE, including actinides or super-heavy elements, or late-time enhanced heating rate from $^{254}$Ca \citep{Holmbeck23} that are not included in these models will likely alter these colors, the color evolution provides an ideal avenue to constrain atomic data and abundances. 

This is in sharp contrast to even the most massive (M = 0.1 M$_\odot$) and lanthanide-rich ($\log_{10}$ (\Xlan) = -2) models using Atomic Dataset \texttt{Autostructure}, which are insufficiently bright at 4.5 $\mu$m by approximately 4 magnitudes. NLTE radiative transfer will be critical to accurately modeling the color evolution out to such late times and has the potential to meaningfully constrain the atomic data of heavy metals. This has already been done with some success, as \cite{Hotokezaka22} identified a handful of candidate species capable of producing the bright 4.5 $\mu$m flux without 3.6 $\mu$m flux such as Se III, W III, Os III, Rh II, Rh III, and Ce IV using experimentally identified lines in the NIST database. Interestingly, the lack of a similar 4.5 $\mu$m - 3.6 $\mu$m color in the KN candidate associated with GRB 230307A \citep{Levan24} could be indicative of substantial composition differences among KNe and have implications for galactic r-process nucleosynthesis.

The inability of the models presented here to reproduce the IR light curve of AT2017gfo points to a tension in KN modeling where it is necessary to produce a bright enough IR transient at late times, but not so bright that the early time IR is over-produced. There are multiple possible avenues through which this issue could be resolved, including a radial gradient of \Xlan, NLTE effects, a more realistic 3D ejecta structure, or variations in r-process heating. Critically, this tension indicates that with idealized 1D models there is no Atomic Dataset that will perfectly fit the data.

Table \ref{Tab:BestFit} shows that the parameter most impacted by changing atomic datasets is the heavy metal content of the ejecta; the estimates range from the smallest amount of heavy metals using Atomic Dataset \texttt{LASER} (4$\times 10^{-6}$ M$_\odot$) to the most with Atomic Dataset \texttt{Autostructure} (2.2$\times 10^{-5}$ M$_\odot$), differing by a factor of $\sim6$. Despite this high level of variance in lanthanide mass between models, the bolometric luminosity is well approximated by the models at $\delta t \gtrsim 1$\,d. The lanthanide-rich component varies by $\sim 0.75$ orders of magnitude while the lanthanide-poor component can vary much more substantially by 5 orders of magnitude. However, at such low \Xlan, the models are largely insensitive to changes until $\log_{10}$(\Xlan) $\sim -4$ as the lanthanide bound-bound opacities become subdominant which is likely the cause of such large uncertainty. 

Mass and velocity are less impacted by changes in Atomic Dataset, with agreement among the models within $\pm 0.005$ M$_\odot$, and velocity within $\pm 0.1$c. Total mass is even less impacted across the Atomic Datasets, as it remains constant across all `best-fitting' models. The overall level of agreement in mass and velocity may be a result of optical filters dominating the fit, as there were five optical filters and only three IR filters.

\begin{deluxetable*}{lccccccccc}[h!]
\tabletypesize{\scriptsize}

\setlength{\tabcolsep}{10pt}
\tablecolumns{9} 
\tablewidth{24pc}
\vspace{-1cm}
\tablecaption{``Best-fitting" model matching ejecta parameters from our $\chi^2$ fitting of GW170817 data. Uncertainties are based on grid spacing. 
\label{Tab:BestFit}}
\tablehead{\colhead{Dataset} & \colhead{M$_1$} & \colhead{v$_1$} &  \colhead{$\log_{10} $\Xlan$_1$}& \colhead{M$_2$}& \colhead{v$_2$} & \colhead{$\log_{10}$\Xlan$_2$} & \colhead{Total Lanthanide Mass} & \colhead{$\chi^2$}\\[0.0 cm]
 & (10$^{-2}$ M$_\odot$) & (c) &  & (10$^{-2}$ M$_\odot$)  & (c) & & (M$_\odot$) &}

\startdata
\texttt{HULLAC} & 2.5$^{+0.13}_{-0.13}$ & 0.05$^{+0.01}_{-0.01}$ & -3.25$^{+0.13}_{-0.13}$ & 1.5$^{+0.13}_{-0.13}$  & 0.3$^{+0.025}_{-0.025}$ & -9$^{+2}_{-2}$ & $1.5\times10^{-5}$ & 1.12 \\ 
\hline
\texttt{LASER} & 2$^{+0.13}_{-0.13}$ & 0.05$^{+0.01}_{-0.01}$  & -3.75$^{+0.13}_{-0.13}$  & 2$^{+0.13}_{-0.13}$ & 0.2$^{+0.013}_{-0.013}$ & -9$^{+2}_{-2}$ & $4\times10^{-6}$& 1.26 \\ 
\hline
\texttt{Autostructure} & 2$^{+0.13}_{-0.13}$  & 0.05$^{+0.01}_{-0.01}$  & -3$^{+0.13}_{-0.13}$  & 2$^{+0.5}_{-0.5}$  & 0.3$^{+0.05}_{-0.05}$ & -4$^{+0.5}_{-0.5}$ & $2.2\times10^{-5}$ & 0.319 \\
\hline
\enddata
\end{deluxetable*}

\begin{deluxetable*}{lccccccc}[h!]
\tabletypesize{\scriptsize}

\setlength{\tabcolsep}{10pt}
\tablecolumns{7} 
\tablewidth{24pc}
\vspace{-2.5cm}
\tablecaption{``Best-fitting" ejecta parameters from the literature of GW170817 data with 1$\sigma$ errors
$^{\dagger}$ Quantity was fixed in modeling
\label{Tab:BestFitOther}}
\tablehead{\colhead{Source} & \colhead{M$_1$} & \colhead{v$_1$} &  \colhead{$\log_{10} $\Xlan$_1$}& \colhead{M$_2$}& \colhead{v$_2$} & \colhead{$\log_{10}$\Xlan$_2$} \\[0.0 cm]
 & (10$^{-2}$ M$_\odot$) & (c) &  & (10$^{-2}$ M$_\odot$)  & (c) & &}

\startdata
\cite{Coughlin18} - Light Curve & 3.09$^{+1.63}_{-1.92}$ & 0.1$^{+0.08}_{-0.06}$ & -1.61$^{+0.96}_{-1.04}$ &   2.57$^{+0.95}_{-1.07}$& 0.17$^{+0.09}_{-0.1}$ & -4.73$^{+0.41}_{-0.2}$  \\ 
\hline
\cite{Villar17} - 2 Component & 5$^{+0.1}_{-0.1}$ & 0.149$^{+0.001}_{-0.002}$  & $\kappa$ = 3.65$^{+0.09}_{-0.28}$ cm$^2$ g$^{-1}$ & 2.3$^{+0.5}_{-0.1}$ &  0.256$^{+0.005}_{-0.002}$ &  $\kappa^{\dagger}$ = 0.5 cm$^2$ g$^{-1}$   \\ 
\hline
\cite{Ristic23} - Weighted &  1.91$^{+0.3}_{-0.3}$ & 0.2$^{+0.01}_{-0.01}$  & -1.28$^{\dagger}$  & 1.58$^{+0.3}_{-0.5}$  &  0.13 $^{+0.04}_{-0.04}$ & -$\infty^{\dagger}$  \\
\hline
\enddata
\end{deluxetable*}

\vspace{-1.7cm}
\subsubsection{Comparison to Other Works} \label{Subsubsec:Other}

While we stress that our fitting methods are based on model matching and are meant to quantify the error caused by thermalization efficiency prescription and atomic data, it is nonetheless helpful to place the derived numbers in context of those already found for AT2017gfo. 

We consider three different estimates of the parameters from AT2017gfo, listed in Table \ref{Tab:BestFitOther}, from \cite{Villar17}, \cite{Coughlin18}, and \cite{Ristic23}. The ``fits" presented in this work most closely resemble the estimations of \cite{Coughlin18}, with most parameters agreeing within $\sim1\sigma$, though the estimates of \cite{Coughlin18} are often larger. This is to be expected as there are many similarities between the process that went into fitting the light curves; \cite{Coughlin18} used the models from \cite{Kasen17} in their fitting process, which used Atomic Dataset \texttt{Autostructure} and is the closest in agreement. To represent a 2-component KN, \cite{Coughlin18} also summed two 1D models. The slight disagreement in \Xlan\, and $< 1\sigma$ mass larger estimate may also be a result of differences in composition, where the \cite{Kasen17} models that \cite{Coughlin18} use had a flat distribution of non-lanthanide material, as opposed to the solar abundance distribution used in the models presented here.

\cite{Ristic23} uses models produced with Atomic Dataset \texttt{LASER} for their fitting process, and imposes a fixed composition with no lanthanides in the lanthanide-poor component and a $\log_{10}($\Xlan) $\sim -1.28$ lanthanide-rich component. Their mass estimates are smaller than those we derive for Atomic Dataset \texttt{LASER}, though this could be due to their use of the global thermalization-prescription and therefore higher retention of r-process decay energy compared to the local thermalization-prescription used to ``fit" AT2017gfo in this work. Additionally, the fixed higher \Xlan in their lanthanide-rich component will cause more optical light to be reprocessed into IR for a longer time. To compensate, a faster lanthanide-rich component would cause a faster decline in IR bands to match AT2017gfo, possibly resulting in the finding of a higher velocity than our model matching fits. As for the lanthanide-poor component, the lower characteristic velocity ($\Delta v \sim 0.07c$) would generate a longer-lasting optical transient and potentially `replenish' some of the additional optical light that had been reprocessed into IR due to the higher \Xlan in the lanthanide-rich component.

Finally, we compare our results to that of the two component fit from \cite{Villar17}. The total mass of the two-component model in \cite{Villar17} is approximately double that of the model matching estimates presented here, as well as a much faster lanthanide-rich component ($\Delta v \sim 0.1c$) and somewhat slower lanthanide-poor component ($\Delta v \sim 0.05c$). This may be a result of their wavelength-independent opacity models, which would under-predict the opacity at optical and UV wavelengths while over-predicting the opacity at IR wavelengths. As a result, less emission will be reprocessed into the IR, creating a brighter optical light curve and a fainter IR light curve. 
To create a sufficiently bright IR transient, the gray opacity scheme will require a more massive lanthanide-rich component. However, to prevent the model from remaining too bright for too long, the fitting procedure may have favored a higher velocity model.

\subsection{``Error Estimation" Self-Comparison} \label{Subsec:Self}

To more generally quantify how atomic data uncertainties affect KN parameter estimations, we consider each model in a given Atomic Dataset and find a best-fitting model from the other two Atomic Datasets. We use the same grid used to ``fit" the GW170817 data, and only use local thermalization efficiency prescription models. We again use a weighted $\chi^2$ fitting (Equation \ref{Eq:ChiSquare}) to determine the best-fitting model with the weighting to equalize the importance of each band. Each model is fit using absolute magnitudes. Due to the large disagreement in optical data between Atomic Datasets \texttt{HULLAC} and \texttt{LASER} with Atomic Dataset \texttt{Autostructure} as well as avoid biasing due to similarities in the optical of Atomic Datasets \texttt{HULLAC} and \texttt{LASER} from their shared non-lanthanide data, we only fit for $J, H,$ and $Ks$ bands. We apply this fitting method for models with $\log_{10} ($\Xlan$) \geq -4$, as the bound-bound transitions of lanthanides will eventually become subdominant to non-lanthanides at smaller values of \Xlan. 

The baseline model has data sampled at a frequency of 1 day plus a $\delta t$ determined randomly by a Gaussian distribution of $\mu = 0$ and $\sigma = 0.2$ to mimic a nightly observing campaign. Additionally, for each photometric point we add Gaussian noise with standard deviation equal to the scatter of the approximately similar filters in GW170817. 

\begin{figure*}
    \centering
  \includegraphics[width=0.87\textwidth,trim={2cm 9.5cm 3cm 3.5cm},clip]{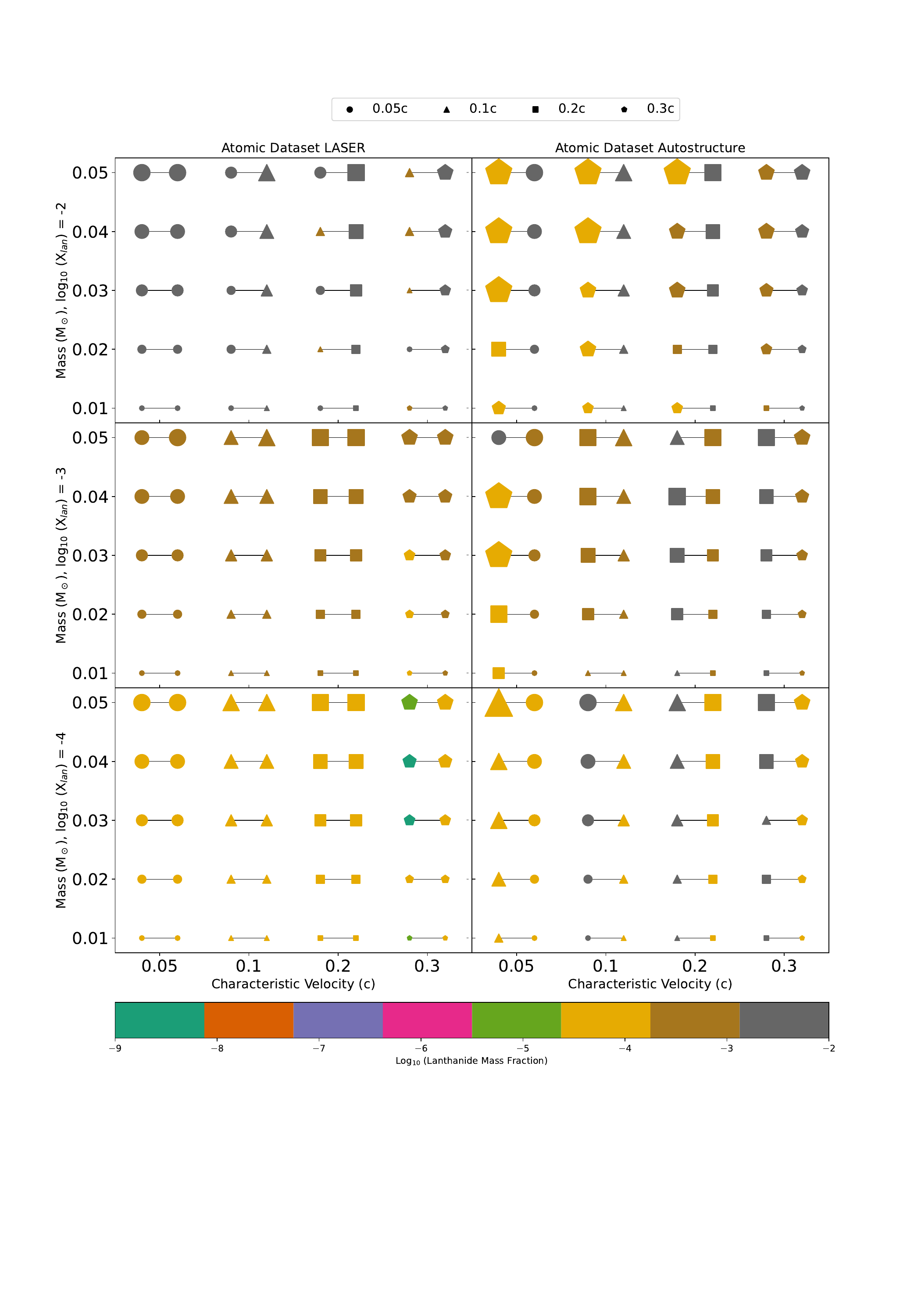}
    \caption{Best-fitting \texttt{LASER} and \texttt{Autostructure} models for models that used Atomic Dataset \texttt{HULLAC} on the left and right, respectively. At each point on the model grid, the left and right symbol connected by a line represent the best-fitting model from the other Atomic Dataset and a symbol representing the true parameters, respectively. Color corresponds to the best-fitting \Xlan, while a circle, triangle, square, and pentagon represent a best-fitting velocity of 0.05, 0.1, 0.2, and 0.3 c, respectively. Size of the marker correlates to the best-fitting mass. Atomic Dataset \texttt{LASER} fits accurately at $\log_{10} ($ \Xlan) = -4, -3 (though at higher velocities a 1 dex discrepancy in \Xlan emerges), while $\log_{10} ($ \Xlan) = -2 masses and velocities tend to be underestimated by 25-40\% and 0.1-0.2c. Atomic Dataset \texttt{Autostructure} often overestimates the mass by as much as a factor of 3 at low velocities, and consistently struggles to fit the correct velocity at all.}
    \label{Fig:TanakaGridFit}
\end{figure*}

\begin{figure*}
    \centering
  \includegraphics[width=0.87\textwidth,trim={2cm 12.5cm 3cm 4.25cm},clip]{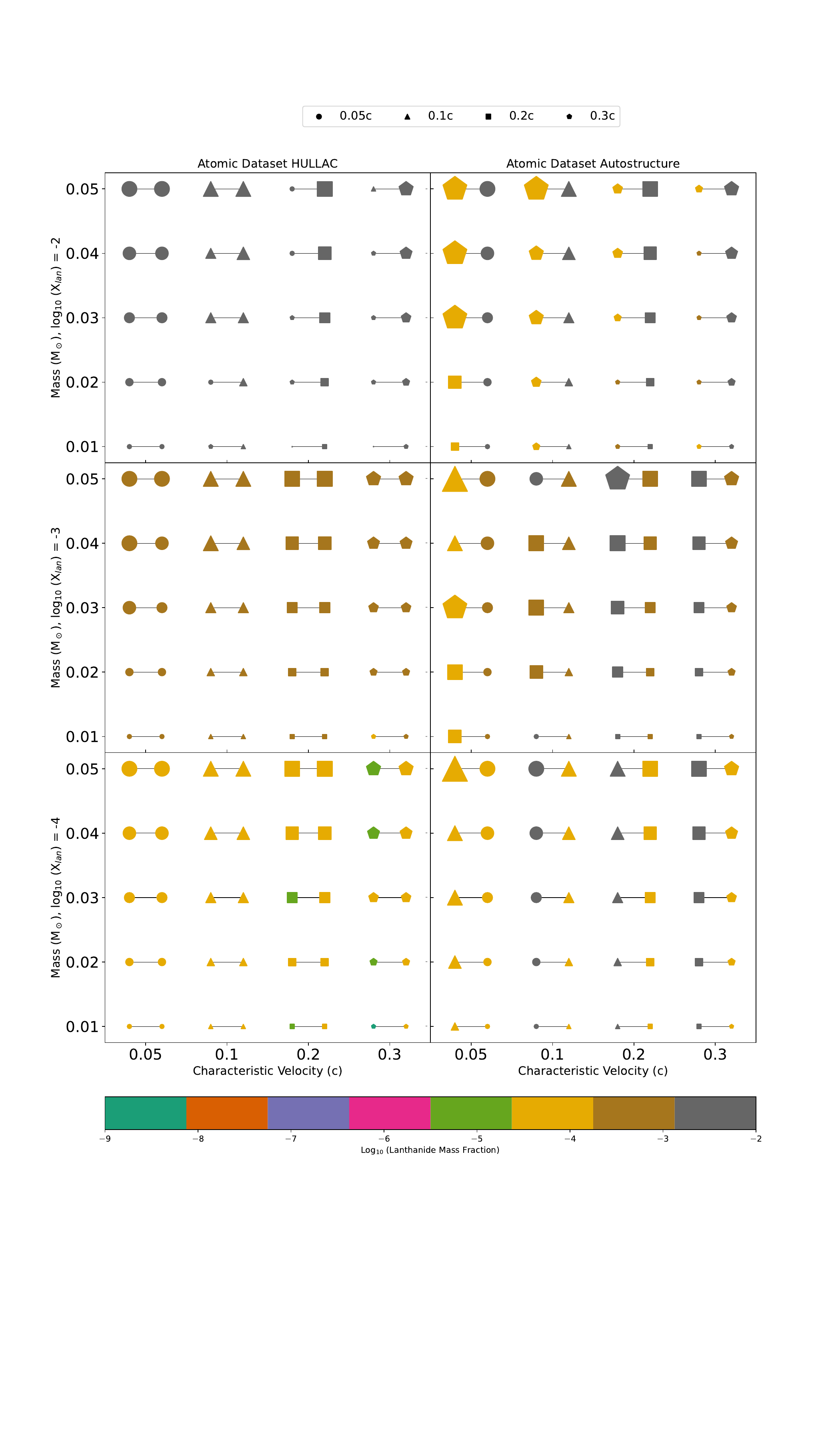}
    \caption{Same as Figure \ref{Fig:TanakaGridFit} except for models that used Atomic Dataset \texttt{LASER}. Similarly to Figure \ref{Fig:TanakaGridFit}, Atomic Dataset \texttt{HULLAC} fits Atomic Dataset \texttt{LASER} well at $\log_{10} ($ \Xlan) = -4, -3 (though at higher velocities a less frequent 1 dex discrepancy in \Xlan emerges). However, at $\log_{10} ($ \Xlan) = -2 Atomic Dataset \texttt{HULLAC} underestimates the mass by a factor of 2 or 3, particularly at higher velocities. Atomic Dataset \texttt{Autostructure} often overestimates the mass by as much as a factor of 3 at low velocities, and consistently struggles to fit the correct velocity at all.}
    \label{Fig:AlamosGridFit}
\end{figure*}

\begin{figure*}
    \centering
  \includegraphics[width=0.87\textwidth,trim={2cm 12cm 3cm 4.25cm},clip]{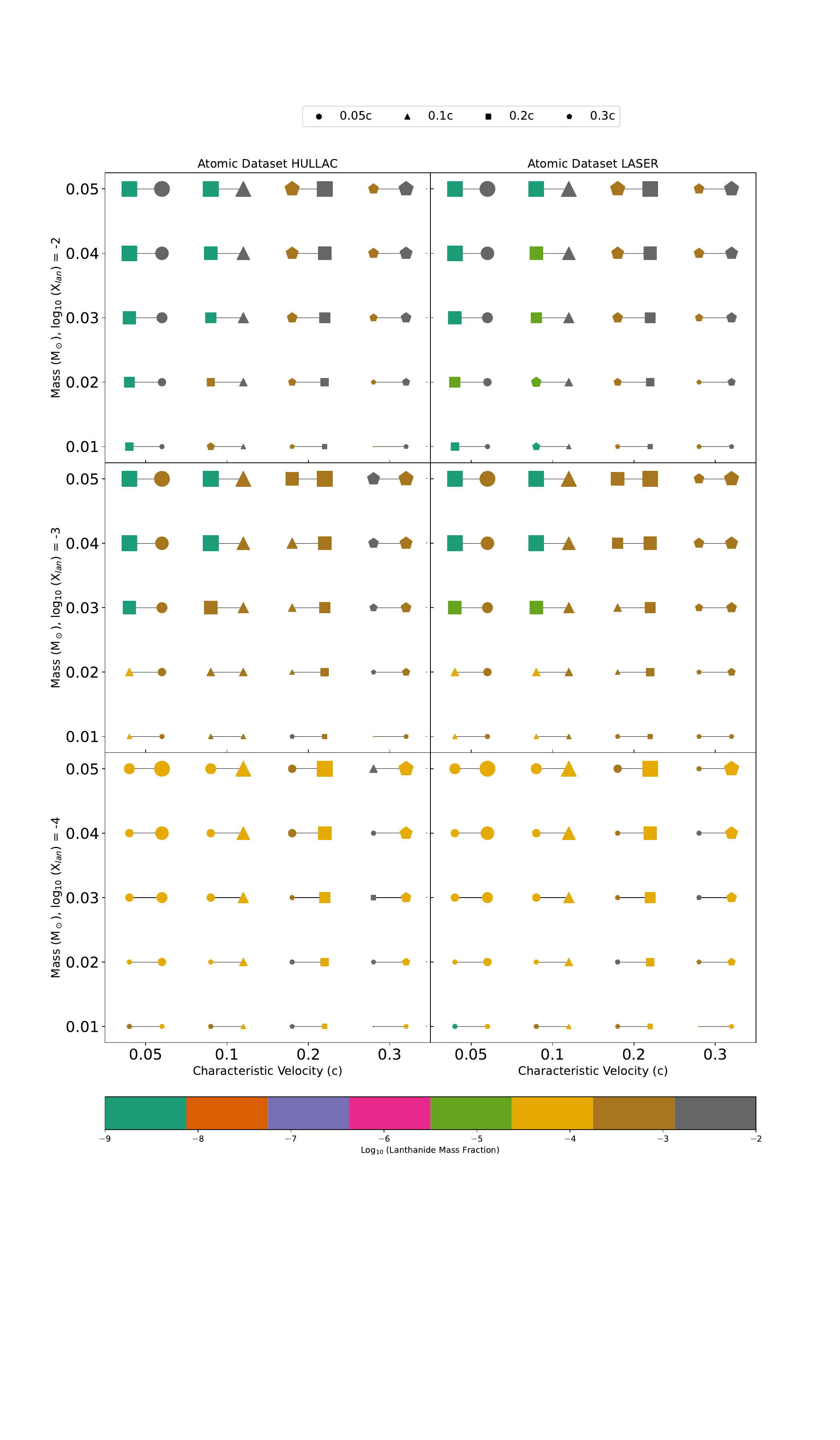}
    \caption{Same as Figure \ref{Fig:TanakaGridFit} except for models that used Atomic Dataset \texttt{Autostructure}. Atomic Datasets \texttt{HULLAC} and \texttt{LASER} have very similar fits: i) require much lower \Xlan and high velocities to match the IR of \texttt{Autostructure} models, ii) tend to overestimate the velocity by 0.1-0.2c, and iii) incorrectly estimate the mass by one grid step (20-50\%).}
    \label{Fig:AutoGridFit}
\end{figure*}

Figures \ref{Fig:TanakaGridFit}, \ref{Fig:AlamosGridFit}, and \ref{Fig:AutoGridFit} show the results of this exercise for each combination of mass and characteristic velocity with $\log_{10} ($ \Xlan $) = -2, -3,$ and $-4$ from the top row to bottom row. When using Atomic Data \texttt{HULLAC} to fit models run with Atomic Dataset \texttt{LASER} and vice versa, there is good agreement in mass, velocity, and \Xlan\, estimation overall for $\log_{10} ($ \Xlan $)$ = -4, -3. However, at higher characteristic velocities like 0.2 and 0.3c, a $\pm 1$ dex uncertainty in \Xlan\, begins to emerge. Similarly, at low velocity-high mass end, a $\pm 0.01$ M$_\odot$ (20-25\%) emerges.

Yet, the highest \Xlan\, fitting reveals a much larger uncertainty with velocity estimations being incorrect by $\sim 0.1-0.2$c and mass estimates off by $\pm 0.02$ M$_\odot$ (25-40\%) for some of the highest mass models. For 8 out of the 20 models, Atomic Dataset \texttt{LASER} requires a smaller \Xlan\, to reproduce a similar IR light curve to those made by Atomic Dataset \texttt{HULLAC}, with most of those being at the higher velocity end. In the other direction, when fitting Atomic Dataset \texttt{LASER} models with those from Atomic Dataset \texttt{HULLAC} always requires the highest \Xlan.

Atomic Dataset \texttt{Autostructure} consistently struggles to fit the light curves from Atomic Datasets \texttt{HULLAC} and \texttt{LASER}, and vice versa. While the mass estimates are the most accurate, they still tend to be incorrect by $\sim 0.02$ M$_\odot$ (25-40\%), and up to $0.05$ M$_\odot$ ($> 100 \%$). Figure \ref{Fig:AutoGridFit} clearly shows a degeneracy between \Xlan\, and velocity, with the lower \Xlan\, models fitting lower velocities better because the IR peaks are more delayed in low \Xlan\, models of Atomic Dataset \texttt{Autostructure}, compared to the much more immediate brightening produced by Atomic Datasets \texttt{HULLAC} and \texttt{LASER}. The faster Atomic Dataset \texttt{Autostructure} models achieve their IR peaks much more rapidly, and so there are better fits at higher \Xlan\,. Despite the major discrepancy in \Xlan\, of the base Atomic Dataset \texttt{Autostructure} model and the ejecta parameters of the Atomic Datasets \texttt{HULLAC} and \texttt{LASER} fits (7-8 dex), the reduced $\chi^2$ is $\sim0.27$, indicating a high quality fit. This discrepancy is reduced in the lower \Xlan\, models, though at the cost of incorrectly estimating the velocity as almost all models are best fit by models of velocity 0.05c. 

Velocity agreement improves at higher velocities and lower \Xlan, but \Xlan\, estimates get worse at higher velocities. Mass estimates tend to consistently be off by $\sim 0.01$ or $0.02$ M$_\odot$ across all Atomic Datasets, though reaching as high as 0.07 M$_\odot$ for a 0.03 M$_\odot$ model. However, spectroscopy is capable of giving a direct measurement of the characteristic velocity and thus break uncertainties making many of the error estimates for velocity moot. In cases where spectroscopy was not obtained or impossible to obtain such as a target that is too faint or a KN that was discovered in the LSST archives, these errors may become applicable. 

\subsection{Impacts on Galactic Nucleosynthesis}

Based on the systematic uncertainties in mass and \Xlan\, estimates in the previous sections, here we propagate the impact of those uncertainties to conclusions about KN r-process as the source of galactic r-process elements. Following a prescription similar to that of \cite{Rosswog17}, we can approximate the amount of lanthanides in the Milky Way, assuming that all lanthanides come from a single type of event, as:




\begin{equation}
    M_{\rm lan} = \mathcal{R} \times \overline{M_{ej}} \times X_{\rm lan} \times \tau_{MW},
\end{equation}

where $M_{\rm lan}$ is the total lanthanide mass in the Milky Way, $\mathcal{R}$ is the rate of mergers per year in the Milky Way, $\overline{M_{ej}}$ is the average ejected mass, and $\tau_{MW}$ is the age of the Milky Way. 

Under this prescription, to achieve the same amount of lanthanides in the Milky Way, a decrease in the compact object merger rate would have to be compensated by a proportional increase in the typical mass of the ejecta and vice versa assuming a constant \Xlan. \cite{Lippuner15} calculate that at most, material undergoing r-process can achieve \Xlan $\sim 35\%$, though the typical low Y$_e$ material is much closer to 10\%, placing an upper limit for that parameter. 

Therefore, the errors presented above (Sections \ref{Subsec:Atomic}, \ref{Subsec:Therm}, \ref{Subsec:170817}, \ref{Subsubsec:Other}, \ref{Subsec:Self}) directly translate into the same percent errors in the required compact object merger rate: a 0.75 dex uncertainty in \Xlan\, translates to roughly a factor of 6 change in the compact object merger rate (assuming the same ejecta mass), and the 25-40\% mass error from atomic data as well as 30\% error from thermalization efficiency would require a factor of $\sim1.3-1.4$ change in compact merger rate. 

To place these errors in context, currently the estimates of compact merger rates are 360 - 1800 events Gpc$^{-3}$ yr$^{-1}$ based on Chandra and XMM-Newton short gamma ray bursts \citep{RoucoEscorial23}, 10 - 1700 BNS merger events Gpc$^{-3}$ yr$^{-1}$ (90\% confidence) based on LIGO rates \citep{Abbott23}, and 244$^{+325}_{-166}$ merger events Gpc$^{-3}$ yr$^{-1}$ based on population synthesis studies from \cite{Kim15} and a conversion factor of $1.16 \times 10^{2}$ Milky Way-like galaxies Mpc$^{-3}$ from \cite{Abadie10}, as discussed in \cite{Rosswog17}. 

While there are still many other sources in error of the compact merger rate such as the opening angle distribution of gamma ray burst jets or the minimum gamma ray luminosity \citep{RoucoEscorial23}, and assumptions about the underlying neutron star population like mass and spin \citep{Abbott23}, among those presented in this work the lanthanide mass fraction has the highest impact.


\section{Conclusions} \label{Sec:Conc}

In this paper, we presented the systematic uncertainties in KN ejecta properties that arise from variance in thermalization efficiency prescriptions and unknown lanthanide atomic data based on fitting light curves of both real KN data from AT2017gfo as well as cross-fitting synthetic light curve models. We presented KN models that span the expected ejecta mass, velocity, and \Xlan\, range using three atomic datasets and two thermalization efficiency prescriptions. We show that the typical errors quoted in parameter estimates (such as a 0.02, 0.05, 0.08 dex in mass error, \citealt{Ristic23,Breschi21,Pang23}, 0.3 dex in \Xlan error, \citealt{Gillanders22}, $< 3\%$ in velocity error \citealt{Breschi21}) do not include the model uncertainties and are significantly smaller than the the 20-50\% uncertainty in mass caused by thermalization-prescription and 25-40\% from atomic data uncertainty. \Xlan\, can vary by 0.75 orders of magnitude for the lanthanide-rich component and as much as 5 orders of magnitude for the lanthanide-poor component by comparing model matching fits of AT2017gfo light curves (the significantly higher uncertainty on the lanthanide-poor component is likely due to the sub-dominance of bound-bound transitions at such low \Xlan). 

Fitting IR light curves with models generated from other Atomic Datasets reveals a typical mass error of 20-40\%, though the error can be as high as a factor of 3. Velocity errors are typically 0.1-0.2c ($\sim 100\%)$, though spectroscopic follow-up would help to eliminate this uncertainty. Similarly, errors in \Xlan\, are more severe at higher \Xlan, with a typical error of $\pm 1$ order of magnitude.

The uncertainties derived in this work imply the total inferred lanthanide production of BNS mergers can vary by a factor of $\sim 6$. Therefore, if BNS merger events are the sole contributor of lanthanides to the r-process enrichment of the Milky Way, the merger rate required to reproduce the abundances seen in the Milky Way have an additional uncertainty factor of $\sim 6$.

While the work presented here identified a few sources of systematic uncertainties, there are still many other systematics that are worthwhile to explore further. NLTE effects, compositional gradients, asymmetries in the ejecta, variable radioactive heating rates, the inclusion of fission and fission fragments, and delayed thermalization may impact the light curves and spectral series substantially. 

Spectral sequences will be able to reduce the impact of some, though not all, of the uncertainties presented in this work. Spectra can provide a much more direct estimate of ejecta velocities, inform viewing angle dependencies, and at late times MIR spectra could constrain individual atomic abundances as well as inform atomic data modeling.

This work highlights the need for progress on both the theoretical and observational front to better constrain underlying KN physics. A nearby KN targeted by facilities such as JWST would provide ample data in the MIR to identify important lines and constrain atomic models, indicating the need for sufficient Target of Opportunity request availability. Further GRMHD simulations of compact object mergers containing a neutron star can constrain ejecta morphology and ejection mechanisms to inform radiative transfer models, and therefore make more realistic assumptions for more accurate models.

\section{Acknowledgements}


DB is partially supported by a NASA Future Investigators in NASA Earth and Space Science and Technology (FINESST) award No. 80NSSC23K1440. R.M. acknowledges support by the National Science Foundation under award No. AST-2221789 and AST-2224255. The TReX team at UC Berkeley is partially funded by the Heising-Simons Foundation under grant 2021-3248 (PI: Margutti).
DK is supported in part by the U.S. Department of Energy, Office of Science, Division of Nuclear Physics, under award numbers DE-SC0004658 and DE-SC0024388, and by the Simons Foundation (award number 622817DK).
This research used the Savio computational cluster resource provided by the Berkeley Research Computing program at the University of California, Berkeley (supported by the UC Berkeley Chancellor, Vice Chancellor for Research, and Chief Information Officer).


%

\vspace{5mm}


\software{numpy \citep{Numpy}, \texttt{sedona} \citep{Kasen06,Roth15}, astropy \citep{astropy:2013,astropy:2018,astropy:2022}, matplotlib \citep{Matplotlib}, h5py
          }

\textit{Data Availability}: All spectra will be made publicly available through Zenodo.

\bibliography{KNe}{}

\begin{thebibliography}{}
\expandafter\ifx\csname natexlab\endcsname\relax\def\natexlab#1{#1}\fi
\providecommand{\url}[1]{\href{#1}{#1}}
\providecommand{\dodoi}[1]{doi:~\href{http://doi.org/#1}{\nolinkurl{#1}}}
\providecommand{\doeprint}[1]{\href{http://ascl.net/#1}{\nolinkurl{http://ascl.net/#1}}}
\providecommand{\doarXiv}[1]{\href{https://arxiv.org/abs/#1}{\nolinkurl{https://arxiv.org/abs/#1}}}

\bibitem[{{Abadie} {et~al.}(2010){Abadie}, {Abbott}, {Abbott}, {Abernathy}, {Accadia}, {Acernese}, {Adams}, {Adhikari}, {Ajith}, {Allen}, {Allen}, {Amador Ceron}, {Amin}, {Anderson}, {Anderson}, {Antonucci}, {Aoudia}, {Arain}, {Araya}, {Aronsson}, {Arun}, {Aso}, {Aston}, {Astone}, {Atkinson}, {Aufmuth}, {Aulbert}, {Babak}, {Baker}, {Ballardin}, {Ballmer}, {Barker}, {Barnum}, {Barone}, {Barr}, {Barriga}, {Barsotti}, {Barsuglia}, {Barton}, {Bartos}, {Bassiri}, {Bastarrika}, {Bauchrowitz}, {Bauer}, {Behnke}, {Beker}, {Belczynski}, {Benacquista}, {Bertolini}, {Betzwieser}, {Beveridge}, {Beyersdorf}, {Bigotta}, {Bilenko}, {Billingsley}, {Birch}, {Birindelli}, {Biswas}, {Bitossi}, {Bizouard}, {Black}, {Blackburn}, {Blackburn}, {Blair}, {Bland}, {Blom}, {Blomberg}, {Boccara}, {Bock}, {Bodiya}, {Bondarescu}, {Bondu}, {Bonelli}, {Bork}, {Born}, {Bose}, {Bosi}, {Boyle}, {Braccini}, {Bradaschia}, {Brady}, {Braginsky}, {Brau}, {Breyer}, {Bridges}, {Brillet}, {Brinkmann}, {Brisson}, {Britzger}, {Brooks}, {Brown},
  {Budzy{\'n}ski}, {Bulik}, {Bulten}, {Buonanno}, {Burguet-Castell}, {Burmeister}, {Buskulic}, {Byer}, {Cadonati}, {Cagnoli}, {Calloni}, {Camp}, {Campagna}, {Campsie}, {Cannizzo}, {Cannon}, {Canuel}, {Cao}, {Capano}, {Carbognani}, {Caride}, {Caudill}, {Cavagli{\`a}}, {Cavalier}, {Cavalieri}, {Cella}, {Cepeda}, {Cesarini}, {Chalermsongsak}, {Chalkley}, {Charlton}, {Chassande Mottin}, {Chelkowski}, {Chen}, {Chincarini}, {Christensen}, {Chua}, {Chung}, {Clark}, {Clark}, {Clayton}, {Cleva}, {Coccia}, {Colacino}, {Colas}, {Colla}, {Colombini}, {Conte}, {Cook}, {Corbitt}, {Corda}, {Cornish}, {Corsi}, {Costa}, {Coulon}, {Coward}, {Coyne}, {Creighton}, {Creighton}, {Cruise}, {Culter}, {Cumming}, {Cunningham}, {Cuoco}, {Dahl}, {Danilishin}, {Dannenberg}, {D'Antonio}, {Danzmann}, {Dari}, {Das}, {Dattilo}, {Daudert}, {Davier}, {Davies}, {Davis}, {Daw}, {Day}, {Dayanga}, {De Rosa}, {DeBra}, {Degallaix}, {del Prete}, {Dergachev}, {DeRosa}, {DeSalvo}, {Devanka}, {Dhurandhar}, {Di Fiore}, {Di Lieto}, {Di Palma}, {Emilio},
  {Di Virgilio}, {D{\'\i}az}, {Dietz}, {Donovan}, {Dooley}, {Doomes}, {Dorsher}, {Douglas}, {Drago}, {Drever}, {Driggers}, {Dueck}, {Dumas}, {Eberle}, {Edgar}, {Edwards}, {Effler}, {Ehrens}, {Engel}, {Etzel}, {Evans}, {Evans}, {Fafone}, {Fairhurst}, {Fan}, {Farr}, {Fazi}, {Fehrmann}, {Feldbaum}, {Ferrante}, {Fidecaro}, {Finn}, {Fiori}, {Flaminio}, {Flanigan}, {Flasch}, {Foley}, {Forrest}, {Forsi}, {Fotopoulos}, {Fournier}, {Franc}, {Frasca}, {Frasconi}, {Frede}, {Frei}, {Frei}, {Freise}, {Frey}, {Fricke}, {Friedrich}, {Fritschel}, {Frolov}, {Fulda}, {Fyffe}, {Gammaitoni}, {Garofoli}, {Garufi}, {Gemme}, {Genin}, {Gennai}, {Gholami}, {Ghosh}, {Giaime}, {Giampanis}, {Giardina}, {Giazotto}, {Gill}, {Goetz}, {Goggin}, {Gonz{\'a}lez}, {Gorodetsky}, {Go{\ss}ler}, {Gouaty}, {Graef}, {Granata}, {Grant}, {Gras}, {Gray}, {Greenhalgh}, {Gretarsson}, {Greverie}, {Grosso}, {Grote}, {Grunewald}, {Guidi}, {Gustafson}, {Gustafson}, {Hage}, {Hall}, {Hallam}, {Hammer}, {Hammond}, {Hanks}, {Hanna}, {Hanson}, {Harms}, {Harry},
  {Harry}, {Harstad}, {Haughian}, {Hayama}, {Heefner}, {Heitmann}, {Hello}, {Heng}, {Heptonstall}, {Hewitson}, {Hild}, {Hirose}, {Hoak}, {Hodge}, {Holt}, {Hosken}, {Hough}, {Howell}, {Hoyland}, {Huet}, {Hughey}, {Husa}, {Huttner}, {Huynh-Dinh}, {Ingram}, {Inta}, {Isogai}, {Ivanov}, {Jaranowski}, {Johnson}, {Jones}, {Jones}, {Jones}, {Ju}, {Kalmus}, {Kalogera}, {Kandhasamy}, {Kanner}, {Katsavounidis}, {Kawabe}, {Kawamura}, {Kawazoe}, {Kells}, {Keppel}, {Khalaidovski}, {Khalili}, {Khazanov}, {Kim}, {Kim}, {King}, {Kinzel}, {Kissel}, {Klimenko}, {Kondrashov}, {Kopparapu}, {Koranda}, {Kowalska}, {Kozak}, {Krause}, {Kringel}, {Krishnamurthy}, {Krishnan}, {Kr{\'o}lak}, {Kuehn}, {Kullman}, {Kumar}, {Kwee}, {Landry}, {Lang}, {Lantz}, {Lastzka}, {Lazzarini}, {Leaci}, {Leong}, {Leonor}, {Leroy}, {Letendre}, {Li}, {Li}, {Lin}, {Lindquist}, {Lockerbie}, {Lodhia}, {Lorenzini}, {Loriette}, {Lormand}, {Losurdo}, {Lu}, {Luan}, {Lubi{\'n}ski}, {Lucianetti}, {L{\"u}ck}, {Lundgren}, {Machenschalk}, {MacInnis}, {Mackowski},
  {Mageswaran}, {Mailand}, {Majorana}, {Mak}, {Man}, {Mandel}, {Mandic}, {Mantovani}, {Marchesoni}, {Marion}, {M{\'a}rka}, {M{\'a}rka}, {Maros}, {Marque}, {Martelli}, {Martin}, {Martin}, {Marx}, {Mason}, {Masserot}, {Matichard}, {Matone}, {Matzner}, {Mavalvala}, {McCarthy}, {McClelland}, {McGuire}, {McIntyre}, {McIvor}, {McKechan}, {Meadors}, {Mehmet}, {Meier}, {Melatos}, {Melissinos}, {Mendell}, {Men{\'e}ndez}, {Mercer}, {Merill}, {Meshkov}, {Messenger}, {Meyer}, {Miao}, {Michel}, {Milano}, {Miller}, {Minenkov}, {Mino}, {Mitra}, {Mitrofanov}, {Mitselmakher}, {Mittleman}, {Moe}, {Mohan}, {Mohanty}, {Mohapatra}, {Moraru}, {Moreau}, {Moreno}, {Morgado}, {Morgia}, {Morioka}, {Mors}, {Mosca}, {Moscatelli}, {Mossavi}, {Mours}, {MowLowry}, {Mueller}, {Mukherjee}, {Mullavey}, {M{\"u}ller-Ebhardt}, {Munch}, {Murray}, {Nash}, {Nawrodt}, {Nelson}, {Neri}, {Newton}, {Nishizawa}, {Nocera}, {Nolting}, {Ochsner}, {O'Dell}, {Ogin}, {Oldenburg}, {O'Reilly}, {O'Shaughnessy}, {Osthelder}, {Ottaway}, {Ottens}, {Overmier},
  {Owen}, {Page}, {Pagliaroli}, {Palladino}, {Palomba}, {Pan}, {Pankow}, {Paoletti}, {Papa}, {Pardi}, {Pareja}, {Parisi}, {Pasqualetti}, {Passaquieti}, {Passuello}, {Patel}, {Pedraza}, {Pekowsky}, {Penn}, {Peralta}, {Perreca}, {Persichetti}, {Pichot}, {Pickenpack}, {Piergiovanni}, {Pietka}, {Pinard}, {Pinto}, {Pitkin}, {Pletsch}, {Plissi}, {Poggiani}, {Postiglione}, {Prato}, {Predoi}, {Price}, {Prijatelj}, {Principe}, {Privitera}, {Prix}, {Prodi}, {Prokhorov}, {Puncken}, {Punturo}, {Puppo}, {Quetschke}, {Raab}, {Rabaste}, {Rabeling}, {Radke}, {Radkins}, {Raffai}, {Rakhmanov}, {Rankins}, {Rapagnani}, {Raymond}, {Re}, {Reed}, {Reed}, {Regimbau}, {Reid}, {Reitze}, {Ricci}, {Riesen}, {Riles}, {Roberts}, {Robertson}, {Robinet}, {Robinson}, {Robinson}, {Rocchi}, {Roddy}, {R{\"o}ver}, {Rogstad}, {Rolland}, {Rollins}, {Romano}, {Romano}, {Romie}, {Rosi{\'n}ska}, {Rowan}, {R{\"u}diger}, {Ruggi}, {Ryan}, {Sakata}, {Sakosky}, {Salemi}, {Sammut}, {Sancho de la Jordana}, {Sandberg}, {Sannibale}, {Santamar{\'\i}a},
  {Santostasi}, {Saraf}, {Sassolas}, {Sathyaprakash}, {Sato}, {Satterthwaite}, {Saulson}, {Savage}, {Schilling}, {Schnabel}, {Schofield}, {Schulz}, {Schutz}, {Schwinberg}, {Scott}, {Scott}, {Searle}, {Seifert}, {Sellers}, {Sengupta}, {Sentenac}, {Sergeev}, {Shaddock}, {Shapiro}, {Shawhan}, {Shoemaker}, {Sibley}, {Siemens}, {Sigg}, {Singer}, {Sintes}, {Skelton}, {Slagmolen}, {Slutsky}, {Smith}, {Smith}, {Smith}, {Somiya}, {Sorazu}, {Speirits}, {Stein}, {Stein}, {Steinlechner}, {Steplewski}, {Stochino}, {Stone}, {Strain}, {Strigin}, {Stroeer}, {Sturani}, {Stuver}, {Summerscales}, {Sung}, {Susmithan}, {Sutton}, {Swinkels}, {Talukder}, {Tanner}, {Tarabrin}, {Taylor}, {Taylor}, {Thomas}, {Thorne}, {Thorne}, {Thrane}, {Th{\"u}ring}, {Titsler}, {Tokmakov}, {Toncelli}, {Tonelli}, {Torres}, {Torrie}, {Tournefier}, {Travasso}, {Traylor}, {Trias}, {Trummer}, {Tseng}, {Ugolini}, {Urbanek}, {Vahlbruch}, {Vaishnav}, {Vajente}, {Vallisneri}, {van den Brand}, {Van Den Broeck}, {van der Putten}, {van der Sluys}, {van Veggel},
  {Vass}, {Vaulin}, {Vavoulidis}, {Vecchio}, {Vedovato}, {Veitch}, {Veitch}, {Veltkamp}, {Verkindt}, {Vetrano}, {Vicer{\'e}}, {Villar}, {Vinet}, {Vocca}, {Vorvick}, {Vyachanin}, {Waldman}, {Wallace}, {Wanner}, {Ward}, {Was}, {Wei}, {Weinert}, {Weinstein}, {Weiss}, {Wen}, {Wen}, {Wessels}, {West}, {Westphal}, {Wette}, {Whelan}, {Whitcomb}, {White}, {Whiting}, {Wilkinson}, {Willems}, {Williams}, {Willke}, {Winkelmann}, {Winkler}, {Wipf}, {Wiseman}, {Woan}, {Wooley}, {Worden}, {Yakushin}, {Yamamoto}, {Yamamoto}, {Yeaton-Massey}, {Yoshida}, {Yu}, {Yvert}, {Zanolin}, {Zhang}, {Zhang}, {Zhao}, {Zotov}, {Zucker}, {Zweizig}, {LIGO Scientific Collaboration}, \& {Virgo Collaboration}}]{Abadie10}
{Abadie}, J., {Abbott}, B.~P., {Abbott}, R., {et~al.} 2010, Classical and Quantum Gravity, 27, 173001, \dodoi{10.1088/0264-9381/27/17/173001}

\bibitem[{{Abbott} {et~al.}(2017){Abbott}, {Abbott}, {Abbott}, {Acernese}, {Ackley}, {Adams}, {Adams}, {Addesso}, {Adhikari}, {Adya}, {Affeldt}, {Afrough}, {Agarwal}, {Agathos}, {Agatsuma}, {Aggarwal}, {Aguiar}, {Aiello}, {Ain}, {Ajith}, {Allen}, {Allen}, {Allocca}, {Altin}, {Amato}, {Ananyeva}, {Anderson}, {Anderson}, {Angelova}, {Antier}, {Appert}, {Arai}, {Araya}, {Areeda}, {Arnaud}, {Arun}, {Ascenzi}, {Ashton}, {Ast}, {Aston}, {Astone}, {Atallah}, {Aufmuth}, {Aulbert}, {AultONeal}, {Austin}, {Avila-Alvarez}, {Babak}, {Bacon}, {Bader}, {Bae}, {Bailes}, {Baker}, {Baldaccini}, {Ballardin}, {Ballmer}, {Banagiri}, {Barayoga}, {Barclay}, {Barish}, {Barker}, {Barkett}, {Barone}, {Barr}, {Barsotti}, {Barsuglia}, {Barta}, {Barthelmy}, {Bartlett}, {Bartos}, {Bassiri}, {Basti}, {Batch}, {Bawaj}, {Bayley}, {Bazzan}, {B{\'e}csy}, {Beer}, {Bejger}, {Belahcene}, {Bell}, {Berger}, {Bergmann}, {Bernuzzi}, {Bero}, {Berry}, {Bersanetti}, {Bertolini}, {Betzwieser}, {Bhagwat}, {Bhandare}, {Bilenko}, {Billingsley}, {Billman},
  {Birch}, {Birney}, {Birnholtz}, {Biscans}, {Biscoveanu}, {Bisht}, {Bitossi}, {Biwer}, {Bizouard}, {Blackburn}, {Blackman}, {Blair}, {Blair}, {Blair}, {Bloemen}, {Bock}, {Bode}, {Boer}, {Bogaert}, {Bohe}, {Bondu}, {Bonilla}, {Bonnand}, {Boom}, {Bork}, {Boschi}, {Bose}, {Bossie}, {Bouffanais}, {Bozzi}, {Bradaschia}, {Brady}, {Branchesi}, {Brau}, {Briant}, {Brillet}, {Brinkmann}, {Brisson}, {Brockill}, {Broida}, {Brooks}, {Brown}, {Brown}, {Brunett}, {Buchanan}, {Buikema}, {Bulik}, {Bulten}, {Buonanno}, {Buskulic}, {Buy}, {Byer}, {Cabero}, {Cadonati}, {Cagnoli}, {Cahillane}, {Calder{\'o}n Bustillo}, {Callister}, {Calloni}, {Camp}, {Canepa}, {Canizares}, {Cannon}, {Cao}, {Cao}, {Capano}, {Capocasa}, {Carbognani}, {Caride}, {Carney}, {Carullo}, {Casanueva Diaz}, {Casentini}, {Caudill}, {Cavagli{\`a}}, {Cavalier}, {Cavalieri}, {Cella}, {Cepeda}, {Cerd{\'a}-Dur{\'a}n}, {Cerretani}, {Cesarini}, {Chamberlin}, {Chan}, {Chao}, {Charlton}, {Chase}, {Chassande-Mottin}, {Chatterjee}, {Chatziioannou}, {Cheeseboro},
  {Chen}, {Chen}, {Chen}, {Cheng}, {Chia}, {Chincarini}, {Chiummo}, {Chmiel}, {Cho}, {Cho}, {Chow}, {Christensen}, {Chu}, {Chua}, {Chua}, {Chung}, {Chung}, {Ciani}, {Ciolfi}, {Cirelli}, {Cirone}, {Clara}, {Clark}, {Clearwater}, {Cleva}, {Cocchieri}, {Coccia}, {Cohadon}, {Cohen}, {Colla}, {Collette}, {Cominsky}, {Constancio}, {Conti}, {Cooper}, {Corban}, {Corbitt}, {Cordero-Carri{\'o}n}, {Corley}, {Cornish}, {Corsi}, {Cortese}, {Costa}, {Coughlin}, {Coughlin}, {Coulon}, {Countryman}, {Couvares}, {Covas}, {Cowan}, {Coward}, {Cowart}, {Coyne}, {Coyne}, {Creighton}, {Creighton}, {Cripe}, {Crowder}, {Cullen}, {Cumming}, {Cunningham}, {Cuoco}, {Dal Canton}, {D{\'a}lya}, {Danilishin}, {D'Antonio}, {Danzmann}, {Dasgupta}, {Da Silva Costa}, {Dattilo}, {Dave}, {Davier}, {Davis}, {Daw}, {Day}, {De}, {DeBra}, {Degallaix}, {De Laurentis}, {Del{\'e}glise}, {Del Pozzo}, {Demos}, {Denker}, {Dent}, {De Pietri}, {Dergachev}, {De Rosa}, {DeRosa}, {De Rossi}, {DeSalvo}, {de Varona}, {Devenson}, {Dhurandhar}, {D{\'\i}az},
  {Dietrich}, {Di Fiore}, {Di Giovanni}, {Di Girolamo}, {Di Lieto}, {Di Pace}, {Di Palma}, {Di Renzo}, {Doctor}, {Dolique}, {Donovan}, {Dooley}, {Doravari}, {Dorrington}, {Douglas}, {Dovale {\'A}lvarez}, {Downes}, {Drago}, {Dreissigacker}, {Driggers}, {Du}, {Ducrot}, {Dudi}, {Dupej}, {Dwyer}, {Edo}, {Edwards}, {Effler}, {Eggenstein}, {Ehrens}, {Eichholz}, {Eikenberry}, {Eisenstein}, {Essick}, {Estevez}, {Etienne}, {Etzel}, {Evans}, {Evans}, {Factourovich}, {Fafone}, {Fair}, {Fairhurst}, {Fan}, {Farinon}, {Farr}, {Farr}, {Fauchon-Jones}, {Favata}, {Fays}, {Fee}, {Fehrmann}, {Feicht}, {Fejer}, {Fernandez-Galiana}, {Ferrante}, {Ferreira}, {Ferrini}, {Fidecaro}, {Finstad}, {Fiori}, {Fiorucci}, {Fishbach}, {Fisher}, {Fitz-Axen}, {Flaminio}, {Fletcher}, {Fong}, {Font}, {Forsyth}, {Forsyth}, {Fournier}, {Frasca}, {Frasconi}, {Frei}, {Freise}, {Frey}, {Frey}, {Fries}, {Fritschel}, {Frolov}, {Fulda}, {Fyffe}, {Gabbard}, {Gadre}, {Gaebel}, {Gair}, {Gammaitoni}, {Ganija}, {Gaonkar}, {Garcia-Quiros}, {Garufi}, {Gateley},
  {Gaudio}, {Gaur}, {Gayathri}, {Gehrels}, {Gemme}, {Genin}, {Gennai}, {George}, {George}, {Gergely}, {Germain}, {Ghonge}, {Ghosh}, {Ghosh}, {Ghosh}, {Giaime}, {Giardina}, {Giazotto}, {Gill}, {Glover}, {Goetz}, {Goetz}, {Gomes}, {Goncharov}, {Gonz{\'a}lez}, {Gonzalez Castro}, {Gopakumar}, {Gorodetsky}, {Gossan}, {Gosselin}, {Gouaty}, {Grado}, {Graef}, {Granata}, {Grant}, {Gras}, {Gray}, {Greco}, {Green}, {Gretarsson}, {Groot}, {Grote}, {Grunewald}, {Gruning}, {Guidi}, {Guo}, {Gupta}, {Gupta}, {Gushwa}, {Gustafson}, {Gustafson}, {Halim}, {Hall}, {Hall}, {Hamilton}, {Hammond}, {Haney}, {Hanke}, {Hanks}, {Hanna}, {Hannam}, {Hannuksela}, {Hanson}, {Hardwick}, {Harms}, {Harry}, {Harry}, {Hart}, {Haster}, {Haughian}, {Healy}, {Heidmann}, {Heintze}, {Heitmann}, {Hello}, {Hemming}, {Hendry}, {Heng}, {Hennig}, {Heptonstall}, {Heurs}, {Hild}, {Hinderer}, {Ho}, {Hoak}, {Hofman}, {Holt}, {Holz}, {Hopkins}, {Horst}, {Hough}, {Houston}, {Howell}, {Hreibi}, {Hu}, {Huerta}, {Huet}, {Hughey}, {Husa}, {Huttner}, {Huynh-Dinh},
  {Indik}, {Inta}, {Intini}, {Isa}, {Isac}, {Isi}, {Iyer}, {Izumi}, {Jacqmin}, {Jani}, {Jaranowski}, {Jawahar}, {Jim{\'e}nez-Forteza}, {Johnson}, {Johnson-McDaniel}, {Jones}, {Jones}, {Jonker}, {Ju}, {Junker}, {Kalaghatgi}, {Kalogera}, {Kamai}, {Kandhasamy}, {Kang}, {Kanner}, {Kapadia}, {Karki}, {Karvinen}, {Kasprzack}, {Kastaun}, {Katolik}, {Katsavounidis}, {Katzman}, {Kaufer}, {Kawabe}, {K{\'e}f{\'e}lian}, {Keitel}, {Kemball}, {Kennedy}, {Kent}, {Key}, {Khalili}, {Khan}, {Khan}, {Khan}, {Khazanov}, {Kijbunchoo}, {Kim}, {Kim}, {Kim}, {Kim}, {Kim}, {Kim}, {Kimbrell}, {King}, {King}, {Kinley-Hanlon}, {Kirchhoff}, {Kissel}, {Kleybolte}, {Klimenko}, {Knowles}, {Koch}, {Koehlenbeck}, {Koley}, {Kondrashov}, {Kontos}, {Korobko}, {Korth}, {Kowalska}, {Kozak}, {Kr{\"a}mer}, {Kringel}, {Krishnan}, {Kr{\'o}lak}, {Kuehn}, {Kumar}, {Kumar}, {Kumar}, {Kuo}, {Kutynia}, {Kwang}, {Lackey}, {Lai}, {Landry}, {Lang}, {Lange}, {Lantz}, {Lanza}, {Larson}, {Lartaux-Vollard}, {Lasky}, {Laxen}, {Lazzarini}, {Lazzaro}, {Leaci},
  {Leavey}, {Lee}, {Lee}, {Lee}, {Lee}, {Lee}, {Lehmann}, {Lenon}, {Leon}, {Leonardi}, {Leroy}, {Letendre}, {Levin}, {Li}, {Linker}, {Littenberg}, {Liu}, {Liu}, {Lo}, {Lockerbie}, {London}, {Lord}, {Lorenzini}, {Loriette}, {Lormand}, {Losurdo}, {Lough}, {Lousto}, {Lovelace}, {L{\"u}ck}, {Lumaca}, {Lundgren}, {Lynch}, {Ma}, {Macas}, {Macfoy}, {Machenschalk}, {MacInnis}, {Macleod}, {Maga{\~n}a Hernandez}, {Maga{\~n}a-Sandoval}, {Maga{\~n}a Zertuche}, {Magee}, {Majorana}, {Maksimovic}, {Man}, {Mandic}, {Mangano}, {Mansell}, {Manske}, {Mantovani}, {Marchesoni}, {Marion}, {M{\'a}rka}, {M{\'a}rka}, {Markakis}, {Markosyan}, {Markowitz}, {Maros}, {Marquina}, {Marsh}, {Martelli}, {Martellini}, {Martin}, {Martin}, {Martynov}, {Marx}, {Mason}, {Massera}, {Masserot}, {Massinger}, {Masso-Reid}, {Mastrogiovanni}, {Matas}, {Matichard}, {Matone}, {Mavalvala}, {Mazumder}, {McCarthy}, {McClelland}, {McCormick}, {McCuller}, {McGuire}, {McIntyre}, {McIver}, {McManus}, {McNeill}, {McRae}, {McWilliams}, {Meacher}, {Meadors},
  {Mehmet}, {Meidam}, {Mejuto-Villa}, {Melatos}, {Mendell}, {Mercer}, {Merilh}, {Merzougui}, {Meshkov}, {Messenger}, {Messick}, {Metzdorff}, {Meyers}, {Miao}, {Michel}, {Middleton}, {Mikhailov}, {Milano}, {Miller}, {Miller}, {Miller}, {Millhouse}, {Milovich-Goff}, {Minazzoli}, {Minenkov}, {Ming}, {Mishra}, {Mitra}, {Mitrofanov}, {Mitselmakher}, {Mittleman}, {Moffa}, {Moggi}, {Mogushi}, {Mohan}, {Mohapatra}, {Molina}, {Montani}, {Moore}, {Moraru}, {Moreno}, {Morisaki}, {Morriss}, {Mours}, {Mow-Lowry}, {Mueller}, {Muir}, {Mukherjee}, {Mukherjee}, {Mukherjee}, {Mukund}, {Mullavey}, {Munch}, {Mu{\~n}iz}, {Muratore}, {Murray}, {Nagar}, {Napier}, {Nardecchia}, {Naticchioni}, {Nayak}, {Neilson}, {Nelemans}, {Nelson}, {Nery}, {Neunzert}, {Nevin}, {Newport}, {Newton}, {Ng}, {Nguyen}, {Nguyen}, {Nichols}, {Nielsen}, {Nissanke}, {Nitz}, {Noack}, {Nocera}, {Nolting}, {North}, {Nuttall}, {Oberling}, {O'Dea}, {Ogin}, {Oh}, {Oh}, {Ohme}, {Okada}, {Oliver}, {Oppermann}, {Oram}, {O'Reilly}, {Ormiston}, {Ortega},
  {O'Shaughnessy}, {Ossokine}, {Ottaway}, {Overmier}, {Owen}, {Pace}, {Page}, {Page}, {Pai}, {Pai}, {Palamos}, {Palashov}, {Palomba}, {Pal-Singh}, {Pan}, {Pan}, {Pang}, {Pang}, {Pankow}, {Pannarale}, {Pant}, {Paoletti}, {Paoli}, {Papa}, {Parida}, {Parker}, {Pascucci}, {Pasqualetti}, {Passaquieti}, {Passuello}, {Patil}, {Patricelli}, {Pearlstone}, {Pedraza}, {Pedurand}, {Pekowsky}, {Pele}, {Penn}, {Perez}, {Perreca}, {Perri}, {Pfeiffer}, {Phelps}, {Piccinni}, {Pichot}, {Piergiovanni}, {Pierro}, {Pillant}, {Pinard}, {Pinto}, {Pirello}, {Pitkin}, {Poe}, {Poggiani}, {Popolizio}, {Porter}, {Post}, {Powell}, {Prasad}, {Pratt}, {Pratten}, {Predoi}, {Prestegard}, {Prijatelj}, {Principe}, {Privitera}, {Prix}, {Prodi}, {Prokhorov}, {Puncken}, {Punturo}, {Puppo}, {P{\"u}rrer}, {Qi}, {Quetschke}, {Quintero}, {Quitzow-James}, {Raab}, {Rabeling}, {Radkins}, {Raffai}, {Raja}, {Rajan}, {Rajbhandari}, {Rakhmanov}, {Ramirez}, {Ramos-Buades}, {Rapagnani}, {Raymond}, {Razzano}, {Read}, {Regimbau}, {Rei}, {Reid}, {Reitze}, {Ren},
  {Reyes}, {Ricci}, {Ricker}, {Rieger}, {Riles}, {Rizzo}, {Robertson}, {Robie}, {Robinet}, {Rocchi}, {Rolland}, {Rollins}, {Roma}, {Romano}, {Romano}, {Romel}, {Romie}, {Rosi{\'n}ska}, {Ross}, {Rowan}, {R{\"u}diger}, {Ruggi}, {Rutins}, {Ryan}, {Sachdev}, {Sadecki}, {Sadeghian}, {Sakellariadou}, {Salconi}, {Saleem}, {Salemi}, {Samajdar}, {Sammut}, {Sampson}, {Sanchez}, {Sanchez}, {Sanchis-Gual}, {Sandberg}, {Sanders}, {Sassolas}, {Sathyaprakash}, {Saulson}, {Sauter}, {Savage}, {Sawadsky}, {Schale}, {Scheel}, {Scheuer}, {Schmidt}, {Schmidt}, {Schnabel}, {Schofield}, {Sch{\"o}nbeck}, {Schreiber}, {Schuette}, {Schulte}, {Schutz}, {Schwalbe}, {Scott}, {Scott}, {Seidel}, {Sellers}, {Sengupta}, {Sentenac}, {Sequino}, {Sergeev}, {Shaddock}, {Shaffer}, {Shah}, {Shahriar}, {Shaner}, {Shao}, {Shapiro}, {Shawhan}, {Sheperd}, {Shoemaker}, {Shoemaker}, {Siellez}, {Siemens}, {Sieniawska}, {Sigg}, {Silva}, {Singer}, {Singh}, {Singhal}, {Sintes}, {Slagmolen}, {Smith}, {Smith}, {Smith}, {Somala}, {Son}, {Sonnenberg}, {Sorazu},
  {Sorrentino}, {Souradeep}, {Spencer}, {Srivastava}, {Staats}, {Staley}, {Steinke}, {Steinlechner}, {Steinlechner}, {Steinmeyer}, {Stevenson}, {Stone}, {Stops}, {Strain}, {Stratta}, {Strigin}, {Strunk}, {Sturani}, {Stuver}, {Summerscales}, {Sun}, {Sunil}, {Suresh}, {Sutton}, {Swinkels}, {Szczepa{\'n}czyk}, {Tacca}, {Tait}, {Talbot}, {Talukder}, {Tanner}, {T{\'a}pai}, {Taracchini}, {Tasson}, {Taylor}, {Taylor}, {Tewari}, {Theeg}, {Thies}, {Thomas}, {Thomas}, {Thomas}, {Thorne}, {Thorne}, {Thrane}, {Tiwari}, {Tiwari}, {Tokmakov}, {Toland}, {Tonelli}, {Tornasi}, {Torres-Forn{\'e}}, {Torrie}, {T{\"o}yr{\"a}}, {Travasso}, {Traylor}, {Trinastic}, {Tringali}, {Trozzo}, {Tsang}, {Tse}, {Tso}, {Tsukada}, {Tsuna}, {Tuyenbayev}, {Ueno}, {Ugolini}, {Unnikrishnan}, {Urban}, {Usman}, {Vahlbruch}, {Vajente}, {Valdes}, {Vallisneri}, {van Bakel}, {van Beuzekom}, {van den Brand}, {Van Den Broeck}, {Vander-Hyde}, {van der Schaaf}, {van Heijningen}, {van Veggel}, {Vardaro}, {Varma}, {Vass}, {Vas{\'u}th}, {Vecchio}, {Vedovato},
  {Veitch}, {Veitch}, {Venkateswara}, {Venugopalan}, {Verkindt}, {Vetrano}, {Vicer{\'e}}, {Viets}, {Vinciguerra}, {Vine}, {Vinet}, {Vitale}, {Vo}, {Vocca}, {Vorvick}, {Vyatchanin}, {Wade}, {Wade}, {Wade}, {Walet}, {Walker}, {Wallace}, {Walsh}, {Wang}, {Wang}, {Wang}, {Wang}, {Wang}, {Ward}, {Warner}, {Was}, {Watchi}, {Weaver}, {Wei}, {Weinert}, {Weinstein}, {Weiss}, {Wen}, {Wessel}, {We{\ss}els}, {Westerweck}, {Westphal}, {Wette}, {Whelan}, {Whitcomb}, {Whiting}, {Whittle}, {Wilken}, {Williams}, {Williams}, {Williamson}, {Willis}, {Willke}, {Wimmer}, {Winkler}, {Wipf}, {Wittel}, {Woan}, {Woehler}, {Wofford}, {Wong}, {Worden}, {Wright}, {Wu}, {Wysocki}, {Xiao}, {Yamamoto}, {Yancey}, {Yang}, {Yap}, {Yazback}, {Yu}, {Yu}, {Yvert}, {Zadro{\.Z}ny}, {Zanolin}, {Zelenova}, {Zendri}, {Zevin}, {Zhang}, {Zhang}, {Zhang}, {Zhang}, {Zhao}, {Zhou}, {Zhou}, {Zhu}, {Zhu}, {Zimmerman}, {Zucker}, {Zweizig}, {LIGO Scientific Collaboration}, \& {Virgo Collaboration}}]{Abbott17}
{Abbott}, B.~P., {Abbott}, R., {Abbott}, T.~D., {et~al.} 2017, \prl, 119, 161101, \dodoi{10.1103/PhysRevLett.119.161101}

\bibitem[{{Abbott} {et~al.}(2023){Abbott}, {Abbott}, {Acernese}, {Ackley}, {Adams}, {Adhikari}, {Adhikari}, {Adya}, {Affeldt}, {Agarwal}, {Agathos}, {Agatsuma}, {Aggarwal}, {Aguiar}, {Aiello}, {Ain}, {Ajith}, {Akutsu}, {de Alarc{\'o}n}, {Akcay}, {Albanesi}, {Allocca}, {Altin}, {Amato}, {Anand}, {Anand}, {Ananyeva}, {Anderson}, {Anderson}, {Ando}, {Andrade}, {Andres}, {Andri{\'c}}, {Angelova}, {Ansoldi}, {Antelis}, {Antier}, {Antonini}, {Appert}, {Arai}, {Arai}, {Arai}, {Araki}, {Araya}, {Araya}, {Areeda}, {Ar{\`e}ne}, {Aritomi}, {Arnaud}, {Arogeti}, {Aronson}, {Arun}, {Asada}, {Asali}, {Ashton}, {Aso}, {Assiduo}, {Aston}, {Astone}, {Aubin}, {Austin}, {Babak}, {Badaracco}, {Bader}, {Badger}, {Bae}, {Bae}, {Baer}, {Bagnasco}, {Bai}, {Baiotti}, {Baird}, {Bajpai}, {Ball}, {Ballardin}, {Ballmer}, {Balsamo}, {Baltus}, {Banagiri}, {Bankar}, {Barayoga}, {Barbieri}, {Barish}, {Barker}, {Barneo}, {Barone}, {Barr}, {Barsotti}, {Barsuglia}, {Barta}, {Bartlett}, {Barton}, {Bartos}, {Bassiri}, {Basti}, {Bawaj}, {Bayley},
  {Baylor}, {Bazzan}, {B{\'e}csy}, {Bedakihale}, {Bejger}, {Belahcene}, {Benedetto}, {Beniwal}, {Bennett}, {Bentley}, {Benyaala}, {Bergamin}, {Berger}, {Bernuzzi}, {Berry}, {Bersanetti}, {Bertolini}, {Betzwieser}, {Beveridge}, {Bhandare}, {Bhardwaj}, {Bhattacharjee}, {Bhaumik}, {Bilenko}, {Billingsley}, {Bini}, {Birney}, {Birnholtz}, {Biscans}, {Bischi}, {Biscoveanu}, {Bisht}, {Biswas}, {Bitossi}, {Bizouard}, {Blackburn}, {Blair}, {Blair}, {Blair}, {Bobba}, {Bode}, {Boer}, {Bogaert}, {Boldrini}, {Bonavena}, {Bondu}, {Bonilla}, {Bonnand}, {Booker}, {Boom}, {Bork}, {Boschi}, {Bose}, {Bose}, {Bossilkov}, {Boudart}, {Bouffanais}, {Bozzi}, {Bradaschia}, {Brady}, {Bramley}, {Branch}, {Branchesi}, {Brandt}, {Brau}, {Breschi}, {Briant}, {Briggs}, {Brillet}, {Brinkmann}, {Brockill}, {Brooks}, {Brooks}, {Brown}, {Brunett}, {Bruno}, {Bruntz}, {Bryant}, {Bulik}, {Bulten}, {Buonanno}, {Buscicchio}, {Buskulic}, {Buy}, {Byer}, {Cadonati}, {Cagnoli}, {Cahillane}, {Bustillo}, {Callaghan}, {Callister}, {Calloni}, {Cameron},
  {Camp}, {Canepa}, {Canevarolo}, {Cannavacciuolo}, {Cannon}, {Cao}, {Cao}, {Capocasa}, {Capote}, {Carapella}, {Carbognani}, {Carlin}, {Carney}, {Carpinelli}, {Carrillo}, {Carullo}, {Carver}, {Diaz}, {Casentini}, {Castaldi}, {Caudill}, {Cavagli{\`a}}, {Cavalier}, {Cavalieri}, {Ceasar}, {Cella}, {Cerd{\'a}-Dur{\'a}n}, {Cesarini}, {Chaibi}, {Chakravarti}, {Subrahmanya}, {Champion}, {Chan}, {Chan}, {Chan}, {Chan}, {Chan}, {Chandra}, {Chanial}, {Chao}, {Chapman-Bird}, {Charlton}, {Chase}, {Chassande-Mottin}, {Chatterjee}, {Chatterjee}, {Chatterjee}, {Chaturvedi}, {Chaty}, {Chatziioannou}, {Chen}, {Chen}, {Chen}, {Chen}, {Chen}, {Chen}, {Chen}, {Chen}, {Cheng}, {Cheong}, {Cheung}, {Chia}, {Chiadini}, {Chiang}, {Chiarini}, {Chierici}, {Chincarini}, {Chiofalo}, {Chiummo}, {Cho}, {Cho}, {Choudhary}, {Choudhary}, {Christensen}, {Chu}, {Chu}, {Chu}, {Chua}, {Chung}, {Ciani}, {Ciecielag}, {Cie{\'s}lar}, {Cifaldi}, {Ciobanu}, {Ciolfi}, {Cipriano}, {Cirone}, {Clara}, {Clark}, {Clark}, {Clarke}, {Clearwater}, {Clesse},
  {Cleva}, {Coccia}, {Codazzo}, {Cohadon}, {Cohen}, {Cohen}, {Colleoni}, {Collette}, {Colombo}, {Colpi}, {Compton}, {Constancio}, {Conti}, {Cooper}, {Corban}, {Corbitt}, {Cordero-Carri{\'o}n}, {Corezzi}, {Corley}, {Cornish}, {Corre}, {Corsi}, {Cortese}, {Costa}, {Cotesta}, {Coughlin}, {Coulon}, {Countryman}, {Cousins}, {Couvares}, {Coward}, {Cowart}, {Coyne}, {Coyne}, {Creighton}, {Creighton}, {Criswell}, {Croquette}, {Crowder}, {Cudell}, {Cullen}, {Cumming}, {Cummings}, {Cunningham}, {Cuoco}, {Cury{\l}o}, {Dabadie}, {Canton}, {Dall'Osso}, {D{\'a}lya}, {Dana}, {Daneshgaranbajastani}, {D'Angelo}, {Danila}, {Danilishin}, {D'Antonio}, {Danzmann}, {Darsow-Fromm}, {Dasgupta}, {Datrier}, {Datta}, {Dattilo}, {Dave}, {Davier}, {Davies}, {Davis}, {Davis}, {Daw}, {Dean}, {Debra}, {Deenadayalan}, {Degallaix}, {de Laurentis}, {Del{\'e}glise}, {Del Favero}, {de Lillo}, {de Lillo}, {Del Pozzo}, {Demarchi}, {de Matteis}, {D'Emilio}, {Demos}, {Dent}, {Depasse}, {de Pietri}, {De Rosa}, {de Rossi}, {Desalvo}, {de Simone},
  {Dhurandhar}, {D{\'\i}az}, {Diaz-Ortiz}, {Didio}, {Dietrich}, {di Fiore}, {di Fronzo}, {di Giorgio}, {di Giovanni}, {di Giovanni}, {di Girolamo}, {di Lieto}, {Ding}, {di Pace}, {di Palma}, {di Renzo}, {Divakarla}, {Dmitriev}, {Doctor}, {D'Onofrio}, {Donovan}, {Dooley}, {Doravari}, {Dorrington}, {Drago}, {Driggers}, {Drori}, {Ducoin}, {Dupej}, {Durante}, {D'Urso}, {Duverne}, {Dwyer}, {Eassa}, {Easter}, {Ebersold}, {Eckhardt}, {Eddolls}, {Edelman}, {Edo}, {Edy}, {Effler}, {Eguchi}, {Eichholz}, {Eikenberry}, {Eisenmann}, {Eisenstein}, {Ejlli}, {Engelby}, {Enomoto}, {Errico}, {Essick}, {Estell{\'e}s}, {Estevez}, {Etienne}, {Etzel}, {Evans}, {Evans}, {Ewing}, {Fafone}, {Fair}, {Fairhurst}, {Farah}, {Farinon}, {Farr}, {Farr}, {Farrow}, {Fauchon-Jones}, {Favaro}, {Favata}, {Fays}, {Fazio}, {Feicht}, {Fejer}, {Fenyvesi}, {Ferguson}, {Fernandez-Galiana}, {Ferrante}, {Ferreira}, {Fidecaro}, {Figura}, {Fiori}, {Fishbach}, {Fisher}, {Fittipaldi}, {Fiumara}, {Flaminio}, {Floden}, {Fong}, {Font}, {Fornal}, {Forsyth},
  {Franke}, {Frasca}, {Frasconi}, {Frederick}, {Freed}, {Frei}, {Freise}, {Frey}, {Fritschel}, {Frolov}, {Fronz{\'e}}, {Fujii}, {Fujikawa}, {Fukunaga}, {Fukushima}, {Fulda}, {Fyffe}, {Gabbard}, {Gadre}, {Gair}, {Gais}, {Galaudage}, {Gamba}, {Ganapathy}, {Ganguly}, {Gao}, {Gaonkar}, {Garaventa}, {Garc{\'\i}a}, {Garc{\'\i}a-N{\'u}{\~n}ez}, {Garc{\'\i}a-Quir{\'o}s}, {Garufi}, {Gateley}, {Gaudio}, {Gayathri}, {Ge}, {Gemme}, {Gennai}, {George}, {George}, {Gerberding}, {Gergely}, {Gewecke}, {Ghonge}, {Ghosh}, {Ghosh}, {Ghosh}, {Ghosh}, {Giacomazzo}, {Giacoppo}, {Giaime}, {Giardina}, {Gibson}, {Gier}, {Giesler}, {Giri}, {Gissi}, {Glanzer}, {Gleckl}, {Godwin}, {Golomb}, {Goetz}, {Goetz}, {Gohlke}, {Goncharov}, {Gonz{\'a}lez}, {Gopakumar}, {Gosselin}, {Gouaty}, {Gould}, {Grace}, {Grado}, {Granata}, {Granata}, {Grant}, {Gras}, {Grassia}, {Gray}, {Gray}, {Greco}, {Green}, {Green}, {Gretarsson}, {Gretarsson}, {Griffith}, {Griffiths}, {Griggs}, {Grignani}, {Grimaldi}, {Grimm}, {Grote}, {Grunewald}, {Gruning}, {Guerra},
  {Guidi}, {Guimaraes}, {Guix{\'e}}, {Gulati}, {Guo}, {Guo}, {Gupta}, {Gupta}, {Gupta}, {Gustafson}, {Gustafson}, {Guzman}, {Ha}, {Haegel}, {Hagiwara}, {Haino}, {Halim}, {Hall}, {Hamilton}, {Hammond}, {Han}, {Haney}, {Hanks}, {Hanna}, {Hannam}, {Hannuksela}, {Hansen}, {Hansen}, {Hanson}, {Harder}, {Hardwick}, {Haris}, {Harms}, {Harry}, {Harry}, {Hartwig}, {Hasegawa}, {Haskell}, {Hasskew}, {Haster}, {Hattori}, {Haughian}, {Hayakawa}, {Hayama}, {Hayes}, {Healy}, {Heidmann}, {Heidt}, {Heintze}, {Heinze}, {Heinzel}, {Heitmann}, {Hellman}, {Hello}, {Helmling-Cornell}, {Hemming}, {Hendry}, {Heng}, {Hennes}, {Hennig}, {Hennig}, {Hernandez}, {Vivanco}, {Heurs}, {Hild}, {Hill}, {Himemoto}, {Hines}, {Hiranuma}, {Hirata}, {Hirose}, {Hochheim}, {Hofman}, {Hohmann}, {Holcomb}, {Holland}, {Hollows}, {Holmes}, {Holt}, {Holz}, {Hong}, {Hopkins}, {Hough}, {Hourihane}, {Howell}, {Hoy}, {Hoyland}, {Hreibi}, {Hsieh}, {Hsu}, {Huang}, {Huang}, {Huang}, {Huang}, {Huang}, {Huang}, {H{\"u}bner}, {Huddart}, {Hughey}, {Hui}, {Hui},
  {Husa}, {Huttner}, {Huxford}, {Huynh-Dinh}, {Ide}, {Idzkowski}, {Iess}, {Ikenoue}, {Imam}, {Inayoshi}, {Ingram}, {Inoue}, {Ioka}, {Isi}, {Isleif}, {Ito}, {Itoh}, {Iyer}, {Izumi}, {Jaberianhamedan}, {Jacqmin}, {Jadhav}, {Jadhav}, {James}, {Jan}, {Jani}, {Janquart}, {Janssens}, {Janthalur}, {Jaranowski}, {Jariwala}, {Jaume}, {Jenkins}, {Jenner}, {Jeon}, {Jeunon}, {Jia}, {Jin}, {Johns}, {Jones}, {Jones}, {Jones}, {Jones}, {Jones}, {Jonker}, {Ju}, {Jung}, {Jung}, {Junker}, {Juste}, {Kaihotsu}, {Kajita}, {Kakizaki}, {Kalaghatgi}, {Kalogera}, {Kamai}, {Kamiizumi}, {Kanda}, {Kandhasamy}, {Kang}, {Kanner}, {Kao}, {Kapadia}, {Kapasi}, {Karat}, {Karathanasis}, {Karki}, {Kashyap}, {Kasprzack}, {Kastaun}, {Katsanevas}, {Katsavounidis}, {Katzman}, {Kaur}, {Kawabe}, {Kawaguchi}, {Kawai}, {Kawasaki}, {K{\'e}f{\'e}lian}, {Keitel}, {Key}, {Khadka}, {Khalili}, {Khan}, {Khazanov}, {Khetan}, {Khursheed}, {Kijbunchoo}, {Kim}, {Kim}, {Kim}, {Kim}, {Kim}, {Kim}, {Kimball}, {Kimura}, {Kinley-Hanlon}, {Kirchhoff}, {Kissel}, {Kita},
  {Kitazawa}, {Kleybolte}, {Klimenko}, {Knee}, {Knowles}, {Knyazev}, {Koch}, {Koekoek}, {Kojima}, {Kokeyama}, {Koley}, {Kolitsidou}, {Kolstein}, {Komori}, {Kondrashov}, {Kong}, {Kontos}, {Koper}, {Korobko}, {Kotake}, {Kovalam}, {Kozak}, {Kozakai}, {Kozu}, {Kringel}, {Krishnendu}, {Kr{\'o}lak}, {Kuehn}, {Kuei}, {Kuijer}, {Kulkarni}, {Kumar}, {Kumar}, {Kumar}, {Kumar}, {Kume}, {Kuns}, {Kuo}, {Kuo}, {Kuromiya}, {Kuroyanagi}, {Kusayanagi}, {Kuwahara}, {Kwak}, {Lagabbe}, {Laghi}, {Lalande}, {Lam}, {Lamberts}, {Landry}, {Landry}, {Lane}, {Lang}, {Lange}, {Lantz}, {La Rosa}, {Lartaux-Vollard}, {Lasky}, {Laxen}, {Lazzarini}, {Lazzaro}, {Leaci}, {Leavey}, {Lecoeuche}, {Lee}, {Lee}, {Lee}, {Lee}, {Lee}, {Lee}, {Lehmann}, {Lema{\^\i}tre}, {Leonardi}, {Leroy}, {Letendre}, {Levesque}, {Levin}, {Leviton}, {Leyde}, {Li}, {Li}, {Li}, {Li}, {Li}, {Li}, {Lin}, {Lin}, {Lin}, {Lin}, {Lin}, {Linde}, {Linker}, {Linley}, {Littenberg}, {Liu}, {Liu}, {Liu}, {Liu}, {Llamas}, {Llorens-Monteagudo}, {Lo}, {Lockwood}, {Loh}, {London},
  {Longo}, {Lopez}, {Portilla}, {Lorenzini}, {Loriette}, {Lormand}, {Losurdo}, {Lott}, {Lough}, {Lousto}, {Lovelace}, {Lucaccioni}, {L{\"u}ck}, {Lumaca}, {Lundgren}, {Luo}, {Lynam}, {Macas}, {Macinnis}, {MacLeod}, {MacMillan}, {Macquet}, {Hernandez}, {Magazz{\`u}}, {Magee}, {Maggiore}, {Magnozzi}, {Mahesh}, {Majorana}, {Makarem}, {Maksimovic}, {Maliakal}, {Malik}, {Man}, {Mandic}, {Mangano}, {Mango}, {Mansell}, {Manske}, {Mantovani}, {Mapelli}, {Marchesoni}, {Marchio}, {Marion}, {Mark}, {M{\'a}rka}, {M{\'a}rka}, {Markakis}, {Markosyan}, {Markowitz}, {Maros}, {Marquina}, {Marsat}, {Martelli}, {Martin}, {Martin}, {Martinez}, {Martinez}, {Martinez}, {Martinovic}, {Martynov}, {Marx}, {Masalehdan}, {Mason}, {Massera}, {Masserot}, {Massinger}, {Masso-Reid}, {Mastrogiovanni}, {Matas}, {Mateu-Lucena}, {Matichard}, {Matiushechkina}, {Mavalvala}, {McCann}, {McCarthy}, {McClelland}, {McClincy}, {McCormick}, {McCuller}, {McGhee}, {McGuire}, {McIsaac}, {McIver}, {McRae}, {McWilliams}, {Meacher}, {Mehmet}, {Mehta},
  {Meijer}, {Melatos}, {Melchor}, {Mendell}, {Menendez-Vazquez}, {Menoni}, {Mercer}, {Mereni}, {Merfeld}, {Merilh}, {Merritt}, {Merzougui}, {Meshkov}, {Messenger}, {Messick}, {Meyers}, {Meylahn}, {Mhaske}, {Miani}, {Miao}, {Michaloliakos}, {Michel}, {Michimura}, {Middleton}, {Milano}, {Miller}, {Miller}, {Miller}, {Miller}, {Millhouse}, {Mills}, {Milotti}, {Minazzoli}, {Minenkov}, {Mio}, {Mir}, {Miravet-Ten{\'e}s}, {Mishra}, {Mishra}, {Mistry}, {Mitra}, {Mitrofanov}, {Mitselmakher}, {Mittleman}, {Miyakawa}, {Miyamoto}, {Miyazaki}, {Miyo}, {Miyoki}, {Mo}, {Modafferi}, {Moguel}, {Mogushi}, {Mohapatra}, {Mohite}, {Molina}, {Molina-Ruiz}, {Mondin}, {Montani}, {Moore}, {Moraru}, {Morawski}, {More}, {Moreno}, {Moreno}, {Mori}, {Morisaki}, {Moriwaki}, {Morr{\'a}s}, {Mours}, {Mow-Lowry}, {Mozzon}, {Muciaccia}, {Mukherjee}, {Mukherjee}, {Mukherjee}, {Mukherjee}, {Mukherjee}, {Mukund}, {Mullavey}, {Munch}, {Mu{\~n}iz}, {Murray}, {Musenich}, {Muusse}, {Nadji}, {Nagano}, {Nagano}, {Nagar}, {Nakamura}, {Nakano}, {Nakano},
  {Nakashima}, {Nakayama}, {Napolano}, {Nardecchia}, {Narikawa}, {Naticchioni}, {Nayak}, {Nayak}, {Negishi}, {Neil}, {Neilson}, {Nelemans}, {Nelson}, {Nery}, {Neubauer}, {Neunzert}, {Ng}, {Ng}, {Nguyen}, {Nguyen}, {Nguyen}, {Quynh}, {Ni}, {Nichols}, {Nishizawa}, {Nissanke}, {Nitoglia}, {Nocera}, {Norman}, {North}, {Nozaki}, {Siles}, {Nuttall}, {Oberling}, {O'Brien}, {Obuchi}, {O'Dell}, {Oelker}, {Ogaki}, {Oganesyan}, {Oh}, {Oh}, {Oh}, {Ohashi}, {Ohishi}, {Ohkawa}, {Ohme}, {Ohta}, {Okada}, {Okutani}, {Okutomi}, {Olivetto}, {Oohara}, {Ooi}, {Oram}, {O'Reilly}, {Ormiston}, {Ormsby}, {Ortega}, {O'Shaughnessy}, {O'Shea}, {Oshino}, {Ossokine}, {Osthelder}, {Otabe}, {Ottaway}, {Overmier}, {Pace}, {Pagano}, {Page}, {Pagliaroli}, {Pai}, {Pai}, {Palamos}, {Palashov}, {Palomba}, {Pan}, {Pan}, {Panda}, {Pang}, {Pang}, {Pankow}, {Pannarale}, {Pant}, {Panther}, {Paoletti}, {Paoli}, {Paolone}, {Parisi}, {Park}, {Park}, {Parker}, {Pascucci}, {Pasqualetti}, {Passaquieti}, {Passuello}, {Patel}, {Pathak}, {Patricelli},
  {Patron}, {Paul}, {Payne}, {Pedraza}, {Pegoraro}, {Pele}, {Arellano}, {Penn}, {Perego}, {Pereira}, {Pereira}, {Perez}, {P{\'e}rigois}, {Perkins}, {Perreca}, {Perri{\`e}s}, {Petermann}, {Petterson}, {Pfeiffer}, {Pham}, {Phukon}, {Piccinni}, {Pichot}, {Piendibene}, {Piergiovanni}, {Pierini}, {Pierro}, {Pillant}, {Pillas}, {Pilo}, {Pinard}, {Pinto}, {Pinto}, {Piotrzkowski}, {Piotrzkowski}, {Pirello}, {Pitkin}, {Placidi}, {Planas}, {Plastino}, {Pluchar}, {Poggiani}, {Polini}, {Pong}, {Ponrathnam}, {Popolizio}, {Porter}, {Poulton}, {Powell}, {Pracchia}, {Pradier}, {Prajapati}, {Prasai}, {Prasanna}, {Pratten}, {Principe}, {Prodi}, {Prokhorov}, {Prosposito}, {Prudenzi}, {Puecher}, {Punturo}, {Puosi}, {Puppo}, {P{\"u}rrer}, {Qi}, {Quetschke}, {Quitzow-James}, {Raab}, {Raaijmakers}, {Radkins}, {Radulesco}, {Raffai}, {Rail}, {Raja}, {Rajan}, {Ramirez}, {Ramirez}, {Ramos-Buades}, {Rana}, {Rapagnani}, {Rapol}, {Ray}, {Raymond}, {Raza}, {Razzano}, {Read}, {Rees}, {Regimbau}, {Rei}, {Reid}, {Reid}, {Reitze}, {Relton},
  {Renzini}, {Rettegno}, {Reza}, {Rezac}, {Ricci}, {Richards}, {Richardson}, {Richardson}, {Riemenschneider}, {Riles}, {Rinaldi}, {Rink}, {Rizzo}, {Robertson}, {Robie}, {Robinet}, {Rocchi}, {Rodriguez}, {Rolland}, {Rollins}, {Romanelli}, {Romano}, {Romel}, {Romero-Rodr{\'\i}guez}, {Romero-Shaw}, {Romie}, {Ronchini}, {Rosa}, {Rose}, {Rosi{\'n}ska}, {Ross}, {Rowan}, {Rowlinson}, {Roy}, {Roy}, {Roy}, {Rozza}, {Ruggi}, {Ryan}, {Sachdev}, {Sadecki}, {Sadiq}, {Sago}, {Saito}, {Saito}, {Sakai}, {Sakai}, {Sakellariadou}, {Sakuno}, {Salafia}, {Salconi}, {Saleem}, {Salemi}, {Samajdar}, {Sanchez}, {Sanchez}, {Sanchez}, {Sanchis-Gual}, {Sanders}, {Sanuy}, {Saravanan}, {Sarin}, {Sassolas}, {Satari}, {Sathyaprakash}, {Sato}, {Sato}, {Sauter}, {Savage}, {Sawada}, {Sawant}, {Sawant}, {Sayah}, {Schaetzl}, {Scheel}, {Scheuer}, {Schiworski}, {Schmidt}, {Schmidt}, {Schnabel}, {Schneewind}, {Schofield}, {Sch{\"o}nbeck}, {Schulte}, {Schutz}, {Schwartz}, {Scott}, {Scott}, {Seglar-Arroyo}, {Sekiguchi}, {Sekiguchi}, {Sellers},
  {Sengupta}, {Sentenac}, {Seo}, {Sequino}, {Sergeev}, {Setyawati}, {Shaffer}, {Shahriar}, {Shams}, {Shao}, {Sharma}, {Sharma}, {Shawhan}, {Shcheblanov}, {Shibagaki}, {Shikauchi}, {Shimizu}, {Shimoda}, {Shimode}, {Shinkai}, {Shishido}, {Shoda}, {Shoemaker}, {Shoemaker}, {Shyamsundar}, {Sieniawska}, {Sigg}, {Singer}, {Singh}, {Singh}, {Singha}, {Sintes}, {Sipala}, {Skliris}, {Slagmolen}, {Slaven-Blair}, {Smetana}, {Smith}, {Smith}, {Soldateschi}, {Somala}, {Somiya}, {Son}, {Soni}, {Soni}, {Sordini}, {Sorrentino}, {Sorrentino}, {Sotani}, {Soulard}, {Souradeep}, {Sowell}, {Spagnuolo}, {Spencer}, {Spera}, {Srinivasan}, {Srivastava}, {Srivastava}, {Staats}, {Stachie}, {Steer}, {Steinhoff}, {Steinlechner}, {Steinlechner}, {Stevenson}, {Stops}, {Stover}, {Strain}, {Strang}, {Stratta}, {Strunk}, {Sturani}, {Stuver}, {Sudhagar}, {Sudhir}, {Sugimoto}, {Suh}, {Sullivan}, {Summerscales}, {Sun}, {Sun}, {Sunil}, {Sur}, {Suresh}, {Sutton}, {Suzuki}, {Suzuki}, {Swinkels}, {Szczepa{\'n}czyk}, {Szewczyk}, {Tacca}, {Tagoshi},
  {Tait}, {Takahashi}, {Takahashi}, {Takamori}, {Takano}, {Takeda}, {Takeda}, {Talbot}, {Talbot}, {Tanaka}, {Tanaka}, {Tanaka}, {Tanaka}, {Tanaka}, {Tanasijczuk}, {Tanioka}, {Tanner}, {Tao}, {Tao}, {Mart{\'\i}n}, {Taranto}, {Tasson}, {Telada}, {Tenorio}, {Terhune}, {Terkowski}, {Thirugnanasambandam}, {Thomas}, {Thomas}, {Thomas}, {Thompson}, {Thondapu}, {Thorne}, {Thrane}, {Tiwari}, {Tiwari}, {Tiwari}, {Toivonen}, {Toland}, {Tolley}, {Tomaru}, {Tomigami}, {Tomura}, {Tonelli}, {Torres-Forn{\'e}}, {Torrie}, {E Melo}, {T{\"o}yr{\"a}}, {Trapananti}, {Travasso}, {Traylor}, {Trevor}, {Tringali}, {Tripathee}, {Troiano}, {Trovato}, {Trozzo}, {Trudeau}, {Tsai}, {Tsai}, {Tsang}, {Tsang}, {Tsao}, {Tse}, {Tso}, {Tsubono}, {Tsuchida}, {Tsukada}, {Tsuna}, {Tsutsui}, {Tsuzuki}, {Turbang}, {Turconi}, {Tuyenbayev}, {Ubhi}, {Uchikata}, {Uchiyama}, {Udall}, {Ueda}, {Uehara}, {Ueno}, {Ueshima}, {Unnikrishnan}, {Uraguchi}, {Urban}, {Ushiba}, {Utina}, {Vahlbruch}, {Vajente}, {Vajpeyi}, {Valdes}, {Valentini}, {Valsan}, {van Bakel},
  {van Beuzekom}, {van den Brand}, {van den Broeck}, {Vander-Hyde}, {van der Schaaf}, {van Heijningen}, {Vanosky}, {van Putten}, {van Remortel}, {Vardaro}, {Vargas}, {Varma}, {Vas{\'u}th}, {Vecchio}, {Vedovato}, {Veitch}, {Veitch}, {Venneberg}, {Venugopalan}, {Verkindt}, {Verma}, {Verma}, {Veske}, {Vetrano}, {Vicer{\'e}}, {Vidyant}, {Viets}, {Vijaykumar}, {Villa-Ortega}, {Vinet}, {Virtuoso}, {Vitale}, {Vo}, {Vocca}, {von Reis}, {von Wrangel}, {Vorvick}, {Vyatchanin}, {Wade}, {Wade}, {Wagner}, {Walet}, {Walker}, {Wallace}, {Wallace}, {Walsh}, {Wang}, {Wang}, {Wang}, {Ward}, {Warner}, {Was}, {Washimi}, {Washington}, {Watchi}, {Weaver}, {Webster}, {Weinert}, {Weinstein}, {Weiss}, {Weller}, {Wellmann}, {Wen}, {We{\ss}els}, {Wette}, {Whelan}, {White}, {Whiting}, {Whittle}, {Wilken}, {Williams}, {Williams}, {Williamson}, {Willis}, {Willke}, {Wilson}, {Winkler}, {Wipf}, {Wlodarczyk}, {Woan}, {Woehler}, {Wofford}, {Wong}, {Wu}, {Wu}, {Wu}, {Wu}, {Wysocki}, {Xiao}, {Xu}, {Yamada}, {Yamamoto}, {Yamamoto}, {Yamamoto},
  {Yamamoto}, {Yamashita}, {Yamazaki}, {Yang}, {Yang}, {Yang}, {Yang}, {Yang}, {Yap}, {Yeeles}, {Yelikar}, {Ying}, {Yokogawa}, {Yokoyama}, {Yokozawa}, {Yoo}, {Yoshioka}, {Yu}, {Yu}, {Yuzurihara}, {Zadro{\.z}ny}, {Zanolin}, {Zeidler}, {Zelenova}, {Zendri}, {Zevin}, {Zhan}, {Zhang}, {Zhang}, {Zhang}, {Zhang}, {Zhang}, {Zhao}, {Zhao}, {Zhao}, {Zhao}, {Zheng}, {Zhou}, {Zhou}, {Zhu}, {Zhu}, {Zimmerman}, {Zlochower}, {Zucker}, {Zweizig}, {LIGO Scientific Collaboration}, {VIRGO Collaboration}, \& {KAGRA Collaboration}}]{Abbott23}
{Abbott}, R., {Abbott}, T.~D., {Acernese}, F., {et~al.} 2023, Physical Review X, 13, 011048, \dodoi{10.1103/PhysRevX.13.011048}

\bibitem[{{Alexander} {et~al.}(2017){Alexander}, {Berger}, {Fong}, {Williams}, {Guidorzi}, {Margutti}, {Metzger}, {Annis}, {Blanchard}, {Brout}, {Brown}, {Chen}, {Chornock}, {Cowperthwaite}, {Drout}, {Eftekhari}, {Frieman}, {Holz}, {Nicholl}, {Rest}, {Sako}, {Soares-Santos}, \& {Villar}}]{Alexander17}
{Alexander}, K.~D., {Berger}, E., {Fong}, W., {et~al.} 2017, \apjl, 848, L21, \dodoi{10.3847/2041-8213/aa905d}

\bibitem[{{Andreoni} {et~al.}(2017){Andreoni}, {Ackley}, {Cooke}, {Acharyya}, {Allison}, {Anderson}, {Ashley}, {Baade}, {Bailes}, {Bannister}, {Beardsley}, {Bessell}, {Bian}, {Bland}, {Boer}, {Booler}, {Brandeker}, {Brown}, {Buckley}, {Chang}, {Coward}, {Crawford}, {Crisp}, {Crosse}, {Cucchiara}, {Cup{\'a}k}, {de Gois}, {Deller}, {Devillepoix}, {Dobie}, {Elmer}, {Emrich}, {Farah}, {Farrell}, {Franzen}, {Gaensler}, {Galloway}, {Gendre}, {Giblin}, {Goobar}, {Green}, {Hancock}, {Hartig}, {Howell}, {Horsley}, {Hotan}, {Howie}, {Hu}, {Hu}, {James}, {Johnston}, {Johnston-Hollitt}, {Kaplan}, {Kasliwal}, {Keane}, {Kenney}, {Klotz}, {Lau}, {Laugier}, {Lenc}, {Li}, {Liang}, {Lidman}, {Luvaul}, {Lynch}, {Ma}, {Macpherson}, {Mao}, {McClelland}, {McCully}, {M{\"o}ller}, {Morales}, {Morris}, {Murphy}, {Noysena}, {Onken}, {Orange}, {Os{\l}owski}, {Pallot}, {Paxman}, {Potter}, {Pritchard}, {Raja}, {Ridden-Harper}, {Romero-Colmenero}, {Sadler}, {Sansom}, {Scalzo}, {Schmidt}, {Scott}, {Seghouani}, {Shang}, {Shannon}, {Shao},
  {Shara}, {Sharp}, {Sokolowski}, {Sollerman}, {Staff}, {Steele}, {Sun}, {Suntzeff}, {Tao}, {Tingay}, {Towner}, {Thierry}, {Trott}, {Tucker}, {V{\"a}is{\"a}nen}, {Krishnan}, {Walker}, {Wang}, {Wang}, {Wayth}, {Whiting}, {Williams}, {Williams}, {Wolf}, {Wu}, {Wu}, {Yang}, {Yuan}, {Zhang}, {Zhou}, \& {Zovaro}}]{Andreoni17}
{Andreoni}, I., {Ackley}, K., {Cooke}, J., {et~al.} 2017, \pasa, 34, e069, \dodoi{10.1017/pasa.2017.65}

\bibitem[{{Andreoni} {et~al.}(2022{\natexlab{a}}){Andreoni}, {Margutti}, {Salafia}, {Parazin}, {Villar}, {Coughlin}, {Yoachim}, {Mortensen}, {Brethauer}, {Smartt}, {Kasliwal}, {Alexander}, {Anand}, {Berger}, {Bernardini}, {Bianco}, {Blanchard}, {Bloom}, {Brocato}, {Bulla}, {Cartier}, {Cenko}, {Chornock}, {Copperwheat}, {Corsi}, {D'Ammando}, {D'Avanzo}, {H{\'e}l{\`e}ne Datrier}, {Foley}, {Ghirlanda}, {Goobar}, {Grindlay}, {Hajela}, {Holz}, {Karambelkar}, {Kool}, {Lamb}, {Laskar}, {Levan}, {Maguire}, {May}, {Melandri}, {Milisavljevic}, {Miller}, {Nicholl}, {Nissanke}, {Palmese}, {Piranomonte}, {Rest}, {Sagu{\'e}s-Carracedo}, {Siellez}, {Singer}, {Smith}, {Steeghs}, \& {Tanvir}}]{Andreoni22}
{Andreoni}, I., {Margutti}, R., {Salafia}, O.~S., {et~al.} 2022{\natexlab{a}}, \apjs, 260, 18, \dodoi{10.3847/1538-4365/ac617c}

\bibitem[{{Andreoni} {et~al.}(2022{\natexlab{b}}){Andreoni}, {Coughlin}, {Almualla}, {Bellm}, {Bianco}, {Bulla}, {Cucchiara}, {Dietrich}, {Goobar}, {Kool}, {Li}, {Ragosta}, {Sagu{\'e}s-Carracedo}, \& {Singer}}]{Andreoni22a}
{Andreoni}, I., {Coughlin}, M.~W., {Almualla}, M., {et~al.} 2022{\natexlab{b}}, \apjs, 258, 5, \dodoi{10.3847/1538-4365/ac3bae}

\bibitem[{{Arcavi} {et~al.}(2017){Arcavi}, {Hosseinzadeh}, {Howell}, {McCully}, {Poznanski}, {Kasen}, {Barnes}, {Zaltzman}, {Vasylyev}, {Maoz}, \& {Valenti}}]{Arcavi17}
{Arcavi}, I., {Hosseinzadeh}, G., {Howell}, D.~A., {et~al.} 2017, \nat, 551, 64, \dodoi{10.1038/nature24291}

\bibitem[{{Asplund} {et~al.}(2009){Asplund}, {Grevesse}, {Sauval}, \& {Scott}}]{Asplund09}
{Asplund}, M., {Grevesse}, N., {Sauval}, A.~J., \& {Scott}, P. 2009, \araa, 47, 481, \dodoi{10.1146/annurev.astro.46.060407.145222}

\bibitem[{{Astropy Collaboration} {et~al.}(2013){Astropy Collaboration}, {Robitaille}, {Tollerud}, {Greenfield}, {Droettboom}, {Bray}, {Aldcroft}, {Davis}, {Ginsburg}, {Price-Whelan}, {Kerzendorf}, {Conley}, {Crighton}, {Barbary}, {Muna}, {Ferguson}, {Grollier}, {Parikh}, {Nair}, {Unther}, {Deil}, {Woillez}, {Conseil}, {Kramer}, {Turner}, {Singer}, {Fox}, {Weaver}, {Zabalza}, {Edwards}, {Azalee Bostroem}, {Burke}, {Casey}, {Crawford}, {Dencheva}, {Ely}, {Jenness}, {Labrie}, {Lim}, {Pierfederici}, {Pontzen}, {Ptak}, {Refsdal}, {Servillat}, \& {Streicher}}]{astropy:2013}
{Astropy Collaboration}, {Robitaille}, T.~P., {Tollerud}, E.~J., {et~al.} 2013, \aap, 558, A33, \dodoi{10.1051/0004-6361/201322068}

\bibitem[{{Astropy Collaboration} {et~al.}(2018){Astropy Collaboration}, {Price-Whelan}, {Sip{\H{o}}cz}, {G{\"u}nther}, {Lim}, {Crawford}, {Conseil}, {Shupe}, {Craig}, {Dencheva}, {Ginsburg}, {Vand erPlas}, {Bradley}, {P{\'e}rez-Su{\'a}rez}, {de Val-Borro}, {Aldcroft}, {Cruz}, {Robitaille}, {Tollerud}, {Ardelean}, {Babej}, {Bach}, {Bachetti}, {Bakanov}, {Bamford}, {Barentsen}, {Barmby}, {Baumbach}, {Berry}, {Biscani}, {Boquien}, {Bostroem}, {Bouma}, {Brammer}, {Bray}, {Breytenbach}, {Buddelmeijer}, {Burke}, {Calderone}, {Cano Rodr{\'\i}guez}, {Cara}, {Cardoso}, {Cheedella}, {Copin}, {Corrales}, {Crichton}, {D'Avella}, {Deil}, {Depagne}, {Dietrich}, {Donath}, {Droettboom}, {Earl}, {Erben}, {Fabbro}, {Ferreira}, {Finethy}, {Fox}, {Garrison}, {Gibbons}, {Goldstein}, {Gommers}, {Greco}, {Greenfield}, {Groener}, {Grollier}, {Hagen}, {Hirst}, {Homeier}, {Horton}, {Hosseinzadeh}, {Hu}, {Hunkeler}, {Ivezi{\'c}}, {Jain}, {Jenness}, {Kanarek}, {Kendrew}, {Kern}, {Kerzendorf}, {Khvalko}, {King}, {Kirkby}, {Kulkarni},
  {Kumar}, {Lee}, {Lenz}, {Littlefair}, {Ma}, {Macleod}, {Mastropietro}, {McCully}, {Montagnac}, {Morris}, {Mueller}, {Mumford}, {Muna}, {Murphy}, {Nelson}, {Nguyen}, {Ninan}, {N{\"o}the}, {Ogaz}, {Oh}, {Parejko}, {Parley}, {Pascual}, {Patil}, {Patil}, {Plunkett}, {Prochaska}, {Rastogi}, {Reddy Janga}, {Sabater}, {Sakurikar}, {Seifert}, {Sherbert}, {Sherwood-Taylor}, {Shih}, {Sick}, {Silbiger}, {Singanamalla}, {Singer}, {Sladen}, {Sooley}, {Sornarajah}, {Streicher}, {Teuben}, {Thomas}, {Tremblay}, {Turner}, {Terr{\'o}n}, {van Kerkwijk}, {de la Vega}, {Watkins}, {Weaver}, {Whitmore}, {Woillez}, {Zabalza}, \& {Astropy Contributors}}]{astropy:2018}
{Astropy Collaboration}, {Price-Whelan}, A.~M., {Sip{\H{o}}cz}, B.~M., {et~al.} 2018, \aj, 156, 123, \dodoi{10.3847/1538-3881/aabc4f}

\bibitem[{{Astropy Collaboration} {et~al.}(2022){Astropy Collaboration}, {Price-Whelan}, {Lim}, {Earl}, {Starkman}, {Bradley}, {Shupe}, {Patil}, {Corrales}, {Brasseur}, {N{"o}the}, {Donath}, {Tollerud}, {Morris}, {Ginsburg}, {Vaher}, {Weaver}, {Tocknell}, {Jamieson}, {van Kerkwijk}, {Robitaille}, {Merry}, {Bachetti}, {G{"u}nther}, {Aldcroft}, {Alvarado-Montes}, {Archibald}, {B{'o}di}, {Bapat}, {Barentsen}, {Baz{'a}n}, {Biswas}, {Boquien}, {Burke}, {Cara}, {Cara}, {Conroy}, {Conseil}, {Craig}, {Cross}, {Cruz}, {D'Eugenio}, {Dencheva}, {Devillepoix}, {Dietrich}, {Eigenbrot}, {Erben}, {Ferreira}, {Foreman-Mackey}, {Fox}, {Freij}, {Garg}, {Geda}, {Glattly}, {Gondhalekar}, {Gordon}, {Grant}, {Greenfield}, {Groener}, {Guest}, {Gurovich}, {Handberg}, {Hart}, {Hatfield-Dodds}, {Homeier}, {Hosseinzadeh}, {Jenness}, {Jones}, {Joseph}, {Kalmbach}, {Karamehmetoglu}, {Ka{l}uszy{'n}ski}, {Kelley}, {Kern}, {Kerzendorf}, {Koch}, {Kulumani}, {Lee}, {Ly}, {Ma}, {MacBride}, {Maljaars}, {Muna}, {Murphy}, {Norman}, {O'Steen},
  {Oman}, {Pacifici}, {Pascual}, {Pascual-Granado}, {Patil}, {Perren}, {Pickering}, {Rastogi}, {Roulston}, {Ryan}, {Rykoff}, {Sabater}, {Sakurikar}, {Salgado}, {Sanghi}, {Saunders}, {Savchenko}, {Schwardt}, {Seifert-Eckert}, {Shih}, {Jain}, {Shukla}, {Sick}, {Simpson}, {Singanamalla}, {Singer}, {Singhal}, {Sinha}, {Sip{H{o}}cz}, {Spitler}, {Stansby}, {Streicher}, {{{S}}umak}, {Swinbank}, {Taranu}, {Tewary}, {Tremblay}, {Val-Borro}, {Van Kooten}, {Vasovi{'c}}, {Verma}, {de Miranda Cardoso}, {Williams}, {Wilson}, {Winkel}, {Wood-Vasey}, {Xue}, {Yoachim}, {Zhang}, {Zonca}, \& {Astropy Project Contributors}}]{astropy:2022}
{Astropy Collaboration}, {Price-Whelan}, A.~M., {Lim}, P.~L., {et~al.} 2022, \apj, 935, 167, \dodoi{10.3847/1538-4357/ac7c74}

\bibitem[{{Badnell}(2011)}]{Badnell11}
{Badnell}, N.~R. 2011, Computer Physics Communications, 182, 1528, \dodoi{10.1016/j.cpc.2011.03.023}

\bibitem[{{Bar-Shalom} {et~al.}(2001){Bar-Shalom}, {Klapisch}, \& {Oreg}}]{Bar-Shalom01}
{Bar-Shalom}, A., {Klapisch}, M., \& {Oreg}, J. 2001, \jqsrt, 71, 169, \dodoi{10.1016/S0022-4073(01)00066-8}

\bibitem[{{Barnes} {et~al.}(2016){Barnes}, {Kasen}, {Wu}, \& {Mart{\'\i}nez-Pinedo}}]{Barnes16}
{Barnes}, J., {Kasen}, D., {Wu}, M.-R., \& {Mart{\'\i}nez-Pinedo}, G. 2016, \apj, 829, 110, \dodoi{10.3847/0004-637X/829/2/110}

\bibitem[{{Breschi} {et~al.}(2021){Breschi}, {Perego}, {Bernuzzi}, {Del Pozzo}, {Nedora}, {Radice}, \& {Vescovi}}]{Breschi21}
{Breschi}, M., {Perego}, A., {Bernuzzi}, S., {et~al.} 2021, \mnras, 505, 1661, \dodoi{10.1093/mnras/stab1287}

\bibitem[{{Bulla}(2023)}]{Bulla24}
{Bulla}, M. 2023, \mnras, 520, 2558, \dodoi{10.1093/mnras/stad232}

\bibitem[{{Chornock} {et~al.}(2017){Chornock}, {Berger}, {Kasen}, {Cowperthwaite}, {Nicholl}, {Villar}, {Alexander}, {Blanchard}, {Eftekhari}, {Fong}, {Margutti}, {Williams}, {Annis}, {Brout}, {Brown}, {Chen}, {Drout}, {Farr}, {Foley}, {Frieman}, {Fryer}, {Herner}, {Holz}, {Kessler}, {Matheson}, {Metzger}, {Quataert}, {Rest}, {Sako}, {Scolnic}, {Smith}, \& {Soares-Santos}}]{Chornock17}
{Chornock}, R., {Berger}, E., {Kasen}, D., {et~al.} 2017, \apjl, 848, L19, \dodoi{10.3847/2041-8213/aa905c}

\bibitem[{{Coughlin} {et~al.}(2018){Coughlin}, {Dietrich}, {Doctor}, {Kasen}, {Coughlin}, {Jerkstrand}, {Leloudas}, {McBrien}, {Metzger}, {O'Shaughnessy}, \& {Smartt}}]{Coughlin18}
{Coughlin}, M.~W., {Dietrich}, T., {Doctor}, Z., {et~al.} 2018, \mnras, 480, 3871, \dodoi{10.1093/mnras/sty2174}

\bibitem[{{Coulter} {et~al.}(2017){Coulter}, {Foley}, {Kilpatrick}, {Drout}, {Piro}, {Shappee}, {Siebert}, {Simon}, {Ulloa}, {Kasen}, {Madore}, {Murguia-Berthier}, {Pan}, {Prochaska}, {Ramirez-Ruiz}, {Rest}, \& {Rojas-Bravo}}]{Coulter17}
{Coulter}, D.~A., {Foley}, R.~J., {Kilpatrick}, C.~D., {et~al.} 2017, Science, 358, 1556, \dodoi{10.1126/science.aap9811}

\bibitem[{{Cowperthwaite} {et~al.}(2019){Cowperthwaite}, {Villar}, {Scolnic}, \& {Berger}}]{Cowperthwaite19}
{Cowperthwaite}, P.~S., {Villar}, V.~A., {Scolnic}, D.~M., \& {Berger}, E. 2019, \apj, 874, 88, \dodoi{10.3847/1538-4357/ab07b6}

\bibitem[{{Cowperthwaite} {et~al.}(2017){Cowperthwaite}, {Berger}, {Villar}, {Metzger}, {Nicholl}, {Chornock}, {Blanchard}, {Fong}, {Margutti}, {Soares-Santos}, {Alexander}, {Allam}, {Annis}, {Brout}, {Brown}, {Butler}, {Chen}, {Diehl}, {Doctor}, {Drout}, {Eftekhari}, {Farr}, {Finley}, {Foley}, {Frieman}, {Fryer}, {Garc{\'\i}a-Bellido}, {Gill}, {Guillochon}, {Herner}, {Holz}, {Kasen}, {Kessler}, {Marriner}, {Matheson}, {Neilsen}, {Quataert}, {Palmese}, {Rest}, {Sako}, {Scolnic}, {Smith}, {Tucker}, {Williams}, {Balbinot}, {Carlin}, {Cook}, {Durret}, {Li}, {Lopes}, {Louren{\c{c}}o}, {Marshall}, {Medina}, {Muir}, {Mu{\~n}oz}, {Sauseda}, {Schlegel}, {Secco}, {Vivas}, {Wester}, {Zenteno}, {Zhang}, {Abbott}, {Banerji}, {Bechtol}, {Benoit-L{\'e}vy}, {Bertin}, {Buckley-Geer}, {Burke}, {Capozzi}, {Carnero Rosell}, {Carrasco Kind}, {Castander}, {Crocce}, {Cunha}, {D'Andrea}, {da Costa}, {Davis}, {DePoy}, {Desai}, {Dietrich}, {Drlica-Wagner}, {Eifler}, {Evrard}, {Fernandez}, {Flaugher}, {Fosalba}, {Gaztanaga}, {Gerdes},
  {Giannantonio}, {Goldstein}, {Gruen}, {Gruendl}, {Gutierrez}, {Honscheid}, {Jain}, {James}, {Jeltema}, {Johnson}, {Johnson}, {Kent}, {Krause}, {Kron}, {Kuehn}, {Nuropatkin}, {Lahav}, {Lima}, {Lin}, {Maia}, {March}, {Martini}, {McMahon}, {Menanteau}, {Miller}, {Miquel}, {Mohr}, {Neilsen}, {Nichol}, {Ogando}, {Plazas}, {Roe}, {Romer}, {Roodman}, {Rykoff}, {Sanchez}, {Scarpine}, {Schindler}, {Schubnell}, {Sevilla-Noarbe}, {Smith}, {Smith}, {Sobreira}, {Suchyta}, {Swanson}, {Tarle}, {Thomas}, {Thomas}, {Troxel}, {Vikram}, {Walker}, {Wechsler}, {Weller}, {Yanny}, \& {Zuntz}}]{Cowperthwaite17}
{Cowperthwaite}, P.~S., {Berger}, E., {Villar}, V.~A., {et~al.} 2017, \apjl, 848, L17, \dodoi{10.3847/2041-8213/aa8fc7}

\bibitem[{{D{\'\i}az} {et~al.}(2017){D{\'\i}az}, {Macri}, {Garcia Lambas}, {Mendes de Oliveira}, {Nilo Castell{\'o}n}, {Ribeiro}, {S{\'a}nchez}, {Schoenell}, {Abramo}, {Akras}, {Alcaniz}, {Artola}, {Beroiz}, {Bonoli}, {Cabral}, {Camuccio}, {Castillo}, {Chavushyan}, {Coelho}, {Colazo}, {Costa-Duarte}, {Cuevas Larenas}, {DePoy}, {Dom{\'\i}nguez Romero}, {Dultzin}, {Fern{\'a}ndez}, {Garc{\'\i}a}, {Girardini}, {Gon{\c{c}}alves}, {Gon{\c{c}}alves}, {Gurovich}, {Jim{\'e}nez-Teja}, {Kanaan}, {Lares}, {Lopes de Oliveira}, {L{\'o}pez-Cruz}, {Marshall}, {Melia}, {Molino}, {Padilla}, {Pe{\~n}uela}, {Placco}, {Qui{\~n}ones}, {Ram{\'\i}rez Rivera}, {Renzi}, {Riguccini}, {R{\'\i}os-L{\'o}pez}, {Rodriguez}, {Sampedro}, {Schneiter}, {Sodr{\'e}}, {Starck}, {Torres-Flores}, {Tornatore}, \& {Zadro{\.z}ny}}]{Diaz17}
{D{\'\i}az}, M.~C., {Macri}, L.~M., {Garcia Lambas}, D., {et~al.} 2017, \apjl, 848, L29, \dodoi{10.3847/2041-8213/aa9060}

\bibitem[{{Drout} {et~al.}(2017){Drout}, {Piro}, {Shappee}, {Kilpatrick}, {Simon}, {Contreras}, {Coulter}, {Foley}, {Siebert}, {Morrell}, {Boutsia}, {Di Mille}, {Holoien}, {Kasen}, {Kollmeier}, {Madore}, {Monson}, {Murguia-Berthier}, {Pan}, {Prochaska}, {Ramirez-Ruiz}, {Rest}, {Adams}, {Alatalo}, {Ba{\~n}ados}, {Baughman}, {Beers}, {Bernstein}, {Bitsakis}, {Campillay}, {Hansen}, {Higgs}, {Ji}, {Maravelias}, {Marshall}, {Moni Bidin}, {Prieto}, {Rasmussen}, {Rojas-Bravo}, {Strom}, {Ulloa}, {Vargas-Gonz{\'a}lez}, {Wan}, \& {Whitten}}]{Drout17}
{Drout}, M.~R., {Piro}, A.~L., {Shappee}, B.~J., {et~al.} 2017, Science, 358, 1570, \dodoi{10.1126/science.aaq0049}

\bibitem[{{Eichler} {et~al.}(1989){Eichler}, {Livio}, {Piran}, \& {Schramm}}]{Eichler89}
{Eichler}, D., {Livio}, M., {Piran}, T., \& {Schramm}, D.~N. 1989, \nat, 340, 126, \dodoi{10.1038/340126a0}

\bibitem[{{Evans} {et~al.}(2017){Evans}, {Cenko}, {Kennea}, {Emery}, {Kuin}, {Korobkin}, {Wollaeger}, {Fryer}, {Madsen}, {Harrison}, {Xu}, {Nakar}, {Hotokezaka}, {Lien}, {Campana}, {Oates}, {Troja}, {Breeveld}, {Marshall}, {Barthelmy}, {Beardmore}, {Burrows}, {Cusumano}, {D'A{\`\i}}, {D'Avanzo}, {D'Elia}, {de Pasquale}, {Even}, {Fontes}, {Forster}, {Garcia}, {Giommi}, {Grefenstette}, {Gronwall}, {Hartmann}, {Heida}, {Hungerford}, {Kasliwal}, {Krimm}, {Levan}, {Malesani}, {Melandri}, {Miyasaka}, {Nousek}, {O'Brien}, {Osborne}, {Pagani}, {Page}, {Palmer}, {Perri}, {Pike}, {Racusin}, {Rosswog}, {Siegel}, {Sakamoto}, {Sbarufatti}, {Tagliaferri}, {Tanvir}, \& {Tohuvavohu}}]{Evans17}
{Evans}, P.~A., {Cenko}, S.~B., {Kennea}, J.~A., {et~al.} 2017, Science, 358, 1565, \dodoi{10.1126/science.aap9580}

\bibitem[{{Fontes} {et~al.}(2020){Fontes}, {Fryer}, {Hungerford}, {Wollaeger}, \& {Korobkin}}]{Fontes20}
{Fontes}, C.~J., {Fryer}, C.~L., {Hungerford}, A.~L., {Wollaeger}, R.~T., \& {Korobkin}, O. 2020, \mnras, 493, 4143, \dodoi{10.1093/mnras/staa485}

\bibitem[{{Freiburghaus} {et~al.}(1999){Freiburghaus}, {Rosswog}, \& {Thielemann}}]{Freiburghaus99}
{Freiburghaus}, C., {Rosswog}, S., \& {Thielemann}, F.~K. 1999, \apjl, 525, L121, \dodoi{10.1086/312343}

\bibitem[{{Fryer} {et~al.}(2024){Fryer}, {Hungerford}, {Wollaeger}, {Miller}, {De}, {Fontes}, {Korobkin}, {Kedia}, {Ristic}, \& {O'Shaughnessy}}]{Fryer24}
{Fryer}, C.~L., {Hungerford}, A.~L., {Wollaeger}, R.~T., {et~al.} 2024, \apj, 961, 9, \dodoi{10.3847/1538-4357/ad1036}

\bibitem[{{Gillanders} {et~al.}(2022){Gillanders}, {Smartt}, {Sim}, {Bauswein}, \& {Goriely}}]{Gillanders22}
{Gillanders}, J.~H., {Smartt}, S.~J., {Sim}, S.~A., {Bauswein}, A., \& {Goriely}, S. 2022, \mnras, 515, 631, \dodoi{10.1093/mnras/stac1258}

\bibitem[{{Goldstein} {et~al.}(2017){Goldstein}, {Veres}, {Burns}, {Briggs}, {Hamburg}, {Kocevski}, {Wilson-Hodge}, {Preece}, {Poolakkil}, {Roberts}, {Hui}, {Connaughton}, {Racusin}, {von Kienlin}, {Dal Canton}, {Christensen}, {Littenberg}, {Siellez}, {Blackburn}, {Broida}, {Bissaldi}, {Cleveland}, {Gibby}, {Giles}, {Kippen}, {McBreen}, {McEnery}, {Meegan}, {Paciesas}, \& {Stanbro}}]{Goldstein17}
{Goldstein}, A., {Veres}, P., {Burns}, E., {et~al.} 2017, \apjl, 848, L14, \dodoi{10.3847/2041-8213/aa8f41}

\bibitem[{{Hallinan} {et~al.}(2017){Hallinan}, {Corsi}, {Mooley}, {Hotokezaka}, {Nakar}, {Kasliwal}, {Kaplan}, {Frail}, {Myers}, {Murphy}, {De}, {Dobie}, {Allison}, {Bannister}, {Bhalerao}, {Chandra}, {Clarke}, {Giacintucci}, {Ho}, {Horesh}, {Kassim}, {Kulkarni}, {Lenc}, {Lockman}, {Lynch}, {Nichols}, {Nissanke}, {Palliyaguru}, {Peters}, {Piran}, {Rana}, {Sadler}, \& {Singer}}]{Hallinan17}
{Hallinan}, G., {Corsi}, A., {Mooley}, K.~P., {et~al.} 2017, Science, 358, 1579, \dodoi{10.1126/science.aap9855}

\bibitem[{Harris {et~al.}(2020)Harris, Millman, van~der Walt, Gommers, Virtanen, Cournapeau, Wieser, Taylor, Berg, Smith, Kern, Picus, Hoyer, van Kerkwijk, Brett, Haldane, del R{\'{i}}o, Wiebe, Peterson, G{\'{e}}rard-Marchant, Sheppard, Reddy, Weckesser, Abbasi, Gohlke, \& Oliphant}]{Numpy}
Harris, C.~R., Millman, K.~J., van~der Walt, S.~J., {et~al.} 2020, Nature, 585, 357, \dodoi{10.1038/s41586-020-2649-2}

\bibitem[{{Hillier} \& {Lanz}(2001)}]{Hillier01}
{Hillier}, D.~J., \& {Lanz}, T. 2001, in Astronomical Society of the Pacific Conference Series, Vol. 247, Spectroscopic Challenges of Photoionized Plasmas, ed. G.~{Ferland} \& D.~W. {Savin}, 343

\bibitem[{{Holmbeck} {et~al.}(2023){Holmbeck}, {Barnes}, {Lund}, {Sprouse}, {McLaughlin}, \& {Mumpower}}]{Holmbeck23}
{Holmbeck}, E.~M., {Barnes}, J., {Lund}, K.~A., {et~al.} 2023, \apjl, 951, L13, \dodoi{10.3847/2041-8213/acd9cb}

\bibitem[{{Hotokezaka} {et~al.}(2015){Hotokezaka}, {Piran}, \& {Paul}}]{Hotokezaka15}
{Hotokezaka}, K., {Piran}, T., \& {Paul}, M. 2015, Nature Physics, 11, 1042, \dodoi{10.1038/nphys3574}

\bibitem[{{Hotokezaka} {et~al.}(2022){Hotokezaka}, {Tanaka}, {Kato}, \& {Gaigalas}}]{Hotokezaka22}
{Hotokezaka}, K., {Tanaka}, M., {Kato}, D., \& {Gaigalas}, G. 2022, \mnras, 515, L89, \dodoi{10.1093/mnrasl/slac071}

\bibitem[{{Hotokezaka} {et~al.}(2023){Hotokezaka}, {Tanaka}, {Kato}, \& {Gaigalas}}]{Hotokezaka23}
---. 2023, \mnras, 526, L155, \dodoi{10.1093/mnrasl/slad128}

\bibitem[{Hu {et~al.}(2017)Hu, Wu, Andreoni, {B. Ashley}, Cooke, Cui, Du, Dai, Gu, Hu, Lu, Li, Li, Liang, Liu, Ma, Shang, Sun, Suntzeff, Tao, Uddin, Wang, Wang, Wen, Xiao, Xu, Yang, Yang, Yuan, Zhou, Zhang, Zhou, \& Zhu}]{Hu17}
Hu, L., Wu, X., Andreoni, I., {et~al.} 2017, Science Bulletin, 62, 1433, \dodoi{https://doi.org/10.1016/j.scib.2017.10.006}

\bibitem[{Hunter(2007)}]{Matplotlib}
Hunter, J.~D. 2007, Computing in Science \& Engineering, 9, 90, \dodoi{10.1109/MCSE.2007.55}

\bibitem[{{Ji} {et~al.}(2016){Ji}, {Frebel}, {Chiti}, \& {Simon}}]{Ji16}
{Ji}, A.~P., {Frebel}, A., {Chiti}, A., \& {Simon}, J.~D. 2016, \nat, 531, 610, \dodoi{10.1038/nature17425}

\bibitem[{{Kasen} {et~al.}(2013){Kasen}, {Badnell}, \& {Barnes}}]{Kasen13}
{Kasen}, D., {Badnell}, N.~R., \& {Barnes}, J. 2013, \apj, 774, 25, \dodoi{10.1088/0004-637X/774/1/25}

\bibitem[{{Kasen} {et~al.}(2017){Kasen}, {Metzger}, {Barnes}, {Quataert}, \& {Ramirez-Ruiz}}]{Kasen17}
{Kasen}, D., {Metzger}, B., {Barnes}, J., {Quataert}, E., \& {Ramirez-Ruiz}, E. 2017, \nat, 551, 80, \dodoi{10.1038/nature24453}

\bibitem[{{Kasen} {et~al.}(2006){Kasen}, {Thomas}, \& {Nugent}}]{Kasen06}
{Kasen}, D., {Thomas}, R.~C., \& {Nugent}, P. 2006, \apj, 651, 366, \dodoi{10.1086/506190}

\bibitem[{{Kasliwal} {et~al.}(2017){Kasliwal}, {Nakar}, {Singer}, {Kaplan}, {Cook}, {Van Sistine}, {Lau}, {Fremling}, {Gottlieb}, {Jencson}, {Adams}, {Feindt}, {Hotokezaka}, {Ghosh}, {Perley}, {Yu}, {Piran}, {Allison}, {Anupama}, {Balasubramanian}, {Bannister}, {Bally}, {Barnes}, {Barway}, {Bellm}, {Bhalerao}, {Bhattacharya}, {Blagorodnova}, {Bloom}, {Brady}, {Cannella}, {Chatterjee}, {Cenko}, {Cobb}, {Copperwheat}, {Corsi}, {De}, {Dobie}, {Emery}, {Evans}, {Fox}, {Frail}, {Frohmaier}, {Goobar}, {Hallinan}, {Harrison}, {Helou}, {Hinderer}, {Ho}, {Horesh}, {Ip}, {Itoh}, {Kasen}, {Kim}, {Kuin}, {Kupfer}, {Lynch}, {Madsen}, {Mazzali}, {Miller}, {Mooley}, {Murphy}, {Ngeow}, {Nichols}, {Nissanke}, {Nugent}, {Ofek}, {Qi}, {Quimby}, {Rosswog}, {Rusu}, {Sadler}, {Schmidt}, {Sollerman}, {Steele}, {Williamson}, {Xu}, {Yan}, {Yatsu}, {Zhang}, \& {Zhao}}]{Kasliwal17}
{Kasliwal}, M.~M., {Nakar}, E., {Singer}, L.~P., {et~al.} 2017, Science, 358, 1559, \dodoi{10.1126/science.aap9455}

\bibitem[{{Kasliwal} {et~al.}(2022){Kasliwal}, {Kasen}, {Lau}, {Perley}, {Rosswog}, {Ofek}, {Hotokezaka}, {Chary}, {Sollerman}, {Goobar}, \& {Kaplan}}]{Kasliwal22}
{Kasliwal}, M.~M., {Kasen}, D., {Lau}, R.~M., {et~al.} 2022, \mnras, 510, L7, \dodoi{10.1093/mnrasl/slz007}

\bibitem[{{Kawaguchi} {et~al.}(2018){Kawaguchi}, {Shibata}, \& {Tanaka}}]{Kawaguchi18}
{Kawaguchi}, K., {Shibata}, M., \& {Tanaka}, M. 2018, \apjl, 865, L21, \dodoi{10.3847/2041-8213/aade02}

\bibitem[{{Kim} {et~al.}(2015){Kim}, {Perera}, \& {McLaughlin}}]{Kim15}
{Kim}, C., {Perera}, B. B.~P., \& {McLaughlin}, M.~A. 2015, \mnras, 448, 928, \dodoi{10.1093/mnras/stu2729}

\bibitem[{{Lattimer} \& {Schramm}(1974)}]{Lattimer&Schramm74}
{Lattimer}, J.~M., \& {Schramm}, D.~N. 1974, \apjl, 192, L145, \dodoi{10.1086/181612}

\bibitem[{{Lattimer} \& {Schramm}(1976)}]{Lattimer&Schramm76}
---. 1976, \apj, 210, 549, \dodoi{10.1086/154860}

\bibitem[{{Levan} {et~al.}(2024){Levan}, {Gompertz}, {Salafia}, {Bulla}, {Burns}, {Hotokezaka}, {Izzo}, {Lamb}, {Malesani}, {Oates}, {Ravasio}, {Rouco Escorial}, {Schneider}, {Sarin}, {Schulze}, {Tanvir}, {Ackley}, {Anderson}, {Brammer}, {Christensen}, {Dhillon}, {Evans}, {Fausnaugh}, {Fong}, {Fruchter}, {Fryer}, {Fynbo}, {Gaspari}, {Heintz}, {Hjorth}, {Kennea}, {Kennedy}, {Laskar}, {Leloudas}, {Mandel}, {Martin-Carrillo}, {Metzger}, {Nicholl}, {Nugent}, {Palmerio}, {Pugliese}, {Rastinejad}, {Rhodes}, {Rossi}, {Saccardi}, {Smartt}, {Stevance}, {Tohuvavohu}, {van der Horst}, {Vergani}, {Watson}, {Barclay}, {Bhirombhakdi}, {Breedt}, {Breeveld}, {Brown}, {Campana}, {Chrimes}, {D'Avanzo}, {D'Elia}, {De Pasquale}, {Dyer}, {Galloway}, {Garbutt}, {Green}, {Hartmann}, {Jakobsson}, {Kerry}, {Kouveliotou}, {Langeroodi}, {Le Floc'h}, {Leung}, {Littlefair}, {Munday}, {O'Brien}, {Parsons}, {Pelisoli}, {Sahman}, {Salvaterra}, {Sbarufatti}, {Steeghs}, {Tagliaferri}, {Th{\"o}ne}, {de Ugarte Postigo}, \& {Kann}}]{Levan24}
{Levan}, A.~J., {Gompertz}, B.~P., {Salafia}, O.~S., {et~al.} 2024, \nat, 626, 737, \dodoi{10.1038/s41586-023-06759-1}

\bibitem[{{Lippuner} \& {Roberts}(2015)}]{Lippuner15}
{Lippuner}, J., \& {Roberts}, L.~F. 2015, \apj, 815, 82, \dodoi{10.1088/0004-637X/815/2/82}

\bibitem[{{Lipunov} {et~al.}(2017){Lipunov}, {Gorbovskoy}, {Kornilov}, {. Tyurina}, {Balanutsa}, {Kuznetsov}, {Vlasenko}, {Kuvshinov}, {Gorbunov}, {Buckley}, {Krylov}, {Podesta}, {Lopez}, {Podesta}, {Levato}, {Saffe}, {Mallamachi}, {Potter}, {Budnev}, {Gress}, {Ishmuhametova}, {Vladimirov}, {Zimnukhov}, {Yurkov}, {Sergienko}, {Gabovich}, {Rebolo}, {Serra-Ricart}, {Israelyan}, {Chazov}, {Wang}, {Tlatov}, \& {Panchenko}}]{Lipunov17}
{Lipunov}, V.~M., {Gorbovskoy}, E., {Kornilov}, V.~G., {et~al.} 2017, \apjl, 850, L1, \dodoi{10.3847/2041-8213/aa92c0}

\bibitem[{{Margutti} \& {Chornock}(2021)}]{Margutti&Chornock21}
{Margutti}, R., \& {Chornock}, R. 2021, \araa, 59, 155, \dodoi{10.1146/annurev-astro-112420-030742}

\bibitem[{{Margutti} {et~al.}(2017){Margutti}, {Berger}, {Fong}, {Guidorzi}, {Alexander}, {Metzger}, {Blanchard}, {Cowperthwaite}, {Chornock}, {Eftekhari}, {Nicholl}, {Villar}, {Williams}, {Annis}, {Brown}, {Chen}, {Doctor}, {Frieman}, {Holz}, {Sako}, \& {Soares-Santos}}]{Margutti17}
{Margutti}, R., {Berger}, E., {Fong}, W., {et~al.} 2017, \apjl, 848, L20, \dodoi{10.3847/2041-8213/aa9057}

\bibitem[{{Margutti} {et~al.}(2018){Margutti}, {Cowperthwaite}, {Doctor}, {Mortensen}, {Pankow}, {Salafia}, {Villar}, {Alexander}, {Annis}, {Andreoni}, {Baldeschi}, {Balmaverde}, {Berger}, {Bernardini}, {Berry}, {Bianco}, {Blanchard}, {Brocato}, {Carnerero}, {Cartier}, {Cenko}, {Chornock}, {Chomiuk}, {Copperwheat}, {Coughlin}, {Coppejans}, {Corsi}, {D'Ammando}, {Datrier}, {D'Avanzo}, {Dimitriadis}, {Drout}, {Foley}, {Fong}, {Fox}, {Ghirlanda}, {Goldstein}, {Grindlay}, {Guidorzi}, {Haiman}, {Hendry}, {Holz}, {Hung}, {Inserra}, {Jones}, {Kalogera}, {Kilpatrick}, {Lamb}, {Laskar}, {Levan}, {Mason}, {Maguire}, {Melandri}, {Milisavljevic}, {Miller}, {Narayan}, {Nielsen}, {Nicholl}, {Nissanke}, {Nugent}, {Pan}, {Pasham}, {Paterson}, {Piranomonte}, {Racusin}, {Rest}, {Righi}, {Sand}, {Seaman}, {Scolnic}, {Siellez}, {Singer}, {Szkody}, {Smith}, {Steeghs}, {Sullivan}, {Tanvir}, {Terreran}, {Trimble}, {Valenti}, {LSST Transient}, \& {Variable Stars Collaboration}}]{Margutti18}
{Margutti}, R., {Cowperthwaite}, P., {Doctor}, Z., {et~al.} 2018, arXiv e-prints, arXiv:1812.04051, \dodoi{10.48550/arXiv.1812.04051}

\bibitem[{{Nakar}(2020)}]{Nakar20}
{Nakar}, E. 2020, \physrep, 886, 1, \dodoi{10.1016/j.physrep.2020.08.008}

\bibitem[{{Narayan} {et~al.}(1992){Narayan}, {Paczynski}, \& {Piran}}]{Narayan92}
{Narayan}, R., {Paczynski}, B., \& {Piran}, T. 1992, \apjl, 395, L83, \dodoi{10.1086/186493}

\bibitem[{{Pang} {et~al.}(2023){Pang}, {Dietrich}, {Coughlin}, {Bulla}, {Tews}, {Almualla}, {Barna}, {Kiendrebeogo}, {Kunert}, {Mansingh}, {Reed}, {Sravan}, {Toivonen}, {Antier}, {VandenBerg}, {Heinzel}, {Nedora}, {Salehi}, {Sharma}, {Somasundaram}, \& {Van Den Broeck}}]{Pang23}
{Pang}, P. T.~H., {Dietrich}, T., {Coughlin}, M.~W., {et~al.} 2023, Nature Communications, 14, 8352, \dodoi{10.1038/s41467-023-43932-6}

\bibitem[{{Pian} {et~al.}(2017){Pian}, {D'Avanzo}, {Benetti}, {Branchesi}, {Brocato}, {Campana}, {Cappellaro}, {Covino}, {D'Elia}, {Fynbo}, {Getman}, {Ghirlanda}, {Ghisellini}, {Grado}, {Greco}, {Hjorth}, {Kouveliotou}, {Levan}, {Limatola}, {Malesani}, {Mazzali}, {Melandri}, {M{\o}ller}, {Nicastro}, {Palazzi}, {Piranomonte}, {Rossi}, {Salafia}, {Selsing}, {Stratta}, {Tanaka}, {Tanvir}, {Tomasella}, {Watson}, {Yang}, {Amati}, {Antonelli}, {Ascenzi}, {Bernardini}, {Bo{\"e}r}, {Bufano}, {Bulgarelli}, {Capaccioli}, {Casella}, {Castro-Tirado}, {Chassande-Mottin}, {Ciolfi}, {Copperwheat}, {Dadina}, {De Cesare}, {di Paola}, {Fan}, {Gendre}, {Giuffrida}, {Giunta}, {Hunt}, {Israel}, {Jin}, {Kasliwal}, {Klose}, {Lisi}, {Longo}, {Maiorano}, {Mapelli}, {Masetti}, {Nava}, {Patricelli}, {Perley}, {Pescalli}, {Piran}, {Possenti}, {Pulone}, {Razzano}, {Salvaterra}, {Schipani}, {Spera}, {Stamerra}, {Stella}, {Tagliaferri}, {Testa}, {Troja}, {Turatto}, {Vergani}, \& {Vergani}}]{Pian17}
{Pian}, E., {D'Avanzo}, P., {Benetti}, S., {et~al.} 2017, \nat, 551, 67, \dodoi{10.1038/nature24298}

\bibitem[{{Pozanenko} {et~al.}(2018){Pozanenko}, {Barkov}, {Minaev}, {Volnova}, {Mazaeva}, {Moskvitin}, {Krugov}, {Samodurov}, {Loznikov}, \& {Lyutikov}}]{Pozanenko18}
{Pozanenko}, A.~S., {Barkov}, M.~V., {Minaev}, P.~Y., {et~al.} 2018, \apjl, 852, L30, \dodoi{10.3847/2041-8213/aaa2f6}

\bibitem[{{Qian} \& {Wasserburg}(2007)}]{Qian&Wasserburg07}
{Qian}, Y.~Z., \& {Wasserburg}, G.~J. 2007, \physrep, 442, 237, \dodoi{10.1016/j.physrep.2007.02.006}

\bibitem[{{Radice} \& {Bernuzzi}(2023)}]{Radice23}
{Radice}, D., \& {Bernuzzi}, S. 2023, \apj, 959, 46, \dodoi{10.3847/1538-4357/ad0235}

\bibitem[{{Radice} {et~al.}(2020){Radice}, {Bernuzzi}, \& {Perego}}]{Radice20}
{Radice}, D., {Bernuzzi}, S., \& {Perego}, A. 2020, Annual Review of Nuclear and Particle Science, 70, 95, \dodoi{10.1146/annurev-nucl-013120-114541}

\bibitem[{Ralchenko {et~al.}(2021)Ralchenko, Olsen, Fontes, Fryer, Hungerford, Wollaeger, \& Korobkin}]{NIST-LANL}
Ralchenko, Y., Olsen, K., Fontes, C., {et~al.} 2021, NIST-LANL Lanthanide/Actinide Opacity Database,  National Institute of Standards and Technology, \dodoi{10.18434/MDS2-2375}

\bibitem[{{Risti{\'c}} {et~al.}(2023){Risti{\'c}}, {O'Shaughnessy}, {Villar}, {Wollaeger}, {Korobkin}, {Fryer}, {Fontes}, \& {Kedia}}]{Ristic23}
{Risti{\'c}}, M., {O'Shaughnessy}, R., {Villar}, V.~A., {et~al.} 2023, Physical Review Research, 5, 043106, \dodoi{10.1103/PhysRevResearch.5.043106}

\bibitem[{{Rosswog} {et~al.}(2017){Rosswog}, {Feindt}, {Korobkin}, {Wu}, {Sollerman}, {Goobar}, \& {Martinez-Pinedo}}]{Rosswog17}
{Rosswog}, S., {Feindt}, U., {Korobkin}, O., {et~al.} 2017, Classical and Quantum Gravity, 34, 104001, \dodoi{10.1088/1361-6382/aa68a9}

\bibitem[{{Rosswog} {et~al.}(1999){Rosswog}, {Liebend{\"o}rfer}, {Thielemann}, {Davies}, {Benz}, \& {Piran}}]{Rosswog99}
{Rosswog}, S., {Liebend{\"o}rfer}, M., {Thielemann}, F.~K., {et~al.} 1999, \aap, 341, 499, \dodoi{10.48550/arXiv.astro-ph/9811367}

\bibitem[{{Rosswog} {et~al.}(2018){Rosswog}, {Sollerman}, {Feindt}, {Goobar}, {Korobkin}, {Wollaeger}, {Fremling}, \& {Kasliwal}}]{Rosswog18}
{Rosswog}, S., {Sollerman}, J., {Feindt}, U., {et~al.} 2018, \aap, 615, A132, \dodoi{10.1051/0004-6361/201732117}

\bibitem[{{Roth} \& {Kasen}(2015)}]{Roth15}
{Roth}, N., \& {Kasen}, D. 2015, \apjs, 217, 9, \dodoi{10.1088/0067-0049/217/1/9}

\bibitem[{{Rouco Escorial} {et~al.}(2023){Rouco Escorial}, {Fong}, {Berger}, {Laskar}, {Margutti}, {Schroeder}, {Rastinejad}, {Cornish}, {Popp}, {Lally}, {Nugent}, {Paterson}, {Metzger}, {Chornock}, {Alexander}, {Cendes}, \& {Eftekhari}}]{RoucoEscorial23}
{Rouco Escorial}, A., {Fong}, W., {Berger}, E., {et~al.} 2023, \apj, 959, 13, \dodoi{10.3847/1538-4357/acf830}

\bibitem[{{Sarin} \& {Rosswog}(2024)}]{Sarin24}
{Sarin}, N., \& {Rosswog}, S. 2024, arXiv e-prints, arXiv:2404.07271, \dodoi{10.48550/arXiv.2404.07271}

\bibitem[{{Savchenko} {et~al.}(2017){Savchenko}, {Ferrigno}, {Kuulkers}, {Bazzano}, {Bozzo}, {Brandt}, {Chenevez}, {Courvoisier}, {Diehl}, {Domingo}, {Hanlon}, {Jourdain}, {von Kienlin}, {Laurent}, {Lebrun}, {Lutovinov}, {Martin-Carrillo}, {Mereghetti}, {Natalucci}, {Rodi}, {Roques}, {Sunyaev}, \& {Ubertini}}]{Savchenko17}
{Savchenko}, V., {Ferrigno}, C., {Kuulkers}, E., {et~al.} 2017, \apjl, 848, L15, \dodoi{10.3847/2041-8213/aa8f94}

\bibitem[{{Shappee} {et~al.}(2017){Shappee}, {Simon}, {Drout}, {Piro}, {Morrell}, {Prieto}, {Kasen}, {Holoien}, {Kollmeier}, {Kelson}, {Coulter}, {Foley}, {Kilpatrick}, {Siebert}, {Madore}, {Murguia-Berthier}, {Pan}, {Prochaska}, {Ramirez-Ruiz}, {Rest}, {Adams}, {Alatalo}, {Ba{\~n}ados}, {Baughman}, {Bernstein}, {Bitsakis}, {Boutsia}, {Bravo}, {Di Mille}, {Higgs}, {Ji}, {Maravelias}, {Marshall}, {Placco}, {Prieto}, \& {Wan}}]{Shappee17}
{Shappee}, B.~J., {Simon}, J.~D., {Drout}, M.~R., {et~al.} 2017, Science, 358, 1574, \dodoi{10.1126/science.aaq0186}

\bibitem[{{Shingles} {et~al.}(2023){Shingles}, {Collins}, {Vijayan}, {Fl{\"o}rs}, {Just}, {Leck}, {Xiong}, {Bauswein}, {Mart{\'\i}nez-Pinedo}, \& {Sim}}]{Shingles23}
{Shingles}, L.~J., {Collins}, C.~E., {Vijayan}, V., {et~al.} 2023, \apjl, 954, L41, \dodoi{10.3847/2041-8213/acf29a}

\bibitem[{{Simmerer} {et~al.}(2004){Simmerer}, {Sneden}, {Cowan}, {Collier}, {Woolf}, \& {Lawler}}]{Simmerer04}
{Simmerer}, J., {Sneden}, C., {Cowan}, J.~J., {et~al.} 2004, \apj, 617, 1091, \dodoi{10.1086/424504}

\bibitem[{{Smartt} {et~al.}(2017){Smartt}, {Chen}, {Jerkstrand}, {Coughlin}, {Kankare}, {Sim}, {Fraser}, {Inserra}, {Maguire}, {Chambers}, {Huber}, {Kr{\"u}hler}, {Leloudas}, {Magee}, {Shingles}, {Smith}, {Young}, {Tonry}, {Kotak}, {Gal-Yam}, {Lyman}, {Homan}, {Agliozzo}, {Anderson}, {Angus}, {Ashall}, {Barbarino}, {Bauer}, {Berton}, {Botticella}, {Bulla}, {Bulger}, {Cannizzaro}, {Cano}, {Cartier}, {Cikota}, {Clark}, {De Cia}, {Della Valle}, {Denneau}, {Dennefeld}, {Dessart}, {Dimitriadis}, {Elias-Rosa}, {Firth}, {Flewelling}, {Fl{\"o}rs}, {Franckowiak}, {Frohmaier}, {Galbany}, {Gonz{\'a}lez-Gait{\'a}n}, {Greiner}, {Gromadzki}, {Guelbenzu}, {Guti{\'e}rrez}, {Hamanowicz}, {Hanlon}, {Harmanen}, {Heintz}, {Heinze}, {Hernandez}, {Hodgkin}, {Hook}, {Izzo}, {James}, {Jonker}, {Kerzendorf}, {Klose}, {Kostrzewa-Rutkowska}, {Kowalski}, {Kromer}, {Kuncarayakti}, {Lawrence}, {Lowe}, {Magnier}, {Manulis}, {Martin-Carrillo}, {Mattila}, {McBrien}, {M{\"u}ller}, {Nordin}, {O'Neill}, {Onori}, {Palmerio}, {Pastorello},
  {Patat}, {Pignata}, {Podsiadlowski}, {Pumo}, {Prentice}, {Rau}, {Razza}, {Rest}, {Reynolds}, {Roy}, {Ruiter}, {Rybicki}, {Salmon}, {Schady}, {Schultz}, {Schweyer}, {Seitenzahl}, {Smith}, {Sollerman}, {Stalder}, {Stubbs}, {Sullivan}, {Szegedi}, {Taddia}, {Taubenberger}, {Terreran}, {van Soelen}, {Vos}, {Wainscoat}, {Walton}, {Waters}, {Weiland}, {Willman}, {Wiseman}, {Wright}, {Wyrzykowski}, \& {Yaron}}]{Smartt17}
{Smartt}, S.~J., {Chen}, T.~W., {Jerkstrand}, A., {et~al.} 2017, \nat, 551, 75, \dodoi{10.1038/nature24303}

\bibitem[{{Sneppen} \& {Watson}(2023)}]{Sneppen23a}
{Sneppen}, A., \& {Watson}, D. 2023, \aap, 675, A194, \dodoi{10.1051/0004-6361/202346421}

\bibitem[{{Sneppen} {et~al.}(2023){Sneppen}, {Watson}, {Gillanders}, \& {Heintz}}]{Sneppen23b}
{Sneppen}, A., {Watson}, D., {Gillanders}, J.~H., \& {Heintz}, K.~E. 2023, arXiv e-prints, arXiv:2312.02258, \dodoi{10.48550/arXiv.2312.02258}

\bibitem[{{Soares-Santos} {et~al.}(2017){Soares-Santos}, {Holz}, {Annis}, {Chornock}, {Herner}, {Berger}, {Brout}, {Chen}, {Kessler}, {Sako}, {Allam}, {Tucker}, {Butler}, {Palmese}, {Doctor}, {Diehl}, {Frieman}, {Yanny}, {Lin}, {Scolnic}, {Cowperthwaite}, {Neilsen}, {Marriner}, {Kuropatkin}, {Hartley}, {Paz-Chinch{\'o}n}, {Alexander}, {Balbinot}, {Blanchard}, {Brown}, {Carlin}, {Conselice}, {Cook}, {Drlica-Wagner}, {Drout}, {Durret}, {Eftekhari}, {Farr}, {Finley}, {Foley}, {Fong}, {Fryer}, {Garc{\'\i}a-Bellido}, {Gill}, {Gruendl}, {Hanna}, {Kasen}, {Li}, {Lopes}, {Louren{\c{c}}o}, {Margutti}, {Marshall}, {Matheson}, {Medina}, {Metzger}, {Mu{\~n}oz}, {Muir}, {Nicholl}, {Quataert}, {Rest}, {Sauseda}, {Schlegel}, {Secco}, {Sobreira}, {Stebbins}, {Villar}, {Vivas}, {Walker}, {Wester}, {Williams}, {Zenteno}, {Zhang}, {Abbott}, {Abdalla}, {Banerji}, {Bechtol}, {Benoit-L{\'e}vy}, {Bertin}, {Brooks}, {Buckley-Geer}, {Burke}, {Carnero Rosell}, {Carrasco Kind}, {Carretero}, {Castander}, {Crocce}, {Cunha}, {D'Andrea},
  {da Costa}, {Davis}, {Desai}, {Dietrich}, {Doel}, {Eifler}, {Fernandez}, {Flaugher}, {Fosalba}, {Gaztanaga}, {Gerdes}, {Giannantonio}, {Goldstein}, {Gruen}, {Gschwend}, {Gutierrez}, {Honscheid}, {Jain}, {James}, {Jeltema}, {Johnson}, {Johnson}, {Kent}, {Krause}, {Kron}, {Kuehn}, {Kuhlmann}, {Lahav}, {Lima}, {Maia}, {March}, {McMahon}, {Menanteau}, {Miquel}, {Mohr}, {Nichol}, {Nord}, {Ogando}, {Petravick}, {Plazas}, {Romer}, {Roodman}, {Rykoff}, {Sanchez}, {Scarpine}, {Schubnell}, {Sevilla-Noarbe}, {Smith}, {Smith}, {Suchyta}, {Swanson}, {Tarle}, {Thomas}, {Thomas}, {Troxel}, {Vikram}, {Wechsler}, {Weller}, {Dark Energy Survey}, \& {Dark Energy Camera GW-EM Collaboration}}]{Soares-Santos17}
{Soares-Santos}, M., {Holz}, D.~E., {Annis}, J., {et~al.} 2017, \apjl, 848, L16, \dodoi{10.3847/2041-8213/aa9059}

\bibitem[{{Sobolev}(1960)}]{Sobolev60}
{Sobolev}, V.~V. 1960, {Moving Envelopes of Stars}, \dodoi{10.4159/harvard.9780674864658}

\bibitem[{{Sugita} {et~al.}(2018){Sugita}, {Kawai}, {Nakahira}, {Negoro}, {Serino}, {Mihara}, {Yamaoka}, \& {Nakajima}}]{Sugita18}
{Sugita}, S., {Kawai}, N., {Nakahira}, S., {et~al.} 2018, \pasj, 70, 81, \dodoi{10.1093/pasj/psy076}

\bibitem[{{Symbalisty} \& {Schramm}(1982)}]{Symbalisty&Schramm82}
{Symbalisty}, E., \& {Schramm}, D.~N. 1982, \aplett, 22, 143

\bibitem[{{Tanaka} {et~al.}(2020){Tanaka}, {Kato}, {Gaigalas}, \& {Kawaguchi}}]{Tanaka20}
{Tanaka}, M., {Kato}, D., {Gaigalas}, G., \& {Kawaguchi}, K. 2020, \mnras, 496, 1369, \dodoi{10.1093/mnras/staa1576}

\bibitem[{{Tanaka} {et~al.}(2018){Tanaka}, {Kato}, {Gaigalas}, {Rynkun}, {Rad{\v{z}}i{\={u}}t{\.{e}}}, {Wanajo}, {Sekiguchi}, {Nakamura}, {Tanuma}, {Murakami}, \& {Sakaue}}]{Tanaka18}
{Tanaka}, M., {Kato}, D., {Gaigalas}, G., {et~al.} 2018, \apj, 852, 109, \dodoi{10.3847/1538-4357/aaa0cb}

\bibitem[{{Tanvir} {et~al.}(2017){Tanvir}, {Levan}, {Gonz{\'a}lez-Fern{\'a}ndez}, {Korobkin}, {Mandel}, {Rosswog}, {Hjorth}, {D'Avanzo}, {Fruchter}, {Fryer}, {Kangas}, {Milvang-Jensen}, {Rosetti}, {Steeghs}, {Wollaeger}, {Cano}, {Copperwheat}, {Covino}, {D'Elia}, {de Ugarte Postigo}, {Evans}, {Even}, {Fairhurst}, {Figuera Jaimes}, {Fontes}, {Fujii}, {Fynbo}, {Gompertz}, {Greiner}, {Hodosan}, {Irwin}, {Jakobsson}, {J{\o}rgensen}, {Kann}, {Lyman}, {Malesani}, {McMahon}, {Melandri}, {O'Brien}, {Osborne}, {Palazzi}, {Perley}, {Pian}, {Piranomonte}, {Rabus}, {Rol}, {Rowlinson}, {Schulze}, {Sutton}, {Th{\"o}ne}, {Ulaczyk}, {Watson}, {Wiersema}, \& {Wijers}}]{Tanvir17}
{Tanvir}, N.~R., {Levan}, A.~J., {Gonz{\'a}lez-Fern{\'a}ndez}, C., {et~al.} 2017, \apjl, 848, L27, \dodoi{10.3847/2041-8213/aa90b6}

\bibitem[{{Troja} {et~al.}(2017){Troja}, {Piro}, {van Eerten}, {Wollaeger}, {Im}, {Fox}, {Butler}, {Cenko}, {Sakamoto}, {Fryer}, {Ricci}, {Lien}, {Ryan}, {Korobkin}, {Lee}, {Burgess}, {Lee}, {Watson}, {Choi}, {Covino}, {D'Avanzo}, {Fontes}, {Gonz{\'a}lez}, {Khandrika}, {Kim}, {Kim}, {Lee}, {Lee}, {Kutyrev}, {Lim}, {S{\'a}nchez-Ram{\'\i}rez}, {Veilleux}, {Wieringa}, \& {Yoon}}]{Troja17}
{Troja}, E., {Piro}, L., {van Eerten}, H., {et~al.} 2017, \nat, 551, 71, \dodoi{10.1038/nature24290}

\bibitem[{{Utsumi} {et~al.}(2017){Utsumi}, {Tanaka}, {Tominaga}, {Yoshida}, {Barway}, {Nagayama}, {Zenko}, {Aoki}, {Fujiyoshi}, {Furusawa}, {Kawabata}, {Koshida}, {Lee}, {Morokuma}, {Motohara}, {Nakata}, {Ohsawa}, {Ohta}, {Okita}, {Tajitsu}, {Tanaka}, {Terai}, {Yasuda}, {Abe}, {Asakura}, {Bond}, {Miyazaki}, {Sumi}, {Tristram}, {Honda}, {Itoh}, {Itoh}, {Kawabata}, {Morihana}, {Nagashima}, {Nakaoka}, {Ohshima}, {Takahashi}, {Takayama}, {Aoki}, {Baar}, {Doi}, {Finet}, {Kanda}, {Kawai}, {Kim}, {Kuroda}, {Liu}, {Matsubayashi}, {Murata}, {Nagai}, {Saito}, {Saito}, {Sako}, {Sekiguchi}, {Tamura}, {Tanaka}, {Uemura}, \& {Yamaguchi}}]{Utsumi17}
{Utsumi}, Y., {Tanaka}, M., {Tominaga}, N., {et~al.} 2017, \pasj, 69, 101, \dodoi{10.1093/pasj/psx118}

\bibitem[{{Valenti} {et~al.}(2017){Valenti}, {Sand}, {Yang}, {Cappellaro}, {Tartaglia}, {Corsi}, {Jha}, {Reichart}, {Haislip}, \& {Kouprianov}}]{Valenti17}
{Valenti}, S., {Sand}, D.~J., {Yang}, S., {et~al.} 2017, \apjl, 848, L24, \dodoi{10.3847/2041-8213/aa8edf}

\bibitem[{{Villar} {et~al.}(2017){Villar}, {Guillochon}, {Berger}, {Metzger}, {Cowperthwaite}, {Nicholl}, {Alexander}, {Blanchard}, {Chornock}, {Eftekhari}, {Fong}, {Margutti}, \& {Williams}}]{Villar17}
{Villar}, V.~A., {Guillochon}, J., {Berger}, E., {et~al.} 2017, \apjl, 851, L21, \dodoi{10.3847/2041-8213/aa9c84}

\bibitem[{{Wallner} {et~al.}(2015){Wallner}, {Faestermann}, {Feige}, {Feldstein}, {Knie}, {Korschinek}, {Kutschera}, {Ofan}, {Paul}, {Quinto}, {Rugel}, \& {Steier}}]{Wallner15}
{Wallner}, A., {Faestermann}, T., {Feige}, J., {et~al.} 2015, Nature Communications, 6, 5956, \dodoi{10.1038/ncomms6956}

\bibitem[{{Wollaeger} {et~al.}(2018){Wollaeger}, {Korobkin}, {Fontes}, {Rosswog}, {Even}, {Fryer}, {Sollerman}, {Hungerford}, {van Rossum}, \& {Wollaber}}]{Wollaeger18}
{Wollaeger}, R.~T., {Korobkin}, O., {Fontes}, C.~J., {et~al.} 2018, \mnras, 478, 3298, \dodoi{10.1093/mnras/sty1018}

\bibitem[{{Wollaeger} {et~al.}(2021){Wollaeger}, {Fryer}, {Chase}, {Fontes}, {Ristic}, {Hungerford}, {Korobkin}, {O'Shaughnessy}, \& {Herring}}]{Wollaeger21}
{Wollaeger}, R.~T., {Fryer}, C.~L., {Chase}, E.~A., {et~al.} 2021, \apj, 918, 10, \dodoi{10.3847/1538-4357/ac0d03}

\bibitem[{{Zhu} {et~al.}(2021){Zhu}, {Lund}, {Barnes}, {Sprouse}, {Vassh}, {McLaughlin}, {Mumpower}, \& {Surman}}]{Zhu21}
{Zhu}, Y.~L., {Lund}, K.~A., {Barnes}, J., {et~al.} 2021, \apj, 906, 94, \dodoi{10.3847/1538-4357/abc69e}

\end{thebibliography}
\bibliographystyle{aasjournal}



\end{document}